\pdfoutput=1

\documentclass[11pt]{article}
\usepackage{amsmath}
\usepackage{amsfonts,amssymb}
\usepackage{multirow,cite}
\usepackage{float}
\usepackage{scalerel}
\usepackage[dvipsnames]{xcolor}
\usepackage{datetime}
\usepackage{eso-pic}
\usepackage{cancel}
\usepackage{caption}
\usepackage{hyperref}

\topmargin - 0.5 in

\def\a{\alpha}
\def\b{\beta}
\def\d{\delta}
\def\e{\epsilon}
\def\g{\gamma}
\def\s{\sigma}
\def\l{\lambda}
\def\k{\kappa}
\def\om{\omega}
\def\p{\partial}
\def\r{\rightarrow}
\def\D{\Delta}
\def\O{\mathcal{O}}

\def\L{\mathcal{L}}

\newcommand{\bi}{\begin{itemize}}
\newcommand{\ei}{\end{itemize}}

\newcommand{\be}{\begin{equation}}
\newcommand{\ee}{\end{equation}}

\newcommand{\bea}{\begin{eqnarray}}
\newcommand{\eea}{\end{eqnarray}}

\numberwithin{equation}{section}
\makeatletter
\renewcommand{\@seccntformat}[1]{%
  \csname the#1\endcsname.\ }
\makeatother

\title{Infinite $\mathrm{T\bar T}$-like symmetries of compactified LST\vspace{5mm}}

\author{Silvia Georgescu$^{\S, \dag}$ and
Monica Guica$^\S$ \vspace{1mm} \\
\\\vspace{1mm}
${}^\S$\emph{\small Universit\'e Paris-Saclay, CNRS, CEA,} 
\emph{\small Institut de Physique Th\'eorique, 91191 Gif-sur-Yvette, France} \\
${}^\dag$\emph{\small 
CPHT, CNRS, \'Ecole polytechnique, Institut Polytechnique de Paris, 91120 Palaiseau, France}}

\date{}

\begin{document}


\maketitle

\abstract{
\vskip 3mm

\noindent We show that the three-dimensional asymptotically linear dilaton background that arises in the near-horizon decoupling region of  NS5-branes compactified on $T^4$  admits boundary conditions that lead to  an infinite set of symmetries. The associated conserved charges, which implement field-dependent coordinate transformations, are found to be identical  to the corresponding generators in a symmetric product orbifold of $T\bar T$ - deformed CFTs.   Their algebra is a non-linear modification of the $\mathrm{Virasoro \times Virasoro}$ algebra, which precisely coincides with the   algebra  of the ``unrescaled'' symmetry generators in $T\bar T$-deformed CFTs.  This further strengthens a previously proposed link between the single-trace $T\bar T$ deformation and compactified little string theory.  

}

\tableofcontents

\section{Introduction}

One of the most important open problems  in quantum gravity is to generalize the highly successful AdS/CFT correspondence \cite{Maldacena:1997re} to other spacetimes, such as the more realistic asymptotically flat ones. While a fascinating link \cite{Strominger:2013jfa,He:2014laa} has been recently uncovered  between the rich  asymptotic structure of  Minkowski spacetimes \cite{Bondi:1962px,Sachs:1962wk} and properties of flat space scattering amplitudes \cite{Weinberg:1965nx} -  culminating in the celestial holography program \cite{Pasterski:2016qvg,McLoughlin:2022ljp,Raclariu:2021zjz,Pasterski:2021rjz} - the underlying quantum-field-theoretical structure of the holographic dual  to flat space has yet to be understood. 
Basic considerations based on e.g.,  the behaviour of  black hole 
   and entanglement  entropy \cite{Li:2010dr}, as well as the structure of correlation functions \cite{Marolf:2006bk}, suggest the corresponding QFT is non-local.

 Part of the difficulty lies in the lack of tractable string-theoretical examples of flat holography. In this article, we will concentrate on the asymptotically linear dilaton spacetime in string theory \cite{Callan:1991at}, where the
metric is asymptotically flat, but the dilaton has a non-trivial asymptotic profile. This spacetime is obtained from a decoupling limit of the geometry created by  NS5-branes and can be viewed as intermediary between the well-understood anti-de Sitter case and the  challenging asymptotically flat one; for example, the growth  of the black hole entropy with energy is Hagedorn, which is intermediary between the sub-Hagedorn growth in AdS and the super-Hagedorn one in flat space. The conjectured holographic dual to string theory in this background is  known as little string theory (LST) \cite{Seiberg:1997zk,Aharony:1998ub,Aharony:1999ks,Kutasov:2001uf}: a non-local, non-gravitational theory containing strings, which arises in the low-energy decoupling limit of the  NS5-branes.

The behaviour of observables in  LST has traditionally been inferrred from  its holographically dual spacetime. However, more recently \cite{Giveon:2017nie} put forth a very interesting  alternate QFT definition of little string theory, at least when compactified on $T^4$, in terms of a solvable irrelevant deformation of a two-dimensional CFT known as  the $T\bar T$ deformation. 

The  $T\bar T$ deformation \cite{Smirnov:2016lqw,Cavaglia:2016oda} is  a universal deformation of two-dimensional QFTs by an operator that is bilinear in the stress tensor components. Remarkably, this deformation appears well-defined - despite its irrelevant nature - and leads to a theory that is UV complete \cite{Dubovsky:2013ira}, albeit non-local at the scale set by the dimensionful deformation parameter. Moreover, many observables in the deformed theory, such as  the finite-size spectrum and the deformed S-matrix, are entirely determined by their counterparts in the original QFT \cite{Smirnov:2016lqw,Cavaglia:2016oda,Dubovsky:2017cnj,Dubovsky:2018bmo}. For the case of a $T\bar T$ - deformed  \emph{CFT}, the deformed spectrum is known in closed form and can be used to 
determine  the density of states, $e^S$, as a function of the energy

\be
S(E)= 2\pi \sqrt{\frac{c E R}{3} + \frac{\mu c}{6 \pi} E^2}  \label{relenteng}
\ee
where $c$ is the central charge of the undeformed CFT and $\mu$ is the  $T\bar T$ parameter. This expression interpolates between a Cardy regime at low energies and
 Hagedorn behaviour  at high energies, which is suggestive of a  connection to strings. 

The original $T\bar T$ deformation put forth by Smirnov and Zamolodchikov is ``double-trace'' from the perspective of AdS/CFT. As a result, its holographic interpretation is to simply  change the boundary conditions on the  dual AdS$_3$ metric from Dirichlet to mixed \cite{mirage}, without affecting the local curvature. In order for the irrelevant deformation to locally modify the geometry of the dual spacetime, one needs  to consider instead the so-called ``single-trace'' $T\bar T$ deformation \cite{Giveon:2017nie}, which is possible when the undeformed CFT is a symmetric product orbifold (SPO) of some seed CFT. This deformation takes the form 
\be
 \sum_{i=1}^p T_i \bar T_i \label{stttb}
\ee
where $p$ is the number of copies in the SPO, and produces a symmetric product orbifold of $T\bar T$- deformed  CFTs,  many of whose properties are  universally determined  by the original seed \cite{ttbspo}. 

The connection between single-trace $T\bar T$ and LST arises from considering the NS5-F1 system instead of just NS5 \cite{Giveon:1999zm}. In this case, the decoupled background  geometry interpolates between AdS$_3$ in the IR and a three-dimensional linear dilaton spacetime in the UV. Worldsheet string theory is well understood in this entire background \cite{Forste:1994wp}; in particular,  the IR AdS$_3$ region is described by an $SL(2,\mathbb{R})$ WZW model \cite{Maldacena:2000hw}. Holographically, 
it is expected \cite{Argurio:2000tb} that 
 at least the long string sector of this theory should be described  by a symmetric product orbifold with respect to the number of F1 strings, $p$, of a seed CFT of central charge $c=6 k$, where  $k$ is the number  of NS5 branes. In \cite{Giveon:2017nie}, it  has been argued that turning on the single-trace $T\bar T$ deformation \eqref{stttb} in this theory corresponds precisely  to the deformation of the dual background from AdS$_3$ to asymptotically linear dilaton. The $T\bar T$ coupling is proportional to the inverse string tension, $\a'$. 


If correct, this proposal provides an alternate QFT definition of compactified LST   in terms of an essentially solvable theory.  This proposal has passed several non-trivial checks, such as the fact that black hole entropy precisely follows the $T\bar T$ relation \eqref{relenteng} between entropy and energy \cite{Giveon:2017nie,Chakraborty:2020swe}, and that the spectrum of long strings in the deformed massless BTZ background precisely matches the single-trace $T\bar T$ - deformed spectrum \cite{Giveon:2017myj}. It has also made some interesting predictions \cite{Chakraborty:2018kpr,Asrat:2020uib} for the behaviour of entanglement entropy in $T\bar T$ - deformed CFTs. Its ultimate status is, however, somewhat unclear, due to the fact that the full CFT dual to the IR AdS$_3$ is \emph{not} a symmetric product orbifold (see  \cite{Eberhardt:2021vsx} for a recently proposed dual).  More work is thus needed before the full picture emerges.

  %
  %
%

 In this article, we  further strengthen the link between the single-trace $T\bar T$ deformation and compactified LST by showing that the asymptotically linear dilaton background dual to the latter possesses an infinite set of symmetries, whose form and algebra \emph{precisely match} those of a single-trace $T\bar T$ - deformed CFT \cite{ttbspo}.  These symmetries are a straightforward generalization of the infinite symmetries of standard $T\bar T$ - deformed CFTs, which have been first predicted by the holographic analysis of \cite{mirage}, then studied at classical level in \cite{Guica:2020uhm}, proven to exist at full quantum level in \cite{Guica:2021pzy}, and are further explored in \cite{clsqsymmttb}.

More specifically, \cite{mirage}  showed that the asymptotic symmetry algebra of the spacetime dual to $T\bar T$-deformed CFTs - namely,   AdS$_3$ with mixed boundary conditions - is Virasoro $\times$ Virasoro with the same central extension as that of the undeformed CFT.  The asymptotic symmetry generators are parametrised by two arbitrary functions of certain ``field-dependent coordinates'', $u,v$, whose radii depend on the left- and right-moving energies of the background as $R_{u,v} = R + \mu H_{R,L}/\pi$, 
%
%
where $R$ is the radius of the circle on which the theory is defined. The Virasoro generators $L_m, \bar L_m$  correspond to the natural Fourier basis of functions \emph{multiplied} by the field-dependent radii. Writing  $L_m = R_u Q_m$, $\bar L_m = R_v \bar Q_m$, the algebra of the  ``unrescaled'' generators  $Q_m, \bar Q_m$  is a non-linear modification of the Virasoro $\times$ Virasoro algebra


\be
i \{ Q_m, Q_n \} = \frac{1}{R_u} (m-n) Q_{m+n} + \frac{\mu^2 H_R}{\pi^2 R R_H R_u} (m-n) Q_m Q_n + \frac{c}{12}\, \frac{m^3}{ R_u^2} \d_{m+n} \nonumber
\ee
\be
i \{  Q_m, \bar Q_n \} = - \frac{\mu (m-n)}{\pi R R_H} Q_m \bar Q_n \label{ttbalg}
\ee
where $R_H = R + \mu H/\pi$; a similar bracket holds for $\{ \bar Q_m, \bar Q_n \}$. 
%
Even though this  algebra is more complicated  and likely receives quantum corrections to all orders in $\hbar$, the ``unrescaled'' basis of generators: i) is very natural to consider, being just the Fourier basis; in particular, the left/right-moving energies simply correspond to $Q_0, \bar Q_0$ and ii) may be preferred for defining pseudo-local observables in the theory, in analogy to the case of $J^1 \wedge J^2$ and $J\bar T$ - deformed CFTs\footnote{ The Virasoro generators of $T\bar T$ - deformed CFTs are obtained at full quantum level by transporting the ones of the seed CFT along the $T\bar T$ flow \cite{Guica:2021pzy}. The same procedure yields  flowed generators in $J^1 \wedge J^2$ and $J\bar T$ - deformed CFTs, whose algebra is Virasoro 
by construction, but which are explicitly
 %
%
 \emph{different} from 
 the Virasoro  generators of  conformal symmetries  in these theories. 
The      interplay between physical and flowed generators is essential for defining analogues of primary operators in the non-local $J\bar T$-deformed CFTs and for computing their correlation functions \cite{Guica:2021fkv}. 
}.
It is not hard to argue \cite{ttbspo} that for single-trace $T\bar T$ - deformed CFTs, the  algebra of the corresponding ``unrescaled'' generators is still given by the above, but with the replacement $\mu \r \mu/p$. 

The main technical result of  this article is to  show that the asymptotic symmetry algebra of asymptotically linear dilaton spacetimes is precisely \eqref{ttbalg} with $\mu \r \pi \a'/p$. To capture the field-dependence of the asymptotic symmetry generators, we perform the asymptotic analysis around non-trivial black hole solutions in these spacetimes. Due to the strong background dependence of the asymptotic diffeomorphisms, we can only compute the charge algebra \emph{perturbatively} around the black hole backgrounds, where the perturbation corresponds to adding boundary gravitons. The terms we  compute are however sufficient to see the entire structure of \eqref{ttbalg} emerge.

This article is organised as follows. In section \ref{setup}, we review the black hole backgrounds we study, present a consistent truncation to three dimensions that captures them, and recall the relationship between their thermodynamics   and that of single-trace $T\bar T$. In section \ref{asyanalbh}, we classify the general linearized perturbations of these backgrounds and use their symplectic product to motivate a set of boundary conditions  that lead to an infinite set of asymptotic symmetries. We also show that the associated conserved charges take an identical form to those in $T\bar T$.  In section \ref{asychalg}, we present the perturbative  computation of the charge algebra  around the black hole backgrounds. Finally, in appendix \ref{conschdtrttb} we revisit
the calculation of the asymptotic symmetry algebra for the spacetimes dual to double-trace $T\bar T$ - deformed CFTs, explaining how to reproduce both  the algebra \eqref{ttbalg} and the originally found Virasoro using covariant phase space methods. 

\section{Setup \label{setup}}

In this section, we  review the construction of the asymptotically linear dilaton back hole backgrounds from a decoupling limit of string theory, and present a consistent truncation to three dimensions that captures their essential properties.  We then review the link between the thermodynamics of these black holes  and that of $T\bar T$ - deformed CFTs. 

\subsection{The backgrounds}

We start by considering the non-extremal NS5 - F1 string frame solution of type IIB string theory, with $n$ units of momentum along the common $S^1$ \cite{Maldacena:1996ky,Chakraborty:2020swe}

\be
ds^2 = \frac{1}{f_1} \left[  d\s^2 - dt^2 + \frac{r_0^2}{r^2} (\cosh \a_n dt + \sinh \a_n d\s)^2 \right]  +f_5 \left( \frac{dr^2}{f} + r^2 d \Omega_3^2\right) + \sum_{i=1}^4 dx_i^2 \nonumber
\ee
%
\be
H = \frac{ g_s^2 \a' p}{v} \,  d\s\wedge dt \wedge d \left(\frac{1}{ f_1 r^2} \right)+ 2 \a' k \, \omega_{S^3}\;, \;\;\;\;\; e^{2\phi} = g_s^2 \,\frac{ f_5}{f_1 } \label{bckgndsol}
\ee
where $d\Omega_3^2$ and $\om_{S^3}$ are the metric and, respectively, the volume form of the unit $S^3$, and  the various harmonic functions are given by
\be
f= 1 - \frac{r_0^2}{r^2} \;, \;\;\;\;\; f_{1,5,n} = 1 + \frac{r_0^2 \sinh^2 \a_{1,5,n}}{r^2}
\ee
with
\be
 \sinh 2 \a_1 = \frac{2 g_s^2 \a' p}{v r_0^2} \;, \;\;\;\;\; \sinh 2 \a_5 = \frac{2 \a' k}{r_0^2} \;, \;\;\;\;\; \sinh 2 \a_n = \frac{2 g_s^2 \a'^2 n}{R^2 v r_0^2}
\ee
Here, $p$ and $k$ are the number of F1 strings and, respectively,  NS5 branes supporting the solution, $g_s$ is the ten-dimensional string coupling, $R$ is the radius of the $S^1$ parametrized by $\s$ and $v$ is the volume of the $T^4$ in string units, $V_{T^4} = (2\pi)^4 \a'^2 v$. This way of writing the metric  makes manifest the boost that generated  the momentum of the solution.

We now take the standard NS5-brane decoupling limit, where the asymptotic string coupling $g_s \r 0 $, with 
\be
\hat r_0 \equiv \frac{r_0}{g_s } \;, \;\;\;\;\; \hat r \equiv \frac{r}{g_s }
\ee 
held fixed. We also introduce the  null coordinates $U, V = \s \pm t$, which are identified mod $2\pi R$. In this limit, the six dimensional string frame decoupled geometry becomes
\be
ds^2 = \frac{1}{f_1} \left( dU dV + \frac{\hat r_0^2}{4 \hat r^2} (e^{\a_n} dU - e^{-\a_n} d V)^2\right) + \a' k \left( \frac{d\hat r^2}{\hat r^2 - \hat r_0^2} + d \Omega_3^2\right) \;, \;\;\;\;e^{2\phi} = \frac{k \a'}{f_1 \hat r^2 } 
\ee
and 
\be
H = \frac{2p \, e^{2\phi}}{v \sqrt{\a' k^3}} \, \om_3 + 2 k \a' \om_{S^3} \label{decH}
\ee
where $\om_3$ is the volume form on the non-compact part of the space-time. This background interpolates between AdS$_3$ as $\hat r \r 0$ and an asymptotically flat spacetime with a linear dilaton as $\hat r \r \infty$. The AdS$_3$ geometry in the IR can also be obtained from the standard $\a' \r 0$ decoupling limit. As such, it is useful 
 to work in coordinates in which this limit is smooth, by letting
\be
\hat r =  \a' \rho  \;, \;\;\;\;\;\; \hat r_0 =  \a' \rho_0
\ee
 Since the metric only depends on the ratio $\hat r_0/\hat r = \rho_0/\rho$, it will take exactly the same form in terms of the new variables, and so will $\om_3$; only $e^{2\phi}$ will pick up a factor of $(\a')^{-2}$, as will the first term in $H$. Finally, it is  useful to introduce a new coordinate $r$ (different from the previous one) and notation
\be 
\rho =  r + \frac{\rho_0^2}{4  r} \;, \;\;\;\;\;\; L_u \equiv \frac{\rho_0^2\,  e^{2\a_n}}{4} \;, \;\;\;\;  L_v \equiv \frac{\rho_0^2 \, e^{-2\a_n}}{4}
\ee
in terms of which the decoupled metric and dilaton become 
\be
ds^2 = \frac{r^2}{ r^4 + \b r^2/(\a' v ) + L_u L_v} \left(  r^2 dU dV + L_u dU^2 + L_v dV^2 + \frac{L_u L_v}{r^2} dU dV\right) + \a' k \left( \frac{d r^2}{ r^2 } + d \Omega_3^2\right)  \nonumber
\ee
\be
e^{2\phi} =  \frac{k r^2}{ \a' ( r^4 + \b r^2/(v\a') + L_u L_v)}  
\ee
where the quantity $\b$ is defined as 
\be
\frac{\b}{\a' v} \equiv \rho_0^2 \left(\sinh^2 \a_1 + \frac{1}{2}\right) = \frac{\rho_0^2}{2} \sqrt{\left( \frac{2p}{v \rho_0^2 \a'}\right)^2+1} = \frac{1}{ \a' v} \sqrt{ p^2 + 4 \a'^2 v^2 L_u L_v }
\ee
The reason for pulling out a factor of $\a' v$ is that we would like $\b$ to have   a finite limit as $\a' \r 0$. In the second step, we  used the definition of $\a_1$, $\sinh 2 \a_1 = \frac{2p}{v \rho_0^2 \a'}$, to solve for $\a_1$ in terms of $\rho_0$ and the relation $\rho_0^4 = 16 L_u L_v$ to write $\b$ in terms of these variables. 

The energy and momentum of these solutions  are given by  \cite{Chakraborty:2020swe}
\be
E = \frac{R v \rho_0^2}{2} (e^{-2 \a_1 } + \cosh 2 \a_n) = R v \left(L_u + L_v  + \sqrt{\left( \frac{p}{v \a'}\right)^2  + 4 L_u L_v} - \frac{ p}{v \a'}\right) \nonumber
\ee

\be
P = \frac{n}{R} = Rv (L_u - L_v) \label{engmombh}
\ee
where we measure the energy with respect to deformed massless BTZ. One last useful manipulation is to absorb the factor of $v$ above into a redefinition of the radial coordinate and of  $L_u, L_v$. More precisely, we let 
\be
r^2 \r \frac{r'^2}{v} \;, \;\;\;\;\;\; L_u \r \frac{L_u'}{v} \;, \;\;\;\;\; L_v \r \frac{L_v'}{v}
\ee
and then drop the primes. This will simply amount to dropping the factors of $v$ from the metric and from the  formulae for the energy and momentum. To  remove this factor also from the dilaton, one can simply define 
\be
e^{2\phi'} = \frac{1}{v}\,  e^{2\phi} \label{phiresc}
\ee
The decoupling limit $\a' \r 0$ can now be straightforwardly taken, taking into account the fact that the metric scales with an overall factor of $\a'$, as usual. The six-dimensional string coupling reduces to a constant, $g^2_6=k/p$, which can be made as small as desired by taking the number of F1 strings to be large.  To further study  these solutions, it is convenient to introduce a consistent truncation to three dimensions that  captures all of their essential properties.

\subsection{Consistent truncation to three dimensions \label{trunc}}

The six-dimensional string frame action is 
\be
S= \frac{1}{2\kappa_6^2}  \int d^6 x \, \sqrt{g} \, e^{-2\phi} \left( R + 4 \p_\mu \phi \p^\mu \phi - \frac{1}{12} H^2 \right)  \;, \;\;\;\;\; \frac{1}{2\kappa_6^2} = \frac{(2\pi)^4 v \a'^2 }{2 \kappa_{10}^2} = \frac{v}{(2\pi)^3 \a'^2}
\ee
and the associated equations of motion read 
\be
R_{MN} + 2 \nabla_M \nabla_N \phi - \frac{1}{4} H_{MAB} H_N{}^{AB}=0 \;, \;\;\;\;\;\; \Box \phi - 2 (\p \phi)^2 = -\frac{1}{12}  H^2 \;, \;\;\;\;\; d(e^{-2\phi} \star H) =0 \label{stringeom}
\ee
We would like to perform a consistent truncation to three dimensions  on  $S^3$  using the following simple Ansatz, inspired by the  family of solutions we described

\be
ds^2 = ds_3^2 + \ell^2 ds_{S_3}^2 \;, \;\;\;\;\;\; H = 2 \ell^2 \, \omega_{S^3} + b\, e^{2\phi} \,  \omega_{3}  
\ee
where $\ell, b$ are constants and $\phi$ is assumed to be  a function of only the three-dimensional non-compact coordinates, $x^\mu$.  Comparing with the background $H$-field \eqref{decH},  we find

\be
\ell= \sqrt{\a' k} \;, \;\;\;\;\;\; b = 
 \frac{2 p}{ \sqrt{\a' k^3}} = \frac{2p}{k \ell}
\ee
Note there should be an additional factor of $v$ in the denominator of $b$;  we  have removed it by rescaling $e^\phi$ as in \eqref{phiresc}. This rescaling also removes the factor of $v$ from the six-dimensional Newton's constant. However, it does not affect the string equations of motion \eqref{stringeom}, since $\phi$ itself only shifts by a constant.

Thus, $\ell$ and $b$ are fixed parameters, depending only on the number of branes and $\alpha'$, but not on the state-dependent properties of the black hole solutions. 
Performing the truncation is trivial: the Ansatz for the $H$-field automatically solves the last equation in \eqref{stringeom}, while the middle equation reads 

\be
\Box_3 \phi - 2 (\p\phi)^2 = - \frac{1}{2} (a^2 - b^2 e^{4\phi})
\ee
where we have introduced the shorthand $a \equiv  2/\ell$.  The sphere components of the Einstein equations set ${}^{(3)}R_{ab} = \frac{2}{\ell^2} g_{ab}$, and 

\be
{}^{(3)}R_{\mu\nu} + 2 \nabla_\mu \nabla_\nu \phi + \frac{1}{2} b^2 e^{4\phi} g_{\mu\nu} =0
\ee
One can easily check that the black hole solutions  satisfy these equations.  To derive them from an action, we can introduce $\tilde g_{\mu\nu} = e^{-4 \phi} g_{\mu\nu}$, which obeys
\be
\tilde R_{\mu\nu} + \left(a^2 e^{4\phi} - \frac{b^2}{2} e^{8\phi} \right) \tilde g_{\mu\nu} - 4 \p_\mu \phi \p_\nu \phi =0 \;, \;\;\;\;\; \tilde \Box \phi = - \frac{1}{2} e^{4\phi} (a^2 -b^2 e^{4\phi})
\ee
These equations can be derived from the action  
\bea
S &= & \frac{1}{16 \pi G_3}\int d^3 x \sqrt{\tilde g} \left[\tilde R - 4 (\tilde \p\phi)^2 + \left(a^2 - \frac{b^2}{2} e^{4\phi}\right) e^{4\phi}\right] \nonumber\\
&= & \frac{1}{16 \pi G_3} \int d^3 x \sqrt{g}\,  e^{-2\phi} \left(R +  4 (\p \phi)^2 +a^2 - \frac{b^2}{2} e^{4\phi} \right) \label{3dact}
\eea
where 
$16 \pi G_3 = 4\pi{\sqrt{\a' k^{-3}}}  $. One may drop the  factors of $\sqrt{\a'}$ from $a$ (or $\ell ),b$ and $G_3$ by rescaling the metric by  $\a'$, as we will do in  the next section.

\subsection{Relationship to $T\bar T$ - deformed CFTs}

As mentioned in the introduction, 
the finite-size spectrum of a $T\bar T$ - deformed CFT is entirely determined by that of the original CFT. The relationship between the undeformed and deformed energies takes the simple form

\be
E_L^{(0)}  = E_L \left( 1 +   \frac{\mu  E_R}{\pi R}\right) \;, \;\;\;\;\;\; E_R^{(0)}  = E_R \left(1 +  \frac{\mu E_L}{\pi R}\right) \label{reledefud}
\ee
where $E_{L,R} = (E \pm P)/2$ and $E_{L,R}^{(0)}$ are their undeformed counterparts. This relationship can be used to compute the entropy in $T\bar T$ - deformed CFTs, using the fact that the number of energy levels  does not change with the deformation. Consequently, the entropy is still given by Cardy's formula 
\be
S_{Cardy} = 2\pi \sqrt{\frac{c\, E_L^{(0)}R}{6}} + 2\pi \sqrt{\frac{c \, E_R^{(0)}R}{6}} 
\ee
when written in terms of $E_{L,R}^{(0)}$. To obtain the entropy as a function of the physical energies $E_{L,R}$, one simply needs to plug in the expression \eqref{reledefud} into the formula above, obtaining

\be
S_{T\bar T} =  2\pi \sqrt{\frac{c  E_L (R + \mu E_R/\pi) }{6}} + 2\pi \sqrt{\frac{c  E_R(R+ \mu E_L/\pi)}{6}}
\ee
This analysis holds for the standard, double-trace $T\bar T$ deformation. The ``single-trace'' $T\bar T$ deformation relevant to our case corresponds to a symmetric product orbifold, $S_p$,  of $T\bar T$ - deformed CFTs, 
and its (untwisted sector)  entropy is obtained by equipartitioning the energy between the $p$ copies \cite{Giveon:2017nie}
%
\be
S_{T\bar T/S_p}=  2\pi \sqrt{\frac{c  E_L (R\, p + \mu E_R/\pi) }{6}} + 2\pi \sqrt{\frac{c  E_R(R \, p+ \mu E_L/\pi)}{6}}
\ee
where $c$ is the central charge of the seed CFT.

One of the  interesting observations of \cite{Giveon:2017nie} was that the relationship between the entropy and the energy of the asymptotically linear dilaton black holes reviewed in the previous section takes precisely this form, where $p$ is given by the number of F1 strings supporting the background and the central charge of the seed CFT is $c=6k$. Concretely, the entropy of the decoupled black holes is 
\be
S = 2 \pi R v  \ell_s \sqrt{k}  \, \rho_0^2 \cosh \a_1 \cosh \a_n
\ee
To compare with the $T\bar T$ formula, we first compute the single-trace analogue of the relation \eqref{reledefud} between the undeformed and deformed energies\footnote{If the $T\bar T$ interpretation of these backgrounds holds, then the quantity $E_{L,R}^{(0)}$  should be  fixed as the $T\bar T$ coupling, $\a'$, is turned on. Note this is different from turning on the $\a_1$ parameter of the solution via a TsT transformation, as then $\rho_0$ would  be  fixed instead. This can be easily seen by writing $E_{L,R}^{(0)}$ as 
\be
E_{L,R}^{(0)} 
= \frac{R v \rho_0^2}{2} \frac{\cosh^2(\a_1 \pm \a_n)}{\sinh 2\a_1} \;, \;\;\;\;\mbox{with} \;\;\; \sinh 2 \a_1 = \frac{2 p}{v \a' \rho_0^2}\nonumber
\ee

}, using $\mu = \pi \a'$

\be
E_{L,R}^{(0)} \equiv E_{L,R} + \frac{\a'}{R \, p}  E_L E_R= \frac{R p \cosh^2(\a_1 \pm \a_n)}{\a' \sinh^2 2 \a_1}
\ee
and the explicit expression \eqref{engmombh} for the deformed energies. 
%
%
%
%
Thus, the entropy can be written as 

\be
S = 2 \pi \sqrt{  p k E_L R + \a' k E_L E_R }+ 2 \pi \sqrt{  p k E_R R  + \a' k E_L E_R } \label{ldentropy}
\ee
exactly matching that of a symmetric product orbifold\footnote{Note this does not  mean the dual is necessarily a symmetric product orbifold - indeed, the seed CFT we started from is not, so neither should be its deformation. }
%
 of $p$ $T\bar T$ - deformed CFTs, with $\mu = \pi \a' $ and $c_{seed}=6  k$.  
This entropy interpolates between a Cardy regime  at low energies and a Hagedorn one at high energies. Note the number of F1 strings drops out from the high-energy entropy, as expected from the fact that this regime (for which $E_{L,R} \a' >> p R$) is controlled by the physics of the  NS5-branes alone.

Given this interesting connection to $T\bar T$, it is useful to translate the  parameters we have used to describe the black hole solutions to a notation that may be more natural from the $T\bar T$ perspective. This parametrization, which has been extensively used and explored in \cite{mirage,Guica:2020uhm}, relies on the observation that the stress tensor of a $T\bar T$ - deformed CFT only has two independent components off-shell, which can be conveniently written in terms of two functions of the coordinates, $\L$ and $\bar \L$, as in \eqref{stresst}. For the black hole backgrounds, these functions are simply constant. The  energy and momentum of the solutions are given in terms of them by \cite{mirage}

\be
E =  \pi R \,p \, \frac{\L + \bar \L + 2 \mu \L \bar \L}{1-\mu^2 \L \bar \L} \;, \;\;\;\;\; P =  \pi R \, p\,\frac{\L-\bar \L}{1-\mu^2 \L \bar \L} \label{engttb}
\ee
where, as explained in the appendix, $\L, \bar \L$ and $\mu$  are rescaled versions of those in \cite{mirage} and the factor of $p$ is related to the symmetric orbifold. Comparing these expressions with \eqref{engmombh} written in terms  of the rescaled $L_{u,v}$, namely

\be
 E =R( L_u + L_v) + \frac{R}{\a'} \left(\sqrt{ p^2  + 4 \a'^2 L_u L_v } - p\right) \;,\;\;\;\;\;\;  P= R(L_u - L_v) \label{engmombhsols}
\ee
we find the relation between the parameters to be 
\be
L_u = \frac{ \pi p\, \L}{1-\mu^2 \L \bar \L} \;, \;\;\;\;\;\; L_v = \frac{\pi p\, \bar \L}{1-\mu^2 \L \bar \L} \;, \;\;\;\;\;\; \a' = \frac{\mu}{\pi} \label{ttbpar}
\ee
The inverse relations read 
\be
\L = \frac{2 L_u}{\pi(\b+p)} \;, \;\;\;\;\; \bar \L = \frac{2 L_v}{\pi(\b+p)}\;, \;\;\;\;\;\;\mu^2 \L \bar \L = \frac{\b -p}{\b+p}  \;, \;\;\;\;\; \b = \sqrt{p^2 + 4 \a'^2 L_u L_v} \label{invttbpar}
\ee
In the following, we will continue to use the $L_{u,v},\a'$ notation for the analysis of the perturbations of the black hole backgrounds. Then, we will switch to the more natural $T\bar T$ notation  $\L, \bar \L, \mu$ when computing the conserved charges and making the link to single-trace $T\bar T$.


\section{Asymptotic analysis of the black hole backgrounds \label{asyanalbh}}

In the previous section, we have shown that the asymptotically linear dilaton backgrounds of interest can be captured by a consistent truncation of the type IIB action  to three dimensions, which we now quickly summarize. The truncated theory consists of three-dimensional gravity non-trivially coupled to the dilaton, for a total of  one propagating degree of freedom. The equations of motion  read 
\be
{}^{(3)}R_{\mu\nu} + 2 \nabla_\mu \nabla_\nu \phi + \frac{2 p^2}{k^3}  e^{4\phi} g_{\mu\nu} =0 \;, \;\;\;\;\;\;
\Box_3 \phi - 2 (\p\phi)^2 = - \frac{2}{k} \left(1 - \frac{p^2}{k^2} e^{4\phi}\right)
\ee
where, as noted at the end of subsection \eqref{trunc}, we have pulled out an explicit  factor of $\a'$ from the metric
. As a consequence, now $G_3 = (4 k^{3/2})^{-1}$. The black hole solutions are given by
\be
d\bar s^2 = \frac{ r^2}{\a' r^4 + \b r^2 + \a' L_u L_v}  \left(  r^2 dU dV + L_u dU^2 + L_v dV^2 + \frac{L_u L_v}{r^2} dU dV\right) +  k  \frac{d r^2}{ r^2 } \nonumber
\ee
\be
e^{2\bar \phi} =  \frac{k r^2}{\a'  r^4 + \b r^2 +\a' L_u L_v} \;, \;\;\;\;\;\;\;\; \b = \sqrt{p^2 + 4\a'^2 L_u L_v} \label{bhbck}
\ee
where the overbars denote the fact that these are the background values of the fields. 
The goal of this section is to understand the boundary conditions that  linearized perturbations of these backgrounds  satisfy. Of particular interest are the perturbations generated by large diffeomorphisms, as these correspond to the asymptotic symmetries of the system. Our choice of boundary conditions will be guided by the asymptotic vanishing of the symplectic form evaluated on these solutions, computed using the covariant phase space formalism  \cite{Compere:2018aar, Compere:2009dp}, which in turn ensures the conservation and finiteness of the corresponding  charges.

We start with a thorough characterization of  the linearized solutions above the black hole backgrounds \eqref{bhbck}. We  then  compute the symplectic form in the covariant phase space formalism and use it to select a set of boundary conditions.  We subsequently compute the asymptotic conserved charges and compare them to the conserved charges in $T\bar T$ - deformed CFTs.

\subsection{The linearized solutions}

We consider linearized perturbations of the backgrounds \eqref{bhbck} 

\be
g_{\mu\nu} = \bar g_{\mu\nu} + \e \, h_{\mu\nu} \;, \;\;\;\;\;\; \phi = \bar \phi + \e \, \d \phi
\ee
where $\e$ is a small parameter. We expand these perturbations in Fourier modes,  fixing  radial gauge for the string frame metric. In components, we have 

\be
h_{\mu\nu} = e^{-i \om U + i \k V} \left(\begin{array}{ccc} f_1(r) & f_2(r) & 0 \\ f_2(r) & f_3(r) & 0 \\ 0 & 0 & 0 \end{array} \right) \;, \;\;\;\;\;\; \d \phi =  e^{-i \om U + i \k V} f(r)
\ee
 The linearized equations of motion  couple these four functions in a non-trivial way. Below,  we  briefly explain their structure  and how the solutions are obtained.   We assume that $L_{u,v}\neq 0$; a separate analysis would be needed otherwise. 
We then classify the solutions into modes generated by diffeomorphisms and propagating ones;  the mode counting turns out to be   analogous to that in  AdS$_3$ and is summarized in table \ref{tab:modetable}. 

\subsubsection*{Solving the linearized equations of motion}

To disentangle the equations of motion, we first note that the scalar equation only mixes $\d\phi$ with the trace of the metric perturbation, which is proportional to 

\be
h(r) \equiv (r^4+L_u L_v) f_2 - L_u r^2 f_3 - L_v r^2 f_1
\ee
The scalar equation then only contains $f, h$ and their derivatives up to $f''$ and, respectively, $h'$. Another simplifying observation is that a linear combination of the Einstein equations (more precisely,  $L_v E_{uu} - L_u E_{vv}$, where $E_{ab}$ denotes the $ab$ component of the Einstein equation) only depends on the following linear combination of the functions 

\be
g(r) \equiv L_u f_3 - L_v f_1
\ee
and its derivatives up to $g''$ and\footnote{The coefficient of $f$ in this equation is proportional to $(L_u \k^2 -L_v \om^2)$, so $g$ decouples from $f$ at zero frequency.} $f$. Thus, $h$ and $g$ are in principle determined by $f$.
Trading $f_{2,3}$ for $h$, $g$, one can then solve the remaining Einstein equations\footnote{For completeness, we list herein the steps leading to the solution, which were carefully chosen so as to not have to divide by the frequency at any point. After replacing $f_{2,3}$ by $g,h$, we solve the $rr$ Einstein equation for $f_1'$ (as a function of $f_1,h$ and up to third derivatives of $f$  and up to the first derivative of  $g$). Plugging the result back into Einstein equations one finds that a particular linear combination of the $uu$ and $uv$ components is purely a fourth order differential equation for $f$. Subtracting this equation from the $uv$ component  so that the $f''''$ term cancels yields  an expression that involves $f$ and its derivatives up to $f'''$ , plus some terms linear in $f_1, h$ (and $g$), with coefficients that vanish in the zero frequency limit. It turns out this is exactly the form of the $ur$ and $vr$ Einstein equations (which additionally  have factors of  $g'$), so we can add them to the previous equation in such a way that the $f_1, h$ terms cancel. We are left with a third order differential equation for $f$ only, the $g$ and $g'$ terms having cancelled out by themselves. 
One can explicitly check that the previous fourth order equation for $f$ is a linear combination of the third order one and its derivative, so we can just concentrate on the latter.} for $f_1$ and $f$. One finds that $f$ obeys a third order differential equation, which is decoupled from all the other functions.  The asymptotic large $r$ expansion of the coefficients of this equation  (multiplied just by $-4$) reads 

\bea
\left(\frac{  \om \k r^7}{4 L_u L_v} + \frac{ ( \k^2 L_u + \om^2 L_v + \b \om \k/\a') r^5}{4 L_u L_v} + \frac{1}{\a' k} \left( 1 + \frac{k \b \k^2}{4 L_v} + \frac{k \b \om^2}{4L_u}+ k \a' \om \k\right)r^3  +  \O(r)\right) f'''& +& \nonumber \\
\left(\frac{5  \om \k r^6}{4 L_u L_v} + \frac{ (7 \k^2 L_u + 7 \om^2 L_v +5 \b \om \k/\a') r^4}{4 L_u L_v} + \frac{1}{\a' k} \left( 9 + \frac{7k \b \k^2}{4 L_v} + \frac{7k \b \om^2}{4L_u}+ 8 k \a' \om \k\right)r^2  +  \O(r^0)\right) f'' &+& \nonumber \\
\left(\frac{ \om \k (3+ 4 k \a' \om \k) r^5}{4 L_u L_v} + \frac{ (5 \k^2 L_u +5  \om^2 L_v +  \om \k \cdot \a_1(\om, \k)) r^3}{4 L_u L_v} + \frac{1}{\a' k} \bigl( 15 + \a_2(\om,\k)\bigr)r  +  \O(r^{-1})\right) f'& +& \nonumber \\
\left(-\frac{2k \b  \om^2 \k^2 r^2}{ L_u L_v} + \frac{2 ( \b \k^2 L_u +\b \om^2 L_v + \om \k \cdot \a_3(\om,\k)) }{\a' L_u L_v} +   \O(r^{-2})\right) f = 0 \;\;\hspace{1.2cm}\;&& 
\eea 
where $\a_i(\om,\k)$ are known polynomials in the given arguments  and the  background parameters, whose particular form is not very illuminating for the general structure of this equation.

The three independent solutions of this equation behave asymptotically as $r^{-s}$, where the power $s$ is determined by the asymptotic behaviour of the coefficients of $f$ and of its derivatives. Interestingly, the asymptotic behaviour of the solutions changes \emph{discontinuously} as one or both of the (null) frequencies is set to zero. Concretely, we find
\bea
\om, \k \neq 0 \;\; & :& \;\;\;\;\;\; s = 1 \pm \sqrt{1-4 k \a' \omega \kappa} \;, \;\; 0 \nonumber\\[3pt] 
\om =0 \;\; or \;\; \k=0 \;\; &:& \;\;\;\;\;\; s = 0,\;4,\; 0 \nonumber \\[3pt]
\om=\k =0 \;\;&:&  \;\;\;\;\;\; s = 2,\;4,\; 0 \label{discwt}
\eea
Thus, except for one solution with $s=0$ that we will discuss shortly, 
the $\om, \k \r 0$ limit of the remaining weights is \emph{not smooth}, which is a rather puzzling behaviour from the point of view of the dual holographic interpretation. The form of the  asymptotic expansion above makes it clear why this happens at a technical level: the $r \r \infty$ and $\om, \k \r 0$ limits of the equation do not commute. One can easily check this does not represent the generic behaviour of the wave equation upon these backgrounds;  if we studied a probe scalar, for example, or simply ignored the coupling between the dilaton and the background metric, we would find the equation of motion  has a standard expansion near infinity, with a smooth zero frequency limit.  We conclude that it is the particular couplings of the dilaton to the background - as  dictated by string theory -  and the inclusion of  backreaction that conspire to produce a linearized equation of motion with this curious property. 

Another interesting observation is that the black hole energies parametrized by $L_{u,v}$ do not enter the asymptotic formula for the weights, but only appear  at subleading order in the asymptotic solution\footnote{Asymptotically, the  equation for $f$ (normalized so that the coefficient of $f'''$ is $r^3$) takes the form 
 \bea
&& r^3 f''' + \left[ 5 r^2 + \frac{2}{\om \k} (L_u \k^2 + L_v \om^2) + \O(r^{-2}) \right] f'' +  \left[ (3+ 4 k \a' \om \k ) r + \frac{4 \b (1+ k \a' \om \k)}{\a' r} +\right. \nonumber  \\ 
&&\hspace{2.8cm} \left.  + \frac{2 (1+2 \a' k \om \k)}{\om \k r} (L_u \k^2 + L_v \om^2) + \O(r^{-3}) \right] f' -  \left[ \frac{8 k \b \om \k}{r^2} + \O(r^{-4}) \right] f =0\nonumber 
 \eea
 \vskip -2mm
We can thus write the (propagating) solution asymptotically as 
\be
f(r) \sim r^{-s} \left(1- \frac{2 \left[ \a'  s (2   k \a'  \omega \kappa -s) (  L_u \kappa ^2 + L_v \om^2) + 2\omega \k \b  \left(   s  +  (s+2)   k \a' \omega \kappa  \right)\right]}{r^2  \omega \kappa  \a'  (s+2)   \left(4 \kappa  k \a'  \omega +s^2+2 s\right)} + \O(r^{-4}) \right) \;, \;\;\; s \in \{ s_+, s_-\}\nonumber
\ee
}. This property is not \emph{a priori} guaranteed; indeed, previous analyses of the $SL(2,\mathbb{R})$ spectrum of warped AdS$_3$ backgrounds in string theory have found that the weights did generically depend on the black hole temperature \cite{Compere:2014bia}.    
   
Given the solution for $f$ (parametrized by three arbitrary integration constants), one can easily solve for the remaining functions  (namely $g, h$ and $f_1$), for which $f$ and its various derivatives act as an inhomogenous source term. As mentioned above, the equation for $g$ is second order, which will lead to two more integration constants.  The equations that one obtains for $h, f_1$ depend on whether one works at zero or non-zero frequency. In the former case, $f_1,h$ satisfy first-order differential equations, which lead to two more integration constants; in the latter case, they are entirely determined\footnote{The difference comes from solving 
 the remaining $ur$ and $vr$ components of the Einstein equations, which were not discussed in the previous footnote. If the frequencies are zero, then these equations vanish identically, so the previously obtained equations for $h'$, $f_1'$ are the ones that determine them, yielding two additional integration constants. If the frequencies are non-zero,  $\om, \k \neq 0$, then the $ur$ and $vr$ Einstein equations yield a solution for $f_1$ and $h$, which can be shown to be compatible  with the previously discussed first order differential equation.  Thus, in this case one no new integration constants are generated. } by the solutions for $f$ and $g$.

\subsubsection*{Analysis of the solutions}

As is well known, three-dimensional gravity does not have any propagating degrees of freedom; the only propagating field in the consistent truncation is the dilaton. We thus expect to have two asymptotically independent propagating solutions to the equations of motion (one normalizable and one non-normalizable) and a number of pure large diffeomorphism modes, corresponding to the asymptotic data of the boundary gravitons. 

It is easy to check that the two propagating modes we expect correspond to two of the three solutions to the equation for the scalar perturbation, $f$. These solutions are most easily identified for generic $\omega, \kappa$, where they correspond to the first two values of $s$ in \eqref{discwt}. Since the weights, denoted $s_{\pm}$, sum to $2=d$, their most natural holographic interpretation is as the source and the expectation value for   a scalar operator of a momentum-dependent dimension 

\be
 \Delta = 1+ \sqrt{1-4 k \a' \omega \kappa} \label{opdim}
\ee
Such a behaviour is characteristic of irrelevant deformations. The combination $\om \k$ is Lorentz invariant, and an anomalous dimension that is a  function of $\a' \om \k$ is precisely what one expects from the broken scale invariance of the system \cite{Guica:2010sw}.  Note, however, that the $\a' \r 0$ limit of \eqref{opdim} is $2$, so the dual operator does not appear to correspond - as one would have na\"{i}vely expected - to the continuation of the ``single-trace $T\bar T$'' operator  defined in the IR theory, which has dimension $4$. This behaviour continues to hold for perturbations around the asymptotically linear dilaton vacuum, to which the arguments of \cite{Guica:2010sw} truly apply; at a technical level, it is due to the fact that the $r \r \infty$ and $\a'\r 0$ limits of the dilaton equation do not commute. This does not happen for probe fields  \cite{Chakraborty:2020yka,Asrat:2017tzd,Giribet:2017imm} or if one ignores the backreaction of the metric on the dilaton.  A  more
%
thorough analysis of holography in this spacetime is clearly  needed to shed light on this issue, as well as on the discontinuous behaviour of the weights \eqref{discwt}  in a black hole background.



Note that the conformal dimension \eqref{opdim}  becomes imaginary at large frequencies, and the associated oscillatory behaviour at infinity is precisely the one expected  for  fields in asymptotically flat spacetimes.  This regime would again  deserve a more in-depth study. However, in this article we restrict to $\om \kappa < 1/(4 k \a')$ so that the weights are real (in particular, $1\leq s_+ < 2$) and thus we only need to consider Dirichlet boundary conditions at infinity for these modes. 

The  solution for $f$  that corresponds to the $s=0$  asymptotic behaviour   can be induced by a diffeomorphism $\xi = r F_r \p_r$, which acts as $\d \phi = \xi^r \p_r \bar \phi $. This solution  can be worked out explicitly 

 \be
f^{(s=0)}(r) =- \frac{\a' \, (r^4 - L_u L_v)}{\a' r^4 + \a' L_u L_v+r^2 \beta} \cdot F_r 
\ee
and this knowledge  can be used to reduce the third order differential equation for $f$ to a second order one for the propagating modes only\footnote{For this, one writes $f(r) = f^{(s=0)}(r)  \int^r dr' \hat f(r')$ and finds that $ \hat f(r)$ satisfies a second order differential equation, whose leading coefficients are the same as those given above for $f''', f''$ and $f'$ (the subleading coefficients are different).}. It can be explicitly checked that the additional homogenous solutions to the remaining equations can  be induced by a (string frame) radial gauge-preserving diffeomorphism, which takes the form

\bea
\xi_{rad}^{string} & = &  \left[ F_U (U,V) + k\,  \frac{ (r^2 \b + 2\a' L_u L_v) \p_V\! F_r - L_v (\b+2 r^2 \a') \p_U \! F_r }{r^4 - L_u L_v} -2 k \a'\,  \p_V \! F_r  \ln r\right] \p_U + \;\;\;\;\;\;\;\;\;\;\;\;\;\;\;\;\; \nonumber \\
&& \hspace{-1.8 cm}+\; \left[ F_V (U,V) + k\,  \frac{ (r^2 \b + 2\a' L_u L_v) \p_U\! F_r - L_u (\b+2 r^2 \a') \p_V \! F_r }{r^4 - L_u L_v} -2 k \a'\,  \p_U \! F_r  \ln r\right] \p_V  + r F_r (U,V) \p_r \label{xirad}
\eea
%
The action of this diffeomorphism on the asymptotic metric yields 

\be
h_{ij} =_{r\r \infty}  \left[ \frac{1}{\a'} \left(\begin{array}{cc} \p_{{U}} F_V &  \frac{\p_U F_U + \p_V F_V}{2}    \\  \frac{\p_U F_U + \p_V F_V}{2}   & \p_V F_U  \end{array} \right) - 2 k \ln r  \left(\begin{array}{cc} \p_U^2 F_r  &  \p_U \p_V F_r \\  \p_U \p_V F_r  &  \p_V^2 F_r \end{array} \right)   \right] \label{effasymet}
\ee
where the log term comes from solving the non-homogenous differential equation for $h$. Both parantheses receive corrections at $\O(r^{-2})$ and we have ommitted the radial components of the metric perturbation, which are zero in this gauge. 

 From a holographic point of view, it is more natural to parametrize the solutions in terms of the components of the asymptotic metric that they induce, namely $\p_U F_V$ and $\p_V F_U$. Note that, unlike in AdS$_3$, here the $UV$ component of the boundary  metric is not independent of the other two for generic functions $F_{U,V}$. 
%
%
Indeed, if we act with the $\a'\r 0$ limit of \eqref{xirad} on 
 the BTZ black hole (as they are are just the radial-gauge-preserving diffeomorphisms of that background), we find that their action on the diagonal component of the boundary metric 
  is instead 

\be
\d g_{\text{\tiny{$UV$}}}^{(0)} = F_r  +  \frac{\p_U F_U + \p_V F_V}{2}  
\ee
in the standard Fefferman-Graham notation. Thus, the $UV$ component of the boundary metric can be considered as an independent piece of boundary data, which is not the case for asymptotically flat spacetimes\footnote{To obtain a behaviour more similar to AdS, one may consider the Einstein frame metric $\tilde g_{\mu\nu} = e^{-4\phi} g_{\mu\nu}$, or  $e^{-2\phi} g_{\mu\nu}$.}. This additional factor of $F_r$ is important in off-setting the effect of boundary conformal transformations in AdS. Due to the different structure of \eqref{effasymet}, this does not happen in the asymptotically flat case where, apart from the log term, $F_r$ only appears at subleading order in the metric. 

\subsubsection*{Zero-frequency perturbations}

We have so far analysed the general solutions to the equations of motion. 
Let us now concentrate on the case of zero-frequency perturbations ($\om =\kappa =0$) which, as we already noted,  cannot be treated as a subcase of the general case. 


Let us start with the $f$ equation, whose solution can now be found explicitly also for the propagating modes, and reads 
\bea
f(r) & = & - \frac{\a' \, C_r \left(r^4-L_u L_v\right)}{ \a'  L_u L_v+ \a'  r^4+\beta  r^2} + \frac{\phi_{(2)} \left(\beta + 2\a'  r^2\right) }{ 2(\a' L_u L_v+ \a'  r^4+\beta  r^2)} + \nonumber\\
&&  \hspace{0.4 cm}+\;  \frac{ \phi_{(4)} \left[ \b+2 r^2 \a' + \left(\frac{r^4 \b}{L_u L_v} +\b + 4 \a' r^2\right) \left(\frac{r^2}{2\sqrt{L_u L_v}}  \ln \frac{r^2 + \sqrt{L_u L_v}}{r^2-\sqrt{L_u L_v}} -1\right)\right]}{ 2 r^2(\a'  L_u L_v+ \a'  r^4+\beta  r^2)} \label{zfphisol}
\eea
There are three integration constants: $C_r$ - the constant mode of $F_r$ - and the propagating modes $\phi_{(2)}$ and $\phi_{(4)}$, which we have labeled according to their leading large $r$ behaviour. Note this behaviour does not match the zero frequency limit of the general one \eqref{discwt}, which would have predicted modes behaving as $r^0$ and $r^{-2}$ at infinity. Since their asymptotic  weights  do not satisfy the holographic pairing relation (nor are they paired inside the symplectic form, as we will soon show) the modes  $\phi_{(2)}$ and $\phi_{(4)}$ cannot be interpreted  holographically as a source and, respectively, expectation value for some operator,  despite being the only propagating modes at zero frequency.

Having solved for $f$, we can find the solutions for the remaining functions. Whenever possible, we label the integration constants by the components of the asymptotic metric that they induce   
%

\bea
f_1 & = & \frac{C_{uu} (r^4+L_u L_v + \frac{r^2 \b}{2\a'}) + 2 C_{uv} r^2 L_u  + (L_u L_v C_b-\frac{\b C_{vv} L_u}{\a'}) \frac{r^2 }{2L_v}}{\a'  L_u L_v+ \a'  r^4+\beta  r^2} - \frac{2 \a' C_r L_u r^2 ( r^4-L_u L_v ) }{(\a'  L_u L_v+ \a'  r^4+\beta  r^2)^2} - \nonumber\\
&& \hspace{3mm}-\;\frac{r^4 p^2 - \a' r^2 \b L_u L_v + r^6 \b \a' }{2\a' L_v(\a'  L_u L_v+ \a'  r^4+\beta  r^2)^2} \, \phi_{(2)}+ \left(\frac{2}{\a' L_v r^2} + \O(r^{-4})\right) \phi_{(4)}\nonumber \\[2pt]
f_2 &= & \frac{C_{uv} (r^4 + L_u L_v) + C_{uu} L_v r^2 + C_{vv} L_u r^2}{\a'  L_u L_v+ \a'  r^4+\beta  r^2}+ \frac{r^2\left[ \a' (r^4+ L_uL_v) - \frac{\b^2}{2\a'}\right] \phi_{(2)} + C_r r^2 \b (r^4 -L_u L_v)}{(\a'  L_u L_v+ \a'  r^4+\beta  r^2)^2} + \nonumber \\[2pt]
&& \hspace{3mm} + \; \O(r^{-4}) \phi_{(4)}  \nonumber \\[2pt]
f_3 & = & \frac{C_{vv} (r^4+L_u L_v + \frac{r^2 \b}{2\a'}) + 2 C_{uv} r^2 L_v  -(L_u L_v C_b+\frac{\b C_{uu} L_v}{\a'}) \frac{r^2 }{2L_u}}{\a'  L_u L_v+ \a'  r^4+\beta  r^2} - \frac{2 \a' C_r L_v r^2 ( r^4-L_u L_v ) }{(\a'  L_u L_v+ \a'  r^4+\beta  r^2)^2} - \nonumber\\
&& \hspace{3mm}-\;\frac{r^4 p^2 - \a' r^2 \b L_u L_v + r^6 \b \a' }{2\a' L_u(\a'  L_u L_v+ \a'  r^4+\beta  r^2)^2} \, \phi_{(2)}+ \left(\frac{2}{\a' L_u r^2} + \O(r^{-4})\right) \phi_{(4)}
\eea

%
%
%
%
%
\noindent Note that since $C_r = const$,  no log terms are induced. It is easy to see that the $C_{ij}, C_r$ and $C_b$ modes can be generated by acting on the black hole background with the diffeomorphism 

\be
\xi_{rad}^{string,ct} = \big[ C_{vv} V + (C_{uv} +\g)\, U \big]\p_U + \big[ C_{uu} U + ( C_{uv} -\g) V \big] \p_V + C_r r \p_r  \nonumber
\ee

\be
\g=   \frac{C_b}{4}  +\frac{\b}{\alpha'} \left( \frac{C_{uu}}{4 L_u} - \frac{C_{vv}}{4 L_v} \right) \label{zerodiff}
\ee
From here, it is easy to identify the significance of each diffeomorphism-induced mode: $C_r$ is a simple rescaling of the radius, $C_b$ is a boost, $C_{uv}$ is a rescaling of space and time, etc\footnote{By comparison,  the diffeomorphism used to generate constant perturbations of the BTZ black hole is 

\be
\xi_{rad}^{BTZ} = \left(C^{AdS}_{vv} V + \frac{ c_{uu}^{AdS} U}{2 L_u} \right) \p_U + \left(C_{uu}^{AdS} U + \frac{ c_{vv}^{AdS} V}{2 L_v} \right) \p_V + \left( C_{uv}^{AdS} - \frac{ c_{uu}^{AdS}}{4 L_u} - \frac{ c_{vv}^{AdS}}{4 L_v}\right) r  \p_r \nonumber
\ee
where $C_{ij}^{AdS}$ parametrize the AdS boundary metric and $c_{ij}^{AdS}$, the change in the stress tensor expectation value. }. Note that this form of the diffeomorphism breaks down in the vacuum.

We are ultimately interested in classifying these diffeomorphisms as allowed (normalizable), disallowed (non-normalizable) and trivial (no effect on physical observables), where  normalizability is measured by the symplectic norm of the induced metric perturbations. This is the goal of  the next section. In the mean time, a simple diagnostic that may shed light on this question is to study how their coefficients respond to varying the parameters of the background. Since the perturbations are in string frame radial gauge, the identification 
is straightforward

\bea
\d \alpha'\;\;\; & : & \; C_r = - C_{uv} =  \frac{\d \a'}{2\a'} \;, \;\;\;\; \phi_{(2)} = - \frac{2 L_u L_v \d \a'}{\b} \;, \;\;\;\;  C_{uu} = C_{vv} = C_b =\phi_{(4)}=0 ~~~~~~\nonumber\\
\d L_u\;\;\;  & : & \;\phi_{(2)} = - \frac{\a' L_v \d L_u}{\b} \;,\;\;\;\;\;\; C_b = \; \frac{\d L_u}{L_u} \;, \;\;\;\; C_{uu} = C_{vv}=C_{uv} =C_r =\phi_{(4)}=0 \nonumber \\
\d L_v\;\;\;  & : & \;\phi_{(2)} = - \frac{\a' L_u \d L_v}{\b}\;, \;\;\;\;\; C_b = - \frac{ \d L_v}{L_v} \;, \;\;\;\;C_{uu} = C_{vv}=C_{uv} =C_r =\phi_{(4)}=0 
\eea
Since variations of $L_{u,v}$ should definitely be part of the phase space,  we conclude that the modes $\phi_{(2)}$ and $C_b$  should be allowed  by the boundary conditions. It is interesting to note that, contrary to the $AdS_3$ case,  changes in the energy and momentum  are not generated by diffeomorphisms only, but one also needs the propagating mode $\phi_{(2)}$. Since the modes   $C_r$ and $C_{uv}$ vary as the (irrelevant)  coupling constant of the theory is changed, we would expect them to be fixed.   Note that $ \alpha' \phi_{(2)}$ corresponds precisely to the variations of $-\b/2$. 

Another useful exercise is to  compute the left and right-moving energies (i.e., the charges associated to the isometries $\partial_U,\partial_{-V}$) carried by these modes 
\bea E_L & = & 2 R \left[    \frac{\b}{4\alpha'} C_{uu}  + L_u C_{vv} \left( 1- \frac{\b}{4 \alpha' L_v}\right) + L_u C_{uv}+\frac{L_u C_b}{4} - \frac{1}{4}\left(2 + \frac{\b}{\alpha' L_v}\right) \phi_{(2)} + \frac{\phi_{(4)}}{L_v} \right]~~~~~~~~\nonumber \\
 E_R & = & 2 R \left[  \frac{\b}{4\alpha'} C_{vv}+  L_v C_{uu}\left(1-\frac{\b}{4 \alpha' L_u} \right) + L_v C_{uv}  - \frac{L_v C_b}{4} - \frac{1}{4}\left(2 + \frac{\b}{\alpha' L_u}\right) \phi_{(2)} + \frac{\phi_{(4)}}{L_u} \right]  \label{engctpert}
\eea
The charges are finite for all  the modes, as the only   potential quadratic divergence (proportional to $C_r$) cancels between the Einstein \eqref{keinst} and the scalar \eqref{kscal}  contribution to the conserved charges. Thus, the energies do not provide information regarding normalizability of the modes.  Moreover, one notes that all modes  but $C_r$ carry non-trivial energy and momentum, indicating that they are all physical, with the possible exception of $C_r$.  


The table below summarizes the  most general solution to the linearized equations of motion and its  parametrization

\begin{table}[H]
\centering
\begin{tabular}{c|c|c}
\small{frequency} & $\om =\k =0$ & ~~~~ $\om, \k \neq 0$~~~~ \\ \hline
\small{propagating} & $\phi_{(2)},\phi_{(4)}$ &  $\phi_{(s_+)}, \phi_{(s_-)}$ \\ \hline
\small{pure diffeomorphisms} & $C_r,C_{uu},C_{uv},C_{vv},C_b$ &  $F_U,F_V,F_r$ 
\end{tabular} \vskip 2mm
\caption{\label{tab:modetable} \small{Classification of the solutions to the linearized equations of motion.}
}
\end{table}

%

\noindent Note that for the non-zero frequency solutions, the propagating modes must be described in momentum space, whereas the ones generated by diffeomorphisms are  more naturally analysed in position space.

\subsection{Choice of boundary conditions}

Having characterized all the linearized perturbations of the geometries, the next step is to understand  the boundary conditions that these perturbations obey. As we saw, a simple analysis of the asymptotic behaviour of the solutions is insufficient to answer this question, and a more systematic approach is needed. 

In general, in order  for a theory to have a well-defined phase space, one requires that the symplectic  $d-1$ form associated with any two allowed perturbations vanish asymptotically. More precisely, in three dimensions one imposes that

 \be
\boldsymbol \om_{ab} [\Phi, \d_1 \Phi, \d_2 \Phi] = o(r^0), \qquad\boldsymbol \om_{r a} [\Phi, \d_1 \Phi, \d_2 \Phi] = o(r^{-1})\label{phs}
\ee
where $a$ denotes the tangent indices to the boundary located at $r \rightarrow \infty$, $\Phi$ stands  generically for the fields in the theory, $\d \Phi$ are arbitrary variations thereof and the notation $o(r^{-c})$ signifies that the fall-off must be faster than the indicated power of $r$. The first condition above ensures that the symplectic flux is conserved, whereas the second leads  to normalizability of the symplectic form at infinity. The boundary conditions are chosen such that these requirements are satisfied. Since $\boldsymbol \om$ is only defined up to the addition of an exact form,  $\boldsymbol \om \r \boldsymbol \om + d \boldsymbol \om_Y $, where $\boldsymbol \om_Y$ is antisymmetric in the variations,  the boundary conditions only need to be satisfied for some choice of this boundary term.  Note that the analysis of the symplectic form will not necessarily indicate all the modes that need to be fixed, but rather which modes shoud not be allowed simultaneously. 

When one of the perturbations is generated by a diffeomorphism, $\d \phi= \L_\xi \phi$, the symplectic form can be related on-shell to a $d-2$ form $ \boldsymbol k_\xi$

\be
\boldsymbol \om [\Phi, \d_\xi \Phi, \d \Phi] = d \boldsymbol k_\xi [\Phi, \d\Phi]
\ee
whose spatial integral at infinity yields the conserved charge difference \eqref{chdiffbb} associated to the corresponding diffeomorphism.
The boundary conditions \eqref{phs} then ensure that the charges are conserved and, respectively, finite. 
The above-mentioned    ambiguities in the construction of $ \boldsymbol \om$ may contribute to the conserved charges (see e.g. \cite{Compere:2014bia}) and would, in principle, be related to counterterms in the action.

The covariant phase space formalism constructs the symplectic form directly from the action, in our case \eqref{3dact}.  The result for Einstein gravity coupled to a scalar  with kinetic term $- \frac{\varkappa}{2}  (\p \phi)^2$ and an arbitrary potential is 

\be
\boldsymbol \om = \boldsymbol \om_g +\boldsymbol \om_{scal} \equiv \frac{1}{16 \pi G_3} \tilde \e_{\mu\nu\rho} ( \om_g^{\mu} + \om_{scal}^{\mu}) \,  dx^{\nu}\wedge dx^{\rho}
\ee
with 
\be
\om^\mu_g[\tilde g,\delta_1 \tilde g,\delta_2 \tilde g] 
=  \frac{1}{2} \left[ ( 2\tilde \nabla_\l h_1^\mu{}_\rho -\tilde \nabla^\mu h^1_{\l\rho}) h_2^{\l \rho} - \tilde \nabla_\rho h_1 \, h_2^{\rho \mu} + h_1 (\tilde\nabla_\rho h^{\rho\mu}_2 - \tilde \nabla^\mu h_2) - (1 \leftrightarrow 2) \right] 
\ee
where $\tilde g$ is the Einstein frame metric and $h_{1\mu\nu} = \delta_1\tilde g_{\mu\nu}$, $h_{2\mu\nu} = \delta_2 \tilde g_{\mu\nu}$. The scalar contribution to the symplectic form is

\be
\om^\mu_{scal} = - \d_1 (\varkappa \tilde \nabla^\mu \phi) \, \d_2 \phi - \frac{\varkappa}{2} \, h_1  \tilde \nabla^\mu \phi \, \d_2 \phi - (1 \leftrightarrow 2)
\ee
For the action \eqref{3dact},  $\varkappa= 8$. 
Note $ \boldsymbol \om$ is closed on-shell, $ d\boldsymbol \om=0$. In the following, we compute the symplectic form for all the linearized perturbations of the black hole backgrounds  that we have presented. We will perform a separate treatment of the zero and non-zero frequency modes.

\subsubsection*{Symplectic pairing of the zero-frequency modes}
We start by considering two general zero-frequency perturbations, parametrized by two sets of constants $C_{ij}, C_r, C_b , \phi_{(2)} $ and $\phi_{(4)}$, and compute their symplectic form. We find that all but the $UV$ component of the two contributions to $\boldsymbol\om$ vanish. Since $\boldsymbol \om$ is closed on-shell, its  $UV$ component must be $r$ - independent. We indeed notice that a potential quadratic divergence in $r$ (proportional to $ C_{uv} \wedge C_r$) cancels between the gravitational and scalar contributions. The symplectic form reads, up to an overall factor of twice the radius 
\bea
\om_{\text{\tiny{$UV$}}} & = &  \frac{ 1}{4}\,   (C_{uu} L_v - C_{vv} L_u)  \wedge C_b +  C_{uv} \wedge (C_{uu} L_v + C_{vv} L_u) +  \nonumber  \\
&+&  \phi_{(2)} \wedge \left(  C_{uv} - \frac{\b}{4\a' L_u L_v} (C_{uu} L_v + C_{vv} L_u)\right) +  \phi_{(4)} \wedge \, \frac{C_{uu} L_v + C_{vv} L_u}{L_u L_v}  \label{ctsymplf}
\eea
where the wedge notation stands for the antisymmetric product of the two modes $a \wedge b \equiv a^{(1)} b^{(2)} - a^{(2)} b^{(1)} $, with the superscript indicating which of the two perturbations we are referring to. From a holographic perspective,   this formula should provide us with the symplectic pairing between sources and expectation values for the operators of the dual theory that correspond to the fields of the consistent truncation \cite{Papadimitriou:2010as}. 

As discussed previously, the modes $C_b$ and $\phi_{(2)}$ must be  allowed  on the phase space, since they correspond to variations of the energy and momentum of 
 the black hole solutions. We conclude that their coefficient inside the symplectic form must be fixed. This yields the boundary condition

\be
C_{uu}= \frac{2 \alpha' L_u}{\b} C_{uv} \;, \;\;\;\;\;\; C_{vv} = \frac{2 \alpha' L_v}{\b} C_{uv} \label{mixedbc}
\ee
which determines the two off-diagonal components of the asymptotic  metric fluctuation in terms of the diagonal piece.
%
%
Upon making this choice, the second term on the first line of \eqref{ctsymplf} cancels due to the antisymmetry of the wedge product. We are nevertheless left with a  coupling

\be
\phi_{(4)} \wedge \hat C_{uv} \label{leftoverom}
\ee
where the hat on the $C_{uv}$  mode indicates it must be accompanied by $C_{uu}$ and $C_{vv}$ perturbations, whose coefficients are fixed by  the mixed boundary conditions \eqref{mixedbc}.   The diffeomorphism that would generate the $\hat C_{uv}$ mode  is 

\be
\xi \mbox{\footnotesize{$[\hat C_{uv}]$}} =\hat  C_{uv}\left[ \left(U + \frac{2 \alpha' L_v}{\b} V \right) \p_U + \left(V + \frac{2\alpha' L_u}{\b}U\right) \p_V \right] \label{ximct}
\ee
Note this diffeomorphism changes  the radius of the $\s$ circle.

It is not entirely clear which one of the modes in \eqref{leftoverom} should be fixed. 
Both of them are zero on the black hole backgrounds, which is consistent with either a source interpretation, or with an expectation value that happens to vanish in thermal states. It would seem natural to classify  $\phi_{(4)}$ as an allowed mode because it is asymptotically subleading to $\phi_{(2)}$; however, we note from the explicit solution \eqref{zfphisol} that the full solution is divergent on the black hole horizon $r = \sqrt[4]{L_u L_v}$. 
The fact that it is not the zero-frequency limit of any of the propagating modes also does not aid in understanding its holographic interpretation. 
%
 The  $\hat C_{uv}$ mode could in principle be allowed, since it carries finite energy; however, the fact that the energy of the black hole backgrounds  is entirely carried  by the $\phi_{(2)}$ mode, and not $\hat C_{uv}$, is somewhat puzzling\footnote{It is interesting to note that upon imposing the boundary conditions \eqref{mixedbc}, the energies \eqref{engctpert} only depend on the combination $\phi_{(2)} - 4 \a' L_u L_v \hat C_{uv}/\b $.}. 
%
As we show in the next section, the presence of the latter mode is in fact required when perturbing the black hole backgrounds, in order to offset the change in the radius induced by an asymptotic symmetry generator.


As far as the asymptotic charges are concerned, the term \eqref{leftoverom}  would result in the non-conservation of the charge associated with the diffeomorphism \eqref{ximct} on a background where $\phi_{(4)}$ is turned on. However,  $\xi \mbox{\tiny{$[\hat C_{uv}]$}}$  is not an allowed diffeomorphism  upon the constant backgrounds - as it would change the asymptotic radius of the $\s$ circle - and thus we are not led to any inconsistency from the point of view of the asymptotic symmetry group analysis.
The  holographic interpretation of these modes as sources or expectation values of dual operators   deserves, nevertheless, a more in-depth analysis, which we leave   to future work. 
%
%

In conclusion, of the seven zero-frequency modes listed in table \ref{tab:modetable}, we find that $C_{uu}$ and $C_{vv}$ must be fixed in terms of $C_{uv}$ as in \eqref{mixedbc},  $C_b$ and $\phi_{(2)}$ must be allowed, whereas the interpretation of the remaining modes  $ \hat C_{uv}, \phi_{(4)}$ and $C_r$ remains unclear. The fact that $C_r$ does not yet appear in the symplectic form, nor in the energies,  suggests it could be a trivial gauge mode; our subsequent analysis will nevertheless indicate that it is in fact  a source mode, which should be fixed. This interpretation is corroborated by the fact that  from the point of view of the IR AdS$_3$,  the $C_r$ mode corresponds to the source for the dilaton.  


 

\subsubsection*{Symplectic pairing of the zero and non-zero frequency modes}

Let us now study the symplectic pairing between zero-frequency $(\om=\k=0)$ and non-zero frequency ($\om, \k \neq 0$) modes. The latter come in two types: i) those generated by a diffeomorphism of the form \eqref{xirad} and ii) those which are propagating, with asymptotic falloffs given in \eqref{discwt}. 
We first discuss the symplectic pairing of the zero frequency modes with the non-zero frequency ones generated by a diffeomorphism.

Before we start, let us make some remarks about the choice of gauge. The zero-frequency perturbations are naturally described in string frame radial gauge, since then their identification with the change in the black hole parameters is very clear\footnote{This is due to the fact that in string frame, the black holes' metric has $g_{rr} = \frac{1}{r^2}$. In Einstein frame, the radial component depends on the black hole parameters, and thus their variation will not preserve $\tilde g_{rr}$. While it is in principle possible to redefine the radial coordinate so that the new radial metric component is independent of the black hole parameters, in practice the resulting metric is  very difficult to work with.}. On the other hand, the mode generated by the non-zero frequency   diffemorphism   can be in any gauge that is compatible with the boundary conditions, i.e. one that does not exclude \emph{physical} large diffeomorphism modes. As it turns out\footnote{If we work with a diffeomorphism in string frame, the symplectic form has  log divergences that constrain $F_r$ to be at most linear in $U,V$. We also find $r^2$ divergences that depend on the second derivatives of $F_r$, though they vanish if $F_r$ takes the form \eqref{formFr}, $C_{ij}$ satisfies \eqref{mixedbc}  and $C_r =0$. While fixing $F_r =0$ in string frame would resolve these issues, it would also impair our ability to find a central extension to the asymptotic symmetry algebra we discuss in the next section.  },  it is better to consider diffeomorphisms that preserve the radial components of the metric in Einstein frame, rather than string frame. To reach this frame, one simply adds in a ``correcting'' diffeomorphism to \eqref{xirad} so that  $ \d \tilde g_{rr} =0$ in the final result, where $\tilde g_{\mu\nu} = e^{-4\phi} g_{\mu\nu}$ is the Einstein frame metric. Interestingly, this requirement completely changes the asymptotic behaviour of $\xi^r$ and cancels away the log terms present in \eqref{xirad}, which  leads to important simplifications in the calculation of the symplectic form. 

The most general diffeomorphism that preserves the radial metric in Einstein frame reads
\bea
\xi_{rad}^{Einst.} &= & \left( F_U (U,V) + \frac{k  (r^2 \,  \p_V F_r - L_v\,  \p_U F_r)}{r^4-L_u L_v} \right) \p_U  +  \left( F_V (U,V) + \frac{k  (r^2 \, \p_U F_r - L_u \,\p_V F_r)}{r^4-L_u L_v} \right) \p_V  + \nonumber \\
&+& \frac{r^3  F_r (U,V)}{\a' r^4 + r^2 \b + \a' L_u L_v} \p_r \label{xiradeinst}
\eea
where $F_U, F_V$ and $F_r$ are arbitrary at this stage. 
Note that the radial component of this diffeomorphism, as well as  the associated perturbation of the dilaton it induces, is now more subleading at large $r$  than it was for the string frame radial gauge perturbations. This allows for the possibility that it be an allowed mode while its string frame counterpart is not, as we will soon argue is the case. From now on, $F_r$ will exclusively denote the function labeling the radial component of the diffeomorphism in Einstein frame \eqref{xiradeinst}, and not \eqref{xirad}.

We thus compute the symplectic product between perturbations generated by the diffeomorphisms above and arbitrary zero-frequency modes. Unlike when the diffeomorphism was in string frame radial gauge, now all the log terms in the symplectic form  vanish. Moreover, the potential quadratic divergence ($\propto C_r$) in the $\om_{\text{\tiny{$UV$}}}$ component now completely cancels between the metric and the scalar contribution. The finite part of $\omega_{\text{\tiny{$UV$}}}$ has terms proportional to all the parameters of the zero-frequency solution.  Since $C_b$ and $\phi_{(2)}$ parametrize the black hole backgrounds, we must require that their coefficients vanish independently, which constrains the functions appearing in the diffeomorphism \eqref{xiradeinst} as follows
\be
L_u\,  \p_V  F_U = L_v \, \p_U  F_V \;, \;\;\;\;\; \p_U  F_U + \p_V  F_V = \frac{\b}{2\a' L_u L_v} (L_v \p_U  F_V + L_u \p_V  F_U  ) 
\ee
The most general solution to these equations is 

\be
 F_U = f \left(u \right) + \frac{2 \a' L_v}{p+\b} \, \bar f \left(v\right) + c_U \;, \;\;\;\;\;  F_V =  \bar f \left(v \right) + \frac{2 \a' L_u}{p+\b}\, f \left(u\right) +c_V\label{solFUV}
\ee
where $c_{U,V}$ are integration constants and the ``field-dependent coordinates'' $u,v$ are defined as

\be
u \equiv  \frac{(p+\b) \, U+2 \a' L_v V}{2p} \;, \;\;\;\;\;\; v \equiv  \frac{(p+\b)\, V + 2 \a' L_u U}{2p} \label{fdepcoordst}
\ee
where their overall normalization has been chosen for later convenience. Note these
 coordinates $u,v$ have identifications
 \be
 u \sim u + 2 \pi R \, r_u  \;, \;\;\;\;\;\; v \sim v + 2 \pi R \, r_v
 \ee
  where
 \be
 r_u = \frac{\b+p+2\a' L_v}{2p} = 1 + \frac{\a' H_R}{p R} \;, \;\;\;\;\; r_v = \frac{\b+p+2\a' L_u}{2p} = 1 + \frac{\a' H_L}{p R} \label{defruv}
 \ee
where we used \eqref{engmombhsols}. The reason for using the notation $H_{L,R}$ for the left/right-moving energies, rather than the previously used $E_{L,R}$, is to emphasize the operatorial origin  of the field-dependent radius of these coordinates; the two quantities are identical as far as the classical analysis of this article is concerned. Acting with these diffeomorphisms produces a change in the components of the asymptotic metric that precisely obeys \eqref{mixedbc}, as one can easily check, where now $C_{ij}$ is space-time dependent.

Note that for $F_r =0$, these diffeomorphisms \emph{precisely coincide} with the  asymptotic symmetry generators  \eqref{xiV}  of the  AdS$_3$ background with mixed boundary conditions that is holographically dual to double-trace $T\bar T$ - deformed CFTs,  particularized  to constant parameters and with the radial component removed. This background and its associated asymptotic symmetries is reviewed in detail in appendix \ref{conschdtrttb}.  A perturbative expansion of these results, which is more appropriate for comparison with the analysis of the asymptotically linear dilaton backgrounds, is presented in appendix \ref{pertanctb}.


 Evaluated on this solution, the radial symplectic form contains a term  

\be
\phi_{(4)} \frac{( L_u \p_V  F_U + L_v \p_U F_V)}{L_u L_v}
\ee
 that is directly analogous to the term \eqref{leftoverom}  encountered in the analysis of the constant modes. 
  Note that since $\phi_{(4)}$ is constant, this term is a total $\s$ derivative,  so it will not immediately affect the  conservation of the charges associated with the allowed diffemorphisms \eqref{xiradef} upon a background with $\phi_{(4)}$ turned on, as one might  worry. In any case,  these charges vanish except when $f$ is a constant. 
%

%
%
%

Next, we also find a coupling 
 
 \be
 2 \left( k \p_U \p_V F_r - \frac{2}{\alpha'} F_r \right) C_r
 \ee 
If $C_r \neq 0$, then the vanishing of this term would  constrain  $F_r$ to be a Bessel function with argument $U V$, which is not the structure we expect. We thus set $C_r =0$. Note that $C_r=0$ does not imply that $F_r =0$, because the former is the radial mode of a diffeomorphism in string frame, whereas the latter is the radial mode of a diffeomorphism in Einstein frame, which has a different radial dependence. In particular, its asymptotic behaviour  is subleading by a factor of $r^{-2}$ with respect to that of $C_r$. Fixing the string frame constant radial mode $C_r$ is also consistent with keeping $\alpha'$ - here interpreted as the coefficient of the irrelevant deformation - fixed. 
 
Finally,  writing $C_{uu}, C_{vv}$ in terms of $C_{uv}$ using the boundary conditions \eqref{mixedbc} and plugging in  the solution for $F_{U,V}$, we find the term proportional to $\hat C_{uv}$ is 
 
 \be
 \frac{k \left(\alpha'  L_u F_r^{(0,2)}(U,V)+\alpha'  L_v F_r^{(2,0)}(U,V)-\beta  F_r^{(1,1)}(U,V)\right)}{\b } \hat  C_{uv} \label{constrradcomp}
 \ee
If $\hat C_{uv} \neq 0$, then  the most general solution for $F_r$ is a linear combination of  two arbitrary functions of the field-dependent coordinates \label{fdepcoordst} 

\be
F_r = f_r (u) + \bar f_r (v) \label{formFr}
\ee
As we shall see, this constraint also follows  from the symplectic pairing of two diffeomorphisms.

Let us now analyse the tangent components, ${\boldsymbol \om}_{ra}$, of the symplectic form, which should  fall off faster than $1/r$ in order to ensure finiteness of the conserved charges. We find that both components   diverge linearly with $r$, with  divergent terms  that are proportional  with $C_r$ and, respectively, $C_{ij}$ multiplied by various derivatives of $F_{U}$ and $F_V$, and that they are both total $\p_a$ derivatives of the same scalar quantity.  If we  plug in the boundary condition  \eqref{mixedbc} for $C_{uu}, C_{vv}$, the term proportional to $\hat  C_{uv}$ vanishes on the solution \eqref{solFUV} for $F_U, F_V$. As for the other term, we must set $C_r=0$  in order to have it vanish. 

Moving on to the $1/r$ term, we find that it is not zero. However,  it can be written as 

\be
\om_{r a} = \frac{1}{r} \p_a Y^{(1)}  + \ldots \label{exactform}
\ee
before imposing any constraint on the form of the diffeomorphism, and thus the finiteness of the integrated charge is not affected. Such a term can be absorbed into a redefinition of the symplectic form $\boldsymbol \om \r \boldsymbol \om + d \boldsymbol \om_Y$, where the one-form  $\boldsymbol \om_Y = r^{-1}Y^{(1)}  dr$ and $Y^{(1)}$ is local. Such a redefinition does not affect the leading behaviour of $\boldsymbol \om_{ab}$, nor does it affect the conserved charges\footnote{Writing $\boldsymbol \om_Y (\Phi, \d_1 \Phi, \d_2 \Phi) = \d_1 \boldsymbol Y(\Phi, \d_2 \Phi)-\d_2 \boldsymbol Y(\Phi, \d_1 \Phi)$,  the shift in the charge integrand is  $\boldsymbol k_\xi \r  \boldsymbol k_\xi + \d \boldsymbol Y - \xi \cdot d \boldsymbol Y$.
}.


The components of the symplectic form for one zero-frequency mode and one non-zero frequency propagating mode with asymptotic weight $s \in \{s_+, s_-\}$  scale as

\be
\boldsymbol \om_{ab} \sim r^{1-s} \;, \;\;\;\;\;\; \boldsymbol \om_{r a} \sim r^{-s }
\ee
For the range of frequencies we are considering, the asymptotic weights $s_\pm = 1 \pm \sqrt{1- 4 k \a'  \om \k}$ satisfy $s_+ \in [1,2)$ and $s_- \in (0,1] $ . We immediately conclude that $s_+$ should be an allowed mode, and $s_-$ should be  disallowed. As previously mentioned, we do not consider the case where $s_\pm$ become imaginary, where a different analysis is necessary. 

\subsubsection*{Symplectic pairing of  non-zero frequency modes}

Finally, we consider the symplectic pairing for two non-zero frequency modes, which can be either generated by diffeomorphisms or be propagating modes.

We start with two pure diffeomorphism modes and evaluate the symplectic form on the solution \eqref{solFUV}  for $F_U,F_V$, though leaving $F_r$ arbitrary. The boundary condition on $\boldsymbol \om_{ab}$ leads to
 the following constraint
\begin{align}
\partial_u\partial_v F_r=0
\end{align}
Hence, the radial function must take the form \eqref{formFr}.  Note this is the same constraint as that which emerged from \eqref{constrradcomp}, but without having to assume that  $\hat C_{uv}\neq 0$. The $\boldsymbol \om_{ra}$ components of the symplectic behave as $\frac{1}{r}$ asymptotically, but they  again take the form \eqref{exactform} and thus can be  absorbed into a shift of $\boldsymbol \om$ by a total derivative.

Next, we compute the symplectic form for two propagating modes, denoted by their asymptotic behaviour $s_{1,2}$, which can be either normalizable or non-normalizable. We find

\be
\boldsymbol \omega_{ab}(s_1,s_2)\sim r^{2-s_1-s_2}\;, \;\;\;\;\;\;\;
\boldsymbol \omega_{ra}(s_1,s_2)\sim r^{1-s_1-s_2}
\ee
which confirms our identification of $s_+$ as a normalizable mode and $s_-$ as a non-normalizable one.  The symplectic pairing between one normalizable and one non-normalizable mode of equal but opposite frequency (so that $s_1 = s_+$ and $s_2 =s_-$) is

\be
\boldsymbol \om_{ab} (s_-, s_+) = 2 \sqrt{1-4 k \a' \om \k}  \;, \;\;\;\;\; \boldsymbol \om_{ra} =0 
\ee
This precisely agrees with the radial symplectic form for a scalar  field in AdS$_{d+1}$   \cite{Papadimitriou:2010as}
\be
\Omega_{rad} = (d-2\Delta) \int \d \phi_{vev} \wedge \d \phi_{source}
\ee
with $\Delta$ given in \eqref{opdim}.   
%
%
Lastly, the symplectic form for one (normalizable) propagating mode and one pure 
 diffeomorphism  reads
\be
\boldsymbol \om_{ab} (s_+, \xi) \sim r^{2-s_+} \;, \;\;\;\;\;\;\;  \boldsymbol \om_{ra} (s_+, \xi) \sim r^{1-s_+}
\ee 
These falloffs violate the boundary conditions \eqref{phs} for the $s_+ \in [1,2)$ range  we have been considering. However,  
%
%
%
 one can explicitly construct a 1-form $\boldsymbol \om_Y$ such that $\omega=d \boldsymbol \om_Y$, when evaluated on the modes above. Hence, these possibly problematic terms can be absorbed in an exact form. It would of course be intersting to provide   covariant expressions  for all the counterterms that we have added to the symplectic form. 
 

To summarize, the symplectic form analysis fixes the components  $F_{U}, F_V$ of the Einstein frame radial-gauge-preserving diffeomorphisms in terms of two arbitrary functions of the field-dependent coordinates \eqref{fdepcoordst}, which precisely match those in  $T\bar{T}$ - deformed CFTs. The radial function  $F_r$ is also constrained to depend on these coordinates via \eqref{formFr}. It is  an interesting question  whether $F_r$ should  be \emph{completely} determined by $F_{U,V}$, as it is in AdS. While we do suspect this to be the case, our analysis - which is solely based on the symplectic form -  is unable to see such a constraint. The same situation occurs in AdS$_3$, where the symplectic form analysis does not fix the radial component of the allowed diffeomorphisms, which is instead determined by a boundary condition on the asymptotic metric. 

We additionally find that, generically, one of the propagating modes, $\phi_{(s_-)}$, should be fixed, whereas the other, $\phi_{(s_+)}$, should be allowed to fluctuate, as is standard in holography. The constant mode $C_r$ should be fixed, whereas the boundary conditions on the remaining constant modes $\hat  C_{uv}$ and $\phi_{(4)} $ are still unclear. 
These results are summarized  in the following table 

\vskip 1mm 
\begin{table}[H]
\centering
\footnotesize
\label{tab:normtable}
\begin{tabular}{|c|c|c|c|c|c|c|c|c|} 
\hline
frequency   & \multicolumn{4}{c|}{$\om =\k =0$}                                                                & \multicolumn{4}{c|}{$\om, \k \neq 0$}                                          \\ 
\hline
mode                    & $\phi_{(2)}, C_b $       & $C_{uu},C_{vv}$ & $C_r$                  & $\phi_{(4)}, \hat C_{uv} $ & $F_U,F_V$   & $F_r$       & $\phi_{(s_+)}$           & $\phi_{(s_-)}$          \\ 
\hline
\multirow{2}{*}{status} & \multirow{2}{*}{allowed} & fixed in terms  & \multirow{2}{*}{fixed} & \multirow{2}{*}{unclear}   & constrained & constrained & \multirow{2}{*}{allowed} & \multirow{2}{*}{fixed}  \\
                        &                          & of $C_{uv}$     &                        &                            & \eqref{solFUV}      & \eqref{formFr}      &                          &                         \\
\hline
\end{tabular}
\vskip 2mm
\caption{\small{ \label{tab:normtable}  Classification of the linearized perturbations into allowed modes and fixed ones.}}
\end{table}

\subsection{Conserved charges}

In the previous section, we have found that the symplectic form   vanishes for perturbations generated by   Einstein frame radial-gauge-preserving diffeomorphisms of the form 
\bea
\xi_{f, \bar f} &= & \left(  f \left(u \right) + \frac{2 \a' L_v}{p+\b} \, \bar f \left(v\right)  + \frac{k  (r^2 \,  \p_V F_r - L_v\,  \p_U F_r)}{r^4-L_u L_v} +c_U \right) \p_U  +  \left( \bar f \left(v \right) + \frac{2 \a' L_u}{p+\b}\, f \left(u\right) +\right. \nonumber \\
&& \hspace{-1cm} + \; \left. \frac{k  (r^2 \, \p_U F_r - L_u \,\p_V F_r)}{r^4-L_u L_v} +c_V \right) \p_V  +  \frac{r^3  F_r }{\a' r^4 + r^2 \b + \a' L_u L_v} \p_r \label{xiradef}
\eea
where $f(u),\bar{f}(v)$ are arbitrary functions of the field-dependent coordinates \eqref{fdepcoordst}  and
\begin{align}
F_r=f_r(u)+\bar{f}_r(v)
\end{align} 
with $f_r, \bar f_r$ arbitrary so far. This  indicates that  the asymptotic  charges associated with the above diffeomorphisms should be conserved.

 We would now like to compute these   charges using the covariant phase space formalism, in which the  charge difference between two nearby on-shell backgrounds is given by 
 
 \be
 \not{\! \d} \boldsymbol{Q}_\xi  =\oint  \boldsymbol{k}_\xi[\d\Phi,\Phi] \label{chdiffbb}
 \ee
Here, $\d\Phi$ stands for an arbitrary variation of all the background fields  and $\boldsymbol k_\xi$ is a $d-2$ form constructed algorithmically from the action, which depends on the background fields, their variations and the diffeomorphism $\xi$. For the problem at hand, we have 
 
 \be
 \boldsymbol{k}_\xi = \frac{1}{2} \tilde \e_{\mu\nu\rho} K^{\nu\rho}_\xi \;, \;\;\; \;\;\;\;\; K^{\nu\rho}_{\xi} = \frac{1}{8\pi G} (K^{\nu\rho}_{\xi,g} + K^{\nu\rho}_{\xi, scal})
 \ee
with
\be
K^{\mu\nu}_g=  \left(\xi^\nu \tilde \nabla^\mu h - \xi^\nu \tilde \nabla_\s h^{\mu\s} + \xi_\s \tilde \nabla^\nu h^{\mu\s} + \frac{1}{2} h \tilde \nabla^\nu \xi^\mu - h^{\rho \nu}\tilde \nabla_\rho \xi^\mu\right)  \label{keinst}
\ee

\be
K^{\mu\nu}_{scal}=  \varkappa \,\xi^\nu \tilde \nabla^\mu \phi \, \d \phi \label{kscal}
\ee
where $\varkappa=8$ and the ~$\tilde{}$~ indicates that we are working in the Einstein frame. 
This charge difference needs to be integrable in order for the charges to be well-defined, a property that is not automatically guaranteed by the definition \eqref{chdiffbb}, but needs to be checked separately. 
Assuming this is the case, the total conserved charge is obtained by integrating $\cancel{\d} \boldsymbol{Q}_\xi$
 along a path in configuration space between the background of interest and a reference background. We will alternatively be using the notation $\cancel{ \d} Q_{f,\bar f}$ for the charge difference associated with a diffeomorphism of the form \eqref{xiradef} parametrized by the functions $f(u), \bar f(v)$.

In our case, if we take $\d \Phi$ to correspond to the difference between two nearby black hole solutions, we find that all the charges vanish except for the energy and momentum.  These charge differences are given by 

\be \label{integr}
\not{\! \d} Q_{f_0,\bar f_0} = (f_0 + \mu \bar \L \bar f_0 + c_U) \d H_L  - (\bar f_0 + \mu  \L f_0 + c_V ) \d H_R
\ee
where  $f_0, \bar f_0$ are the Fourier zero modes of the respective functions and
we used \eqref{invttbpar} to write the answer in terms of the natural $T\bar T$ parametrization. Given that 

\be
\L = \frac{H_L}{\pi R p + \mu H_R} \;, \;\;\;\;\; \bar \L = \frac{H_R}{\pi R p + \mu H_L}
\ee
we immediately note that  the charges are not  integrable for generic (field-independent) $f_0$ and $\bar f_0$. A natural way to remedy this problem, which 
recovers the standard energies of the background when $f_0=1$ and  $\bar f_0=-1$, is to choose the constants $c_{U,V}$ as

\be
c_U = - \mu \bar \L \bar f_0\;, \;\;\;\;\; c_V = - \mu \L f_0 \label{solcUV}
\ee
case in which the charges become trivially integrable, and read 


\be
Q_f^{(0)} = f_0 H_L= \frac{1}{2} \oint d\s r_u\,  p\, \L  f(u)  \;,\;\; \;\;\;\;\; \bar Q_{\bar f}^{(0)} = -\bar f_0 H_R= - \frac{1}{2} \oint d\s r_v\,  p\, \bar \L  \bar f(v) \label{order0ch}
\ee
where we have used \eqref{engttb} to write the charges in a notation that resembles the $T\bar T$ one \eqref{QL}, and  we note that  for constant perturbations $\p_\s u = r_u$, defined in \eqref{defruv}. The $^{(0)}$ superscript indicates that this is the only contribution to the conserved charges upon the constant backgrounds.

 To obtain conserved charges that are generically non-zero, we first need to perturb the  black hole background  by an arbitrary allowed diffeomorphism $\eta$, which takes the form \eqref{xiradef} for some other functions $h (u), \bar h (v), h_r(u), \bar h_r(v)$.  We obtain the following expression for the difference in the charge  associated with a diffeomorphism $\xi$ on the background generated by $\eta$ 
%
%
\bea
\d_\eta Q_\xi  &=& \frac{1 }{2\pi } \oint d\s  \left[\frac{4p L_u r_u}{p+\b} f( u) h'( u) - \frac{4p L_v r_v}{p+\b} \bar f( v) \bar h'( v) + k r_u  h_{r}''( u) f( u) - k r_v  \bar h_{r}''( v) \bar f( v) \right.+ \nonumber \\ & & \hspace{-1 cm}+\; \left. \frac{k}{2 }(r_u f_{r}'( u)-r_v \bar f_{r}'( v)) \left(h'( u) + \bar h'( v)\right)-\frac{k}{2 }(f_{r}( u)+\bar f_{r}( v))\left(r_u h''( u) - r_v \bar h''( v) \right)\right]
\eea
Integrating by parts, the cross left-right terms cancel, and we are left with 
\bea
\d_\eta Q_\xi 
&= &\frac{1 }{2\pi } \oint d\s  r_u \left(\frac{4p L_u }{p+\b} f( u) h'( u) + k h_{r}''(u) f(u) -k f_{r}''(u) h (u)\right) + RM \nonumber \\
&=& \frac{1}{2} \oint d\s r_u \left(2 p \L f h' + \frac{k}{\pi} h_r'' f - \frac{k}{\pi} f_r'' h\right) + RM \label{leadstchalg}
\eea
where in the second line we have translated the result to 
 the $T\bar T$ parametrization \eqref{ttbpar}. 
 
 Let us now compare this  with the corresponding expression \eqref{delgQf} for double-trace $T\bar T$ - deformed CFTs, specialized to $\L = const.$ 
 
\be
\d_\eta Q_\xi^{\,double-trace} = 
 \frac{1}{2} \oint d\s \p_\s u \left(2 \L f h' - \frac{c}{12 \pi} f h''' \right) \label{chargeoe}
\ee
where we have used the fact that $\ell/8\pi G = c/(12\pi)$ in \eqref{delgQf}, where $c$ is the central charge of the undeformed CFT, and we dropped the winding term, which 
 vanishes on the constant $\L, \bar \L$ black hole backgrounds. 
The  expressions \eqref{leadstchalg} and \eqref{chargeoe} are almost the same, up to a factor of $p$ in the first term - which is related to the symmetric product orbifold - and the fact that the $\L$ - independent terms  in \eqref{leadstchalg} depend on more functions than their $T\bar T$ counterpart.

%

As discussed at the end of the previous subsection,  we would expect that the full set of boundary conditions on the asymptotically linear dilaton backgrounds  includes one that determines the radial function $F_r$ in terms of $f, \bar f$; however, the symplectic form analysis that we have performed is insensitive to this additional boundary condition. Requiring that the central contribution to the charge in \eqref{leadstchalg} match that in symmetric product  orbifolds of $T\bar T$ - deformed CFTs, which is $6 k p$ in our case, allows us to fix the radial function in terms of the other ones as

\be
f_r(u) = -\frac{p}{4} \, f'(u) \;, \;\;\;\;\; \bar f_r (v) = - \frac{p}{4}\, \bar f'(v) \label{choiceFr}
\ee
It would be very interesting if we could independently justify this form of $F_r$ from an asymptotic boundary condition on the fields. 


%
%
%
%

Thus, we find that not only do the asymptotic symmetries of the asymptotically linear dilaton background \emph{precisely match} (for $F_r=0$) those of the AdS$_3$ backgrounds with mixed boundary conditions that are dual to double-trace $T\bar T$ - deformed CFTs, but also the  conserved charges match \emph{exactly} for a particular choice \eqref{choiceFr} of the radial function $F_r$ that is consistent with the results of the symplectic form analysis, but  could not be singled out by it. This coincidence  is quite   remarkable, 
  %
%
given that the background metrics \eqref{bhbck} and \eqref{g24UV} are entirely different, as are the boundary conditions that their fluctuations obey. Even more remarkably,  
this identification continues to hold when the two backgrounds are perturbed, as we show in the next section.

\section{The charge algebra \label{asychalg}}

In this section, we use the covariant phase space formalism to compute the algebra of the conserved charges \emph{perturbatively} above the asymptotically linear dilaton black hole backgrounds and show the result  is identical to the nonlinear $T\bar T$ charge algebra \eqref{ttbalg} to the order we checked. We start by reviewing a few generalities.

The standard definition of the charge algebra in the covariant phase space formalism is  

\be
\{Q_{\xi},Q_{\chi}\} \equiv \d_\chi Q_\xi = \oint \boldsymbol k_\xi (\L_\chi \Phi,\Phi) \label{chalgstdef}
\ee 
where the last term  corresponds to the difference in the charge  associated with some diffeomorphism, $\xi$, between two backgrounds that differ by some other diffeomorphism, $\chi$. This definition relies on the representation theorem \cite{Barnich:2001jy,Compere:2018aar}, which relates the right-hand-side of \eqref{chalgstdef} to the charge associated to the Lie bracket $[\xi, \chi]$ of the two diffeomorphisms 

\be
\d_\chi Q_\xi = Q_{[\xi,\chi]} \label{strepthm}
\ee
assuming the charges are integrable. 
 This conclusion continues to hold when the diffeomorphisms are field-dependent, provided one replaces the Lie bracket with a modified Lie bracket \cite{Barnich:2010eb,Barnich:2010xq}

\be \label{modbr}
[\xi, \chi]_* \equiv [\xi, \chi]_{L.B.} - \d_\xi \chi + \d_\chi \xi
\ee
in the representation theorem. It is important to note that in this formalism the  integrability \cite{Barnich:2007bf} of the conserved charges - which is not guaranteed by the definition \eqref{chdiffbb}, but needs to be checked separately - plays a crucial role in obtaining the correct algebra, especially when the diffeomorphisms are field-dependent. This fact is neatly  exemplified  in our analysis, presented in appendix \ref{conschdtrttb}, of the asymptotic symmetries of the spacetime dual to the double-trace $T\bar T$ deformation,  where integrability is essential for reproducing the correct charge algebra when the generators are rescaled by a field-dependent factor.


\subsection{Perturbative calculation setup }

The aim of this section is to set up the computation of the charge algebra for the asymptotically linear dilaton black hole backgrounds. This is significantly more complicated than for the AdS$_3$ spacetimes with mixed boundary conditions  dual to double-trace $T\bar T$-deformed CFTs because the allowed modes are only known \emph{perturbatively} around these constant backgrounds, whereas in the double-trace case the general, non-linear solution \eqref{g24UV} for the allowed metrics is known.


A quick glance at \eqref{strepthm} suffices to understand  that the allowed diffeomorphisms we have found so far are \emph{insufficient} to determine the full algebra of the conserved charges.    The limitation stems from the fact that  all the non-trivial charges (i.e., except the energy and momentum) evaluate to zero on the constant black hole backgrounds, and thus $Q_{[\xi,\chi]}$ can only receive non-vanishing contributions from the zero mode of the Lie bracket of the two diffeomorphisms. Working in a Fourier basis for the functions $f,g$ that parametrize the two diffeomorphisms, with $f = e^{i m u/R_u}$ and $g = e^{i n u/R_u}$, the computation of the algebra upon the constant backgrounds will only be able to recover the contributions to the $\{ Q_m, Q_n\}$ commutator that have $m+n=0$. 
 
Let us briefly discuss this zeroth-order result before moving on to the more general contributions.  
The computation of the charge variation
$\d_\eta Q_\xi$ associated to two diffeomorphisms $\xi_f$ and $\eta_h$ upon the constant backgrounds has already been performed in \eqref{leadstchalg}, 
and  now we simply need to interpret it
as the commutator $\{ Q_\xi, Q_\eta\}$ of the two conserved charges. 
%
 The result can be made more transparent by an integration by parts, which replaces the  $2 f h'$ factor in the first term of \eqref{leadstchalg}  by $f h'-h f'$. We thus obtain

\be
\{ Q_f, Q_h \} =  Q^{(0)}_{f h'-f' h} - \frac{k}{\pi} \oint d\s r_u f_r'' h \label{O1chalg}
\ee
where the superscript ${}^{(0)}$  indicates that the charge associated with the given diffeomorphism should be evaluated on the constant black hole backgrounds, where only the zero mode of the function contributes. This algebra is similar to a centrally-extended Witt algebra.
%
%
 Further making the  choice $f_r = - p f'/4$ that was discussed  previously and working in the Fourier basis for the generators, the charge algebra we obtain to this order is

\be
\bigg. \{Q_m, Q_{n}\} \bigg|_{m+n=0}=\; \frac{2m}{R_u}  Q_{0} + \frac{6 k p}{12} \cdot \frac{m^3}{ R_u^2}\,  \label{zerothochalg}
\ee
where 
$Q_0 = H_L$ and $R_u = R \, r_u$, with $r_u$ given in \eqref{defruv}.  
 The factors of $R_u$ are due to the fact that the periodicity of the  coordinate $u$ that enters the diffeomorphisms is field-dependent; note that they perfectly match the structure \eqref{ttbalg} expected from $T\bar T$ - deformed  CFTs.  A similar result holds  for the right-movers. The central term has been \emph{ chosen} to match the prediction from a symmetric product orbifold of $T\bar T$ - deformed  CFTs with $c=6kp$. While it would have been far more valuable to have \emph{derived} 
this central extension by finding a natural boundary condition that leads to \eqref{choiceFr}, the fact that were were able to find some form of the radial function that yields precisely this value is 
 still a non-trivial result; for example, such a choice would not have been available had we been working in string frame radial gauge. 
 

 
 Let us now discuss the
  computation of  the charge algebra for  generic $m,n$. For this, we need to consider a perturbed
%
%
%
background that generically has non-zero charges with respect to arbitrary diffeomorphisms belonging to the asymptotic symmetry group. As in the previous section, we can generate such a  perturbed  background by acting with an allowed diffeomorphism, $\eta$, on the constant background, $\bar{g}$
\be
g = \bar g + \e \, \L_\eta \bar g \label{defbck}
\ee
According to our discussion, the charges associated to  allowed diffeomorphisms $\xi_{f,\bar f}$ of such a background satisfy 

\be
Q_{\xi_{f,\bar f}} [g] = \left\{ \begin{array}{ccc} \O(1) & \mbox{for} &  f, \bar f = const.  \vspace{2mm} \\  \O(\e)& \mbox{for} & f,\bar f \neq const.\end{array} \right.
\ee
To compute the charge algebra, we  will subsequently  act with another allowed diffeomorphism, $\chi$, on this deformed background and then, according to the recipe \eqref{chalgstdef}, we will compute the charge difference $\d_\chi Q_\xi$ associated to a third allowed diffeomorphism, $\xi$.  
The terms in the algebra 
 with $m+n \neq 0$ are  visible at $\O(\e)$. Note, however, that the diffeomorphisms $\xi, \chi$ themselves can receive corrections at $\O(\e)$, which will, in principle, contribute to the charge algebra at the same order. Consequently, in order to find the full answer for the charge algebra at $\O(\e)$, we do need to work out the corresponding  correction to the diffeomorphisms. From the point of view of the expected algebra \eqref{ttbalg}, the $\O(\e)$ analysis  reveals the generic linear terms, as well as the non-linear terms with either $m$ or $n$ zero, whereas the generic non-linear terms  appear only at $\O(\e^2)$.  

The $\O(\e)$ correction to the diffeomorphisms \eqref{xiradef} can again be determined using  the symplectic form,  now evaluated  on  the perturbed backgrounds \eqref{defbck}.
 This analysis  does not, however,  completely fix the correction, and 
complementary arguments  need to be invoked.  
One such argument is based on the representation theorem, which relates the $\O(\e)$ modification to the vectors (which appears in the modified Lie bracket \eqref{modbr}, but without the factor of $\e$) to the charge algebra at $\O(1)$. Thus, knowing the latter, we can obtain further constraints on the diffeomorphisms themselves. This argument still does not completely fix the vectors, and a ``minimal continuation'' assumption on the functions that parametrize them is also needed.

The final solution for the asymptotic diffeomorphisms up to $\O(\e)$ \emph{precisely agrees} with the corresponding expression in double-trace $T\bar T$ - deformed CFTs, particularized to perturbations around a constant background, that we work out in appendix \ref{pertanctb}. Finally equipped with the solution for the vectors, we derive the charge algebra to $\O(\e)$ and find a \emph{perfect match} to the charge algebra in double-trace $T\bar T$ - deformed CFTs, computed to the same order.  
Throughout this section, we use the natural $T\bar T$ parametrization \eqref{ttbpar} of the black hole backgrounds, in order to facilitate the comparison with the double-trace results.


\subsection{Allowed diffeomorphisms upon the perturbed backgrounds \label{alloweddiffs}}

As explained, we consider again the perturbed background obtained by acting with an allowed diffeomorphism $\eta_{h,\bar h}$, of the form \eqref{xiradef} with $c_{U,V}$ given by \eqref{solcUV}, on the black hole background $\bar g$. For simplicity, we set the radial component of $\eta$ to zero,  since the symplectic form analysis is not expected to constrain it very much, and keeping  it would significantly complicate the calculations. 
Thus,  $\eta$ is parametrized by two periodic functions,  $h(u)$ and $\bar{h}(v)$, of the field-dependent coordinates \eqref{fdepcoordst} only. In terms of the $T\bar{T}$ notation, it takes the form

\begin{align}
\eta&=\left[h(u)-\mu\bar{\L}(\bar{h}(v)-\bar{h}_0)\right]\p_U+\left[\bar{h}(v)-\mu\L(h(u)-h_0)\right]\p_V \label{xiradsimp}
\end{align}
The goal of this subsection is to determine the first-order correction to the allowed diffeomorphisms   of the perturbed background \eqref{defbck}.

As for the case of the black hole backgrounds, we expect  the answer to be largely fixed by requiring that the symplectic product of the diffeomorphism we search for with the perturbation of the background that is generated by varying the black hole parameters $ L_{u,v}$ vanish asymptotically. The result takes the form
\be
\xi = \xi^{(0)} + \e \, \xi^{(1)} \label{pexpdiff}
\ee
where $\xi^{(0)}$ was previously determined in \eqref{xiradef},  and $\xi^{(1)}$ is to be computed  by imposing the requirements \eqref{phs}. More concretely, this perturbative computation takes the form
\bea
&&\om_{\bar g + \e \L_\eta \bar g} \left(\frac{\d}{\d L_i} (\bar g + \e \L_\eta \bar g), \L_{\xi^{(0)} + \e \xi^{(1)}} (\bar g + \e \L_\eta \bar g) \right) = \om_{\bar g + \e \L_\eta \bar g} \left(\frac{\d \bar g}{\d L_i} , \L_{\xi^{(0)}} \bar g\right) + \e\,  \om_{\bar g} \left(\frac{\d  \L_\eta \bar g}{\d L_i}, \L_{\xi^{(0)}} \bar g\right) + \nonumber\\
& & \hspace{1.2cm}+ \; \e \, \om_{\bar g} \left(\frac{\d \bar g}{\d L_i} , \L_{\xi^{(1)}} \bar g + \L_{\xi^{(0)}} (\L_\eta \bar g)\right) + \O(\e^2)
\eea
where for simplicity we have suppressed the factor of the dilaton from the background notation and  $L_i=L_{u,v}$
. One important remark is that our analysis  determines the metric perturbation only up to terms that have zero symplectic product with $\d_{L_i} \bar g$, such as the constant modes $C_b$ and  $\hat C_{uv}$ discussed in the previous section. In principle, we have the freedom to add such contributions to $\L_{\xi^{(1)}} \bar g$ without violating the requirements on $\boldsymbol \omega$, as long as we do not generate winding for the fixed coordinates $U,V$.

While we perform the computations on the string backgrounds, it is useful to transcribe the results in terms of the natural $T\bar T$ notation \eqref{invttbpar}. 
The vanishing of the symplectic flux through the boundary at $\O(\e)$ imposes
the following constraints on the components of the correction $\xi^{(1)}$ to the allowed diffeomorphisms  
\be 
\p_u \left(\xi^{(1)V} - \mu \L \,  \xi^{(1)U} \right) =  \; 2 \mu \L  \, h' (f'(u) + \bar f'(v)) \nonumber
\ee

\be 
\p_v \left(\xi^{(1)U} - \mu \bar \L \,  \xi^{(1)V} \right) =  \; 2 \mu \bar \L \, \bar h' (f'(u) + \bar f'(v)) \label{ecordere}
\ee
where $u,v$ are the field-dependent coordinates \eqref{fdepcoordst} associated with the unperturbed background. The solution is given by 
\bea
 \xi^{(1)U} \!\!  & = &\!\!\frac{1}{1-\mu^2 \L \bar \L} \left[F_h (u) + \mu \bar \L \bar F_{\bar{h}} (v) + 2 \mu^2 \L \bar \L \left(\int^u  h' f' + (h-h_0) \bar{f}' \right) + 2 \mu \bar \L \left(\int^v \bar{h}' \bar{f}' + (\bar{h} -\bar{h}_0)f'\right) \right]~~~~~~~~ \nonumber \\
 \xi^{(1)V} \!\! & = & \!\! \frac{1}{1-\mu^2 \L \bar \L} \left[\bar{F}_{\bar{h}} (v) + \mu  \L F_h (u) + 2 \mu^2 \L \bar \L \left(\int^v \bar{h}' \bar{f}' + (\bar{h}-\bar{h}_0) f' \right) + 2 \mu  \L \left(\int^u h' f' + (h-h_0) \bar{f}'\right) \right]\label{correction2}
\eea
where $F_h,\bar{F}_{\bar{h}}$ are so far unfixed, background-dependent  functions of the field-dependent coordinates, which represent the continuation of the $\O(1)$ functions $f,\bar{f}$ to the perturbed background generated by $h,\bar{h}$, and the primitives of $h'f'$ and $\bar h' \bar f'$ are integrated over a corresponding field-dependent variable  $\tilde u \, $/\,$\tilde v$  that we have ommitted in order to simplify the notation. These primitives are defined to not contain any  constant Fourier mode, and thus all the possible integration constants that arise are included in the functions $F_h,\bar{F}_{\bar{h}}$; the zero modes of $h, \bar h$ were subtracted for further convenience. The most general solution for the vector field should also include contributions from the zero-frequency modes $C_b$ and $\hat C_{uv}$ with arbitrary coefficients, since including them does  not affect the symplectic form. While these terms will turn out to be essential for obtaining the correct algebra, we can simply add them at the end, without affecting the result of the discussion below.

It is interesting to compare the corrections \eqref{correction2} with the expansion of the allowed diffeomorphisms for the spacetime dual to double-trace $T\bar T$ - deformed CFTs, a computation we perform in appendix \ref{appendixB1}. We find
that the choice 

\be
F_h = 2 \mu^2 \L \bar \L \bigg(\int^u ( h-h_0) f'' + (hf')_{zm} \bigg)\;, \;\;\;\;\; \bar{F}_{\bar{h}} =2 \mu^2 \L \bar \L \bigg(\int^v (\bar{h}-\bar h_0) \bar{f}''+(\bar{h}\bar{f}')_{zm}\bigg) \label{choiceFF}
\ee
- where the subscript stands for the ``zero mode'' of the corresponding expression -
precisely reproduces the expansion \eqref{difper2} of the allowed diffeomorphisms 
in that theory
up to $\O(\e)$ about the background of constant parameters, barring winding terms. However, from the point of view of the asymptotic analysis of the charge algebra in asymptotically linear dilaton backgrounds, this choice needs to be justified from first principles.

For this purpose, we use the representation theorem, which should be satisfied order by order in the perturbation, provided we use the modified Lie bracket \eqref{modbr} that takes into account the field-dependence of vector fields\footnote{As noted in appendix \ref{conschdtrttb}, the modified Lie bracket \eqref{modbr} needs to be further modified in order for the representation theorem to hold for the spacetimes dual to double-trace $T\bar T$ - deformed CFTs. These subtleties are however linked to winding terms, and do not appear at the order in perturbation theory that we are discussing. The representation theorem then (almost) takes its standard modified form, as we show in appendix \eqref{commrepthm}.  }. To ascertain the necessity of this modified bracket, let us consider the representation theorem at $\O(1)$. 
 The correction to the usual Lie bracket is given by the change  in a vector field under the action of another, which  precisely corresponds to the $\O(\e)$ term in \eqref{pexpdiff}. For simplicity, we will focus on the ``left" vector fields, obtained   by setting all the functions of $v$ to zero in \eqref{xiradsimp}. The standard Lie bracket of two such $\O(1)$ vector fields is 

\be
[\xi, \chi]_{L.B} = \frac{1+\mu^2\L\bar{\L}}{1-\mu^2\L\bar{\L}}(fg'-gf')(\partial_U+\mu\L\partial_V)+\frac{\mu^2\L\bar{\L}}{1-\mu^2\L\bar{\L}}(g_0 f'-f_0 g')(\partial_U+\mu\L\partial_V)
\ee
Its  associated conserved charge on the constant backgrounds only receives contributions from the first term, and one can easily check  that it does not agree with  the $\O(1)$ charge algebra that we have already computed in \eqref{O1chalg}. 
 Thus, in order for the representation theorem to be obeyed, the  modification \eqref{modbr} to the Lie bracket is required to be

\begin{align}\label{diferenta}
\delta_{\xi}\chi-\delta_{\chi}\xi=\frac{2\mu^2\mathcal{L}\mathcal{\bar{L}}}{1-\mu^2\mathcal{L}\mathcal{\bar{L}}}(fg'-gf')(\partial_U+\mu\mathcal{L}\partial_V) + \ldots 
\end{align}
where the $\ldots$ stand for terms that integrate to zero upon the constant backgrounds. It seems natural that these terms should take the form  $\sum_{n>0} c_n (f_0 g^{(n)} - g_0 f^{(n)})(\p_U+\mu\L\p_V)$ for some constants $c_n$, as the correction needs to be bilinear in $f,g$, antisymmetric, and generically not posess a zero mode.

At the same time,  $\delta_{\xi}\chi=\chi^{(1)}$ is the correction to $\chi$ on the background generated by $\xi$, which is given in \eqref{correction2}, and vice-versa. 
We therefore have

\be
\delta_{\xi}\chi-\delta_{\chi}\xi=\frac{1}{1-\mu^2\mathcal{L}\mathcal{\bar{L}}}\big(G_f-F_g)(\partial_U+\mu\L\partial_V) \label{diffvectvar}
\ee
Equating the last two expressions immediately yields a constraint on the difference $G_f - F_g$. To further narrow it, note that when, say, $f=const$, both $G_f$ and $F_g$ need to vanish, irrespectively of what $g(u)$ is: $G_f$ vanishes because a constant coordinate shift is an isometry of the black hole  backgrounds, and thus the asymptotic symmetries cannot depend on it, while $F_g$ vanishes because $\xi_f$ is a background-independent diffeomorphism in this case. Taking this into account, 
%
%
%
the $\O(\e)$ information from the Lie bracket fixes  the difference to
\begin{align}\label{constrdiferenta}
G_f-F_g=2\mu^2\L\bar{\L}(fg'-gf'+ g_0 f'-f_0 g')
\end{align}
Applying  the same reasoning for to the right-moving vector fields, we obtain 
\begin{align}
\bar{G}_{\bar{f}}-\bar{F}_{\bar{g}}=2\mu^2\L\bar{\L}(\bar{f}\bar{g}'-\bar{g}\bar{f}'+ \bar{g}_0 \bar{f}'-\bar{f}_0 \bar{g}')
\end{align}
Thus, the zeroth order modified Lie bracket can partly fix the $\O(\e)$ correction to the allowed diffeomorphisms. The fact that the representation theorem does not fully fix these functions is not surprising. A comparison with the analogous computation in standard AdS$_3$ with Brown-Henneaux boundary conditions shows that an assumption that $f$ be minimally continued to higher orders is also necessary. The difference is that for the asymptotically linear dilaton backgrounds, the minimal continuation does require $F_h$ to be non-zero.


Guided by our $T\bar T$ intuition, and  since $u$ depends on the background parameters, the minimal continuation at $\O(\e)$ of $f(u)$ should be due to a change in the argument of the function, which implies that
%
%
$\xi^{(1)U}=f'\delta_h u$. Comparing this with the general solution \eqref{correction2} with all the functions of $v$ set to zero, we identify 
\begin{align}\label{deltau}
\frac{1}{1-\mu^2\L\bar{\L}}\bigg(F_g+2\mu^2\L\bar{\L}\int^u g' f'\bigg)=f'\delta_g u
\end{align}
Using \eqref{constrdiferenta}, it follows that
\begin{align}
g'\delta_f u-f'\delta_g u=\frac{1}{1-\mu^2\L\bar{\L}}(G_f-F_g)=\frac{2\mu^2\L\bar{\L}}{1-\mu^2\L\bar{\L}}\big(fg'-gf'+g_0 f'-f_0 g'\big)
\end{align}
From here we conclude that  $\delta_f u= \frac{2\mu^2\L\bar{\L}(f- f_0)}{1-\mu^2\L\bar{\L}}+\g f'$, where $\g$ can be any constant, reflecting the possibility of adding terms that will drop out from the expression above because of antisymmetry.  The minimal choice, which sets such terms to zero\footnote{In the next subsection, we show such terms do not contribute to the charge algebra.}, plugged into \eqref{deltau}, yields 
\begin{align}
F_g&=2\mu^2\L\bar{\L}\big(f'(g- g_0)-\int^u f'g'\big)=2\mu^2\L\bar{\L}\bigg(\int^u (f'g)'+(f'g)_{zm}-\int^u f'g'- g_0 f'\bigg)=\nonumber\\
&=2\mu^2\L\bar{\L}\big(\int^u (g- g_0) f'' + (f'g)_{zm}\big)
\end{align}
where  we are still using  the convention that our primitives do not contain a zero mode. The result precisely coincides with the value of this function in $T\bar T$ - deformed CFTs,  
which is rather remarkable, given that our  analysis   only used    the $\O(1)$ representation theorem in the  asymptotically linear dilaton backgrounds and the assumption that the $\O(1)$ vector fields should be minimally continued at next order. The right-movers work in an analogous way.


 The full solution for the  correction  to the vector field is\footnote{The leading behaviour of the $\boldsymbol \omega_{ra}$ components of the symplectic form  
 evaluated on this solution is given by
 $1/r$ terms that can be written as \eqref{exactform}, and will thus be ignored.} 

\bea
\xi^{(1)U}\! \! &=&\! \!\frac{2\mu\bar{\L}(\mu\L (h- h_0)+\bar{h}-\bar{h}_0)}{1-\mu^2\L\bar{\L}}f'+\mu\bar{\L}\frac{2\mu\L (h-h_0+\mu\bar{\L}(\bar{h}-\bar{h}_0))}{1-\mu^2\L\bar{\L}}\bar{f}'+2\mu\bar{\L}\int^v\bar{f}'\bar{h}' + \xi^U [\hat C_{uv}, C_b]~~~~~~~~~ \nonumber\\
\xi^{(1)V}\! \! &=&\! \!\frac{2\mu\L(\mu\bar{\L} (\bar{h}-\bar{h}_0)+h-h_0)}{1-\mu^2\L\bar{\L}}\bar{f}'+\mu\L\frac{2\mu\bar{\L} (\bar{h}-\bar{h}_0+\mu\L (h- h_0))}{1-\mu^2\L\bar{\L}}f'+2\mu\L\int^u f'h' + \xi^V [\hat C_{uv}, C_b] \label{finaloediff}
\eea
where, as promised, we have included back the terms proportional to  $ \hat C_{uv}$ and $C_b$ that can be freely added to the solution and read
\be
\xi^U[\hat C_{uv}, C_b] = \bigg(\hat C_{uv}+\frac{C_b}{4}\bigg)U+\frac{2\mu \bar{\L}\, \hat C_{uv} }{1+\mu^2\L\bar{\L}}V \;, \;\;\;\; \xi^V[\hat C_{uv}, C_b] = \frac{2\mu \L \, \hat C_{uv}}{1+\mu^2\L\bar{\L}}U+\bigg(\hat C_{uv}-\frac{C_b}{4}\bigg)V \label{xicuvcb}
\ee
%
Remember that upon the constant backgrounds, diffeomorphisms of this form were disallowed, as they would change the identification of the compact coordinate, $\s$. Here, however, the part of the correction to the allowed diffeomorphisms that \emph{is} fixed by the symplectic form analysis already has winding (due to the terms under the integral in \eqref{finaloediff}).
%
This winding  can be off-set by  choosing the constants $\hat C_{uv}, C_b$ that parametrize \eqref{xicuvcb} in such a way that the periodicity of the coordinates $U,V$ is   overall unaffected. This choice corresponds to

\be
\hat C_{uv} =-\frac{\mu(\bar{\L}\oint dv \bar{f}'\bar{h}'+\L\oint du f'h')(1+\mu^2\L\bar{\L})}{2\pi R(1+\mu \L)(1+\mu\bar{\L})}  \nonumber
\ee

\be
 C_{b} = \frac{4\mu \L  (1+\mu^2 \L \bar \L + 2 \mu \bar \L) \oint du f' h' - 4 \mu \bar \L  (1+\mu^2 \L \bar \L + 2 \mu \L)\oint dv \bar f' \bar h'}{2\pi R(1+\mu \L)(1+\mu \bar \L)} \label{solcuvcb}
\ee
With this choice, the diffeomorphisms \eqref{finaloediff} are \emph{identical} to their double-trace counterparts \eqref{vectorexp}. 

We will henceforth find it convenient to write the full allowed diffeomorphisms up to $\O(\e)$ in the form
\be
\xi = \xi_{(p)} + \xi_{(w)} \label{xidecomp}
\ee
where $\xi_{(p)}$ corresponds to the part of the diffeomorphism  that is fixed by the symplectic form analysis and the representation theorem, while  $\xi_{(w)}$ - which is given by  \eqref{xicuvcb} with the choice \eqref{solcuvcb} for the parameters - will be referred to as a ``compensating'' diffeomorphism, in the sense that its presence is required to off-set the unphysical winding introduced by $\xi_{(p)}$. As is clear from the solution above,  
%
%
 $\xi_{(p)}$ receives contributions at both zeroth and first order in the perturbation, whereas the expansion of $\xi_{(w)}$ starts at $\O(\e)$. It is interesting to rewrite the  compensating diffeomorphisms in terms of the field-dependent coordinates $u,v$, with the result

\be
\xi_w^U = w_f u + \mu \bar \L w_{\bar f} \;, \;\;\;\;\; \xi^V_w = w_{\bar f} v + \mu \L w_f u
\ee
where 
\be
w_f = \frac{2\mu^2\L\bar{\L}\oint du f'h'-2\mu\bar{\L}\oint dv \bar{h}'\bar{f}'}{2\pi R(1+\mu\bar{\L})} \, \e \;, \;\;\;\;\; w_{\bar f} = \frac{2\mu^2\L\bar{\L}\oint dv \bar{f}'\bar{h}'-2\mu\L\oint d uh'f'}{2\pi R(1+\mu \L)} \, \e \label{windingst}
\ee
These expressions for the winding precisely coincide with the corresponding perturbative expansion \eqref{pertwind} of the winding  in double-trace $T\bar{T}$ - deformed CFTs. They can  alternatively be rewritten in the form \eqref{windingffb} by  noting that  $ \e \,\L \oint du f' h' $ and $ \e \,\bar  \L \oint dv \bar f' \bar  h' $ are nothing but the conserved charges  $Q_{f'}/p$ and, respectively, $-\bar Q_{\bar f'}/p$ \eqref{leadstchalg} upon the perturbed background.

Finally, we  need to show, as promised,  that adding these winding terms in the final step of our computation does not affect the previous steps. Had they been included from the beginning, they could have affected  the expression \eqref{diffvectvar} for $\d_\xi \chi-\d_\chi \xi$. However, from the explicit form of the solution above, we see that $C_b,\hat C_{uv}$ are symmetric in $f\leftrightarrow g,\bar{f}\leftrightarrow\bar{g}$. Hence, \eqref{diffvectvar} is not affected by adding the compensating diffeomorphisms, and thus the arguments presented in this section are unaffected.
%



\subsection{Algebra of the conserved charges and match to $T\bar T$}

Armed with the knowledge of the allowed diffeomorphisms up to $\O(\e)$, we are now finally ready to compute the algebra of conserved charges to the same order. For this, it is useful to split the  allowed diffeomorphisms as in \eqref{xidecomp} 
 and to organise the computation into four separate pieces that correspond, schematically, to
\begin{align} \label{algebraoe}
\delta_{\chi_{(p)}+\chi_{(w)}}Q_{\xi_{(p)}+\xi_{(w)}}&=\delta_{\chi_{(p)}}Q_{\xi_{(p)}}+\delta_{\chi_{(w)}}Q_{\xi_{(p)}}+\delta_{\chi_{(p)}}Q_{\xi_{(w)}}+\delta_{\chi_{(w)}}Q_{\xi_{(w)}}
\end{align}
Since the compensating diffeomorphisms are at least $\O(\e)$,  the last term in \eqref{algebraoe} does not contribute to the order we are interested in.

We first concentrate on the algebra of the ``left-movers'', obtained by setting all the periodic functions of $v$ to zero. The diffeomorphism $\xi$ is parametrized by the periodic function $f$, and $\chi$ by\footnote{In principle, one may worry that the variation $\d_\chi \xi$ of the diffeomorphism between the two backgrounds could contribute to the charge difference. One can nevertheless show  (see e.g. \cite{Compere:2015knw}) that this is not the case. } $g$. 
After various integrations by parts, the $\O(\e)$ contribution to the periodic-periodic charge difference is
\begin{align}
\bigg.\d_{\chi_{(p)}}Q_{\xi_{(p)}} \bigg|_{\O(\e)}\!\!\!\!&=p \, \L \! \oint \! d\s r_u(fg'-gf')h'+ \frac{\mu^2\L\bar{\L}}{1-\mu^2\L\bar{\L}}\, p \, \L \! \oint\! d\s r_u(f_0 g'h'+g_0 f'h')- 2\pi R_u p\L w_f g_0 - \nonumber\\
&- \frac{p\L}{2\pi R} \frac{2\mu^2\L\bar{\L}}{1-\mu^2\L\bar{\L}} \oint d\s r_u  f'h' \oint d\s ug'
\end{align}
where we have introduced the field-dependent radius of the $u$ coordinate, $R_u = R r_u$, and similarly $R_v =  R r_v$. 
The first term above is precisely the $\O(\e)$ contribution to $Q_{fg'-gf'}$, which agrees with the zeroth order result \eqref{O1chalg}, but is now generically non-zero. Using \eqref{order0ch}, the zero modes $f_0$, $g_0$ of the functions can be traded, at this order, for the zeroth order conserved charges. The expression can be further simplified  by writing it in terms of the the winding \eqref{windingst}, particularized to $\bar{f}=0$. The final result is
%
%
\begin{align}\label{persgtr}
\d_{\chi_{(p)}}Q_{\xi_{(p)}}&=Q_{fg'-gf'}+ w_g Q_f^{(0)}- w_f Q_g^{(0)}-p\, \L  w_f\oint d\s r_u ug'
\end{align}
This precisely matches the analogous charge difference in the perturbative  double-trace $T\bar{T}$ case, which is computed in appendix \eqref{pertchalg}.

 The second term in \eqref{algebraoe} vanishes for the particular choice \eqref{solcUV} of integration constants in the  $\O(1)$ vector fields that was required by integrability. The details are given in appendix \eqref{pertchalg} for the double-trace $T\bar T$ case, which works identically. Finally, the third term in \eqref{algebraoe} cancels the integral of the non-periodic function in \eqref{persgtr} with no finite contribution, leading to the following final result for the algebra of the left-movers 
\begin{align}
\{ Q_f, Q_g\} \equiv \delta_{g}Q_{f}&=Q_{fg'-gf'}+w_g Q_f^{(0)}-w_f Q_g^{(0)}
\end{align}
which holds  up to  $\O(\e)$. This can be rewritten in terms of the conserved charges as

\vskip 2mm

\be\boxed{
\{ Q_f, Q_g\} =
Q_{fg'-gf'}+\frac{\mu^2 H_R}{ p^2 \pi^2 R R_H} \left(Q_{g'}Q_f^{(0)}-Q_{f'}Q_g^{(0)}\right)} \label{QfQgalg}
\ee
\vskip 2mm
\noindent where we defined $R_H=R(r_u+r_v-1)=R (1+\mu\L)(1+\mu\bar{\L})/(1-\mu^2\L\bar{\L})$. Using a Fourier basis for $f,g$ and taking into account  the fact that $f' = i m f/R_u $, we find precisely the algebra \eqref{ttbalg} with $\mu \r \mu/p$ that was advertised in the introduction. A similar result holds for the right-movers. 

The commutation relations of the charges associated with one left-moving and one right-moving diffeomorphism can be obtained by setting $\bar f(v) =0$ in $\xi$ and $g(u)=0$ in $\chi$. The steps of the computation are explained in appendix \eqref{pertchalg} for the case of the double-trace $T\bar T$ deformation, which turn out to be identical to this order in perturbation theory to our computation in the asymptotically linear dilaton background. The  result is  

\vskip 2mm

\be
\boxed{\{ Q_f, \bar Q_{\bar g} \} =
w_g Q_f^{(0)}-w_{\bar{f}}\bar{Q}_{\bar{g}}^{(0)}=\frac{\mu}{\pi p R R_H}\left(R_v\bar{Q}_{\bar{g}'}Q_f^{(0)}+R_u Q_{f'}\bar{Q}_{\bar{g}}^{(0)}\right)} \label{QfbarQgalg}
\ee

\vskip 2mm
%

\noindent where the windings are again given by \eqref{windingst}, particularized to $\bar f = g =0$.

 Hence, the charge algebra that we obtain at $\O(\e)$ is \emph{non-linear} in the conserved charges $Q_f,\bar{Q}_{\bar f}$. The source of non-linearity is the presence of charge-dependent parameters\footnote{A similar mechanism for obtaining a non-linear charge algebra has recently appeared in \cite{Fuentealba:2021yvo,Fuentealba:2022yqt} for the case of five-dimensional asymptotically flat spacetimes.}, namely the windings which, as discussed in the previous subsection, are required for consistency of the asymptotic coordinate transformations. 
This algebra \emph{precisely matches} the algebra of the ``unrescaled'' generators  in single-trace $T\bar T$ - deformed CFTs, where the factors of $p$ are exactly those required by the symmetric product orbifold \cite{ttbspo}.  We have thus succeeded to establish an additional important link between asymptotically linear dilaton backgrounds and the single-trace $T\bar T$  deformation, by showing that the full doubly-infinite set of extended symmetries of the two theories are \emph{identical} to the order we checked.  

We end this section with a few technical comments. First, we checked that 
%
 the redefinition $\d_f u\mapsto \d_f u+\gamma f'$ that adds a term $\gamma f'h'/(1-\mu^2\L\bar{\L})$ to $F_h$ and could not be fixed by the arguments of the previous section contributes to the  charge difference  \eqref{algebraoe} only   a total derivative term, which vanishes upon integration. We were therefore justified to neglect this term.

Second, it would be interesting to show that we obtain the charge algebra  \eqref{nonlineartt} for \emph{arbitrary} functions $f,g$, rather than just when one of them  is a constant. This requires computing the  $\O(\e^2)$  corrections to the charge algebra. While, in principle, this computation can be performed along the same lines of the $\O(\e)$ one,  it nevertheless becomes extremely cumbersome. 
 One may however wonder  whether the interesting non-linear terms in the algebra, which are proportional to the windings, could not be recovered just from the $\O(\e)$ solution for the  vector fields, given that they are always multiplied by the winding, which starts at $\O(\e)$. This expectation turns out to be too na\"{i}ve, for two reasons: first, the non-linear contributions include terms where the charges are at $\O(1)$ and therefore  the windings are at $\O(\e^2)$; these ``compensating'' windings cannot be computed without knowing the full $O(\e^2)$ correction to the $\xi_{(p)}$  part of the vector fields that is determined from the symplectic form analysis. Second, as can be seen \eqref{vectorexp} for the doube-trace $T\bar{T}$ computation expanded up to $\O(\e^2)$ around the constant backgrounds, the $\xi_{(p)}$ piece of the vector fields receives $\O(\e^2)$ contributions that depend on the  windings at $\O(\e)$ 
 and enter the result of the charge algebra. Thus,  the computation cannot be performed  without explicit knowledge of the $\O(\e^2)$ corrections to the allowed diffeomorphisms.

Finally, it is natural to ask whether the algebra \eqref{QfQgalg}, \eqref{QfbarQgalg} can be linearized by a redefinition of the generators. Guided by the $T\bar{T}$ results, we expect that the rescaled generators $R_u Q_f$ and $R_v \bar Q_{\bar f}$ will simply satisfy a Virasoro $\times $ Virasoro algebra. As explained at length in appendix \ref{conschdtrttb}, the effects of a field-dependent rescaling  of the diffeomorphisms on the associated charge algebra are somewhat non-trivially implemented in the covariant phase space formalism via a different solution to the charge integrability constraints. Concretely, starting from the $\O(1)$ expression \eqref{integr} and temporarily setting $\bar f_0 =0$ for simplicity,  we can easily note that the charges are again integrable for the field-dependent choice

\be
 f_0 =  R_u \tilde f_0  = \left( R + \frac{\mu H_R}{\pi p} \right) \tilde f_0 \;, \;\;\;\;\; c_V= - \mu \L f_0 - \frac{\mu H_L}{\pi p} \tilde f_0 = -  \frac{2\mu H_L}{\pi p}\tilde f_0 \label{newsolint}
\ee
where $\tilde f_0$ is now field-independent, and we have used \eqref{defruv} and the fact that $H_L = \pi p \L R_u$ to manipulate the expressions. A similar choice holds for the right movers. The charges \eqref{integr} associated to the diffeomorphisms $\tilde f_0 R_u \p_U$ and, respectively, $- \tilde{\bar f}_0 R_v \p_V$ become
\bea
\cancel{\d} Q_{\tilde f_0 R_u \p_U}  &=&\; \tilde f_0 R_u \d H_L + \frac{\mu H_L}{\pi p} \tilde f_0 \d H_R = \d (\tilde f_0  R_u H_L) \nonumber \\
\cancel{\d} Q_{-\tilde{\bar f}_0 R_v \p_V} & = &-\tilde{\bar f}_0 R_v \d H_R - \frac{\mu H_R}{\pi p} \tilde{\bar f}_0\d H_L = -\d (\tilde{\bar f}_0  R_v H_R)
\eea
%
The  set of allowed  vector fields that are associated with this new integrability constraint are, to leading order 
\begin{align}
\xi_L&=R_u\left(\tilde f(u) \p_U +\mu\L(\tilde f-2\tilde f_0)\p_V\right) \;, \;\;\;\;\;\;\xi_R=R_v(\tilde{\bar f}(v) \p_V +\mu\bar \L(\tilde{\bar f}-2\tilde{ \bar f}_0)\p_U)
\end{align}
 One can straightforwardly repeat the steps of the previous section and compute the $\O(\e)$ correction to these vector fields. The constraints \eqref{ecordere}
are not affected and the correction takes exactly the same form as before, depending on two functions analogous to $F_h,\bar{F}_{\bar{h}}$. The arguments that determine these functions still hold. 
However, fixing the constant field-dependent piece of the $\O(\e)$ correction to $\xi_{(p)}$   (i.e., the $\O(\e)$ correction to $c_V$ in \eqref{newsolint} above) is somewhat more challenging in this case, as the full non-linear $T\bar T$ answer \eqref{altsolcLf} shows it is only fixed by integrability - a property that we did not check beyond leading order in the asymptotically linear dilaton background. Fortunately, this extra constant 
%
 %
 does not contribute to the charge algebra at $\O(\e)$  - as can also be seen from \eqref{finaldelQ}, and  the final result for the left-moving algebra corresonding to this  rescaled set of  vector fields is   
\begin{align}
\d_{\chi}(R_u Q_{\xi})&=R_u Q_{fg'-gf'}
\end{align}
in perfect agreement with the single-trace $T\bar{T}$ prediction \cite{ttbspo}. A similar result holds for the right-movers, with $R_u$ replaced by $R_v$.

\section{Discussion}

The main result of this article was to show that the asymptotic symmetry  algebra of  asymptotically linear dilaton backgrounds in string theory \emph{precisely matches}  that of single-trace $T\bar T$ - deformed CFTs, at least to the order in perturbation theory to which we have checked. This further strengthens the connection proposed in \cite{Giveon:2017nie} between compactified little string theory and a symmetric product orbifold of $T\bar T$ - deformed CFTs.

This asymptotic symmetry group analysis of the linear dilaton backgrounds was performed in parallel to that of the spacetime dual to double-trace $T\bar T$ - deformed CFTs - namely, AdS$_3$ with mixed boundary conditions \cite{mirage} - using the same formalism. Despite the entirely different asymptotics of these two space-times  and the different boundary conditions for the allowed fluctuations, the asymptotic symmetry generators and conserved charges were found to be virtually identical at each step of the calculation, up to a constant rescaling  related to the symmetric product orbifold. The  only quantity that our study of the   asymptotically linear dilaton background could not fix was the central extension of the asymptotic symmetry algebra, as the symplectic form analysis  we used can recover most, but not all, of the boundary conditions that the perturbations obey. There does exist, however, a choice of boundary conditions leading to \eqref{choiceFr}, which yields precisely the central extension of single-trace $T\bar T$ - deformed CFTs.

The algebra that we have recovered using the natural Fourier basis of generators is a non-linear modification \eqref{QfQgalg} - \eqref{QfbarQgalg} of the Virasoro $\times$ Virasoro algebra. As a result, the algebra itself depends on the background upon which we are evaluating it. This algebra can be linearised via a simple rescaling of the generators by an energy-dependent function, case in which one obtains two commuting copies of the Virasoro algebra, with the same central extension as that of the original CFT.  Note that in the covariant phase space formalism,  the effect of a field-dependent rescaling of the generators on their algebra can only be seen by carefully taking into account the constraints imposed by charge integrability, as we show in great detail in our holographic analysis of the double-trace $T\bar T$ deformation.

For both the case of AdS$_3$ with mixed boundary conditions and the asymptotically linear dilaton backgrounds, the algebra of the rescaled generators agrees with that obtained by transporting the original CFT generators along the double/single-trace $T\bar T$ flow \cite{Guica:2021pzy,ttbspo}.  It is not clear, however, which or whether one of these bases of generators is preferred: while the algebra of the rescaled generators is certainly more standard,  being also a Lie algebra, in the closely-related  $J\bar T$  - deformed CFTs it was found that the natural  ``physical'' symmetry generators -
in the sense that the Ward identities for analogues of primary operators   were defined with respect to them  - 
 were the ones  satisfying the non-linear algebra  \cite{Guica:2021fkv}.

It would be very interesting to compare our results to the asymptotic symmetry group analyses of three-dimensional asymptotically flat spacetimes \cite{Barnich:2006av}. Note that, besides the presence of a linear dilaton, our setup is rather different from the standard BMS$_3$ one, in that the asymptotic geometry is $\mathbb{R}^{1,1} \times S^1$ rather than $\mathbb{R}^{1,2}$ - with the $S^1$ playing an important role in obtaining a non-linear algebra - and we perform our analysis at spatial - rather than null - infinity, as it is natural for this spacetime \cite{Marolf:2006bk}. Since our setup has a well-defined QFT dual with a non-negligible set of observables that can be  computed directly in field theory, finding a link to asymptotically flat spacetimes could lead to interesting insights into their holographic interpretation beyond the pure gravity regime \cite{Barnich:2012rz,Barnich:2013yka}. On a different note, let us mention that  asymptotic symmetry algebras that are non-linear due to a similar mechanism to that discussed herein have been recently discovered in five-dimensional asymptotically flat spacetimes  at spatial infinity \cite{Fuentealba:2021yvo,Fuentealba:2022yqt}. One may naturally wonder whether such non-linear algebras could play a larger role in non-AdS holography.


Another  interesting direction would be to understand the implications of this work for the properties of little string theory. According to the standard holographic dictionary, the asymptotic symmetries we have uncovered are nothing but the extended  symmetries of compactified LST. The question of which basis of symmetry generators to use could, however, become relevant at this point. If we use the rescaled basis, where the symmetry algebra is simply Virasoro with a  central charge that is very likely $6 k p$, the symmetries of the UV theory would be intimately tied to those of the infrared CFT. This  may seem slightly unnatural as, after all,   the IR CFT  upon which the irrelevant deformation is built is a somewhat artificial construction, designed to provide a tractable QFT interpretation of compactified LST, and  we would expect the UV physics to ultimately decouple from the IR one in the deep UV limit. This is 
indeed the case  for the thermodynamic entropy  \eqref{ldentropy}, from which the number of F1 strings drops out in the high-energy  limit $ E >> R p/\a'$.
%
On the other hand,  the algebra of the unrescaled generators,  which is dominated by the non-linear term in  \eqref{QfQgalg} in the high- energy limit mentioned above,  becomes $p$-independent,  and may thus provide a more natural basis to use.  It would be interesting to better understand the significance of this algebra,  both from the perspective of  the high- energy limit of $T\bar{T}$ - deformed CFTs, and that of little string theory.

Apart from the symmetry analysis, our classification of all the linearized perturbations upon the interpolating backgrounds may have interesting implications for the holographic dictionary for these spacetimes. By taking into account the full coupling of the dilaton to the background metric, we uncovered several unusual features of the wave equation, whose holographic interpretation is not immediately obvious. 
 Concretely, upon the black hole backgrounds, we have found that the zero frequency and the $r \r \infty$ limits of the wave equation do not commute, leading to a discontinuity in the asymptotic  weights that one generally associates with dual operator dimensions. Second, we have found that 
 %
 the $r \r \infty$ and $\a' \r 0$ limits of this equation do not commute, with the result that the operator whose momentum-dependent conformal dimension can be read off from the holographic analysis at infinity does not appear to reduce to the single-trace $T\bar T$ operator in the IR limit, as one would have na\"{i}vely expected. These peculiar  behaviours only arise upon considering the particular couplings of the dilaton - as dictated by string theory - and fully including backreaction effects, and would not have been visible in the probe approximation.
 %
%
 It would therefore be quite interesting to perform a more thorough analysis of this system using e.g. holographic renormalization and  clarify these issues.


Finally, it would be worthwhile to better understand the reason that certain observables of the three-dimensional asymptotically linear dilaton backgrounds
match so well the predictions from single-trace $T\bar T$ - deformed CFTs, given that the CFT dual to the IR AdS$_3$ that is being deformed is not itself a symmetric product orbifold \cite{Eberhardt:2021vsx}.  One possibility is that there exists a universality class of ``non-local CFTs'' -  in the sense of \cite{Guica:2021pzy} - that includes the single-trace $T\bar T$ deformation, but whose members are not generically symmetric product orbifolds, where universal quantities such as the entropy-energy relation and the symmetry algebra take the same form as in single-trace $T\bar T$ - deformed CFTs. 
It would be very interesting to further invstigate this possibility, and understand how the appropriate generalizations of symmetric product orbifolds of $T\bar{T}$- deformed CFTs  should be defined.
%


\subsection*{Acknowledgements}

The authors are grateful to Soumangsu Chakraborty, Geoffrey Comp\`ere, Mahdi Godazgar, Dan Jafferis, Nicolas Kovensky, Emil  Martinec, Ruben Monten, Yorgo Pano,   Andrea Puhm, Kostas Skenderis and Andrew Strominger for interesting discussions.  This research was supported in part by  the ERC Starting Grant 679278 Emergent-BH.  The work of S.G. is supported by the PhD track fellowship of the \'Ecole Polytechnique.

\appendix

\section{Conserved charges in double-trace $T\bar T$-deformed CFTs \label{conschdtrttb}}

In this appendix, we revisit the construction of conserved charges in the standard $T\bar T$ - deformed CFTs from a holographic perspective. We closely follow the original work  \cite{mirage}, but now use the covariant phase space formalism to compute the conserved charges. Special attention is paid to the  coordinate periodicities and to integrability, which turn out to be essential for obtaining the correct charge algebra. A field-theoretical derivation of the same charges and algebra is presented in \cite{clsqsymmttb}.

\subsection{Asymptotic symmetries of the dual gravitational background}

As explained at length in \cite{mirage}, the double-trace $T\bar T$ deformation of a holographic CFT corresponds to turning on mixed boundary conditions for the metric of the dual AdS$_3$ spacetime, which take the form
\be
\gamma^{[\mu]}_{\a\b} =  g_{\a\b}^{(0)} - \mu \, g_{\a\b}^{(2)} + \frac{\mu^2}{4} g^{(2)}_{\a\g}\,  g^{(0)\, \g \d} g^{(2)}_{\d\b}  = fixed \label{mixedbcdtr}
\ee
where $g_{\a\b}^{(n)}$ are the Fefferman-Graham coefficients 
 in the asymptotic metric  expansion and  $\g^{[\mu]}_{\a\b}$ is the metric of the $T\bar T$ - deformed CFT.  If we work in pure gravity, then the last term in \eqref{mixedbcdtr} becomes $\mu^2 g^{(4)}$ - the last non-zero term in the Fefferman-Graham expansion for this case \cite{Skenderis:1999nb} -  and thus the $T\bar T$ metric coincides with the induced metric on a surface of constant radial distance  $ \rho_c = - \mu$ from the AdS$_3$ boundary, providing a link to the work of \cite{McGough:2016lol}. 
 
  The phase space of the theory with a flat $T\bar T$ metric corresponds to  the most general gravitational solutions satisfying \eqref{mixedbcdtr} with $\g_{\a\b}^{[\mu]} = \eta_{\a\b}$. In \cite{mirage}, it was shown these are given by

\be
g^{(0)}_{\a\b} dx^\a dx^\b = \frac{(dU+\mu  \bar \L(v) dV)(dV+\mu \L(u) dU)}{(1-\mu^2 \L(u)\bar \L(v))^2} \equiv du dv  \nonumber 
\ee
\vspace{1mm}
\be
g^{(2)}_{\a\b} \, dx^\a dx^\b = \frac{ \left(1+\mu^2 \L(u) \bar \L(v)\right) ( \L(u)\, dU^2 +\bar \L(v) \, dV^2) +4 \mu \L(u) \bar \L(v)\, dU dV }{\left(1-\mu^2 \L(u) \bar \L(v)\right)^2} 
\label{g24UV} \ee
where $\L,\bar \L$ are arbitrary functions of the so-called field-dependent coordinates $u,v$, defined via
%
%
\be
U = u -\mu \int^v \bar \L(v')\,  dv' \;, \;\;\;\;\;\; V = v - \mu \int^u \L(u') \, du'  \label{deffdcoord}
\ee
where $U,V = \s \pm t$, as in the main text. Note our definition of $\L, \bar \L$ differs by a factor of  $4 \pi G \ell$ from that of \cite{mirage}. Also note that now $\L, \bar \L$ can be arbitrary functions of their argument, whereas in the main text the same notation denoted solely their zero modes. The function $\L(u)$ is assumed to be  periodic, and thus its primitive consists of  a  term linear in $u$ that is proportional to the zero mode of $\L$, a  set of non-zero Fourier modes, as well as a possible  constant term that will be denoted $c_{\L}$. The same applies to the primitive of $\bar \L(v)$, where the constant term is denoted $c_{\bar \L}$.

The expectation value of the holographic stress tensor, computed using holographic renormalization in presence of the mixed boundary conditions, is given by 

\be
 T_{\a\b}^{[\mu]} dx^\a dx^\b
 =\frac{\L(u) dU^2 + \bar \L(v) dV^2- 2 \mu \L(u) \bar \L(v) dU dV }{2(1-\mu^2 \L(u)\bar \L(v))}  \label{stresst}
\ee
and one can easily check it is conserved. One may  use it to compute the left and right-moving energies of the background, which are the  conserved charges associated with the null translational symmetries of the $T\bar T$ metric, generated by $\p_U$ and $\p_V$
\be
H_L  = \frac{1}{2} \oint d\s \frac{\L (1+\mu \bar \L)}{1-\mu^2 \L \bar \L}  = \frac{1}{2} \oint d\s \L\, \p_\s u \;,\;\; \;\;\;\; H_R = \frac{1}{2} \oint d\s \frac{\bar\L (1+\mu \L)}{1-\mu^2 \L \bar \L}  = \frac{1}{2} \oint d\s \bar\L \, \p_\s v
 \label{holoHLR}
\ee
It follows that the field-dependent coordinates \eqref{deffdcoord} have energy-dependent periodicities 

\be
2\pi R_u = 2\pi R+2 \mu H_R \;, \;\;\;\;\;\;  2\pi R_v =  2\pi R + 2 \mu H_L \label{fdepradii}
\ee
where $R_{u,v}$ are the field-dependent radii of $u,v$ and $R$ is the radius of the $\s$ circle.

The asymptotic symmetries of the above set of backgrounds - which, by the standard holographic correspondence, are associated to  the extended symmetries of the $T\bar T$ - deformed CFTs - are generated by the most general diffeomorphisms that preserve the boundary conditions \eqref{mixedbcdtr}. Working in radial gauge, one finds 

\be
\xi^U = f(u)+\mu \bar \L_{\bar f}(v)+ \frac{\ell(\rho+\mu)(\rho \bar \L f''(u) - \bar f''(v))}{8\pi G(1-\rho^2 \L \bar \L)}\nonumber 
\ee

\be
\xi^V = \bar f(v)+\mu \L_f(u) + \frac{\ell(\rho+\mu)(\rho \L \bar f''(v) - f''(u))}{8\pi G(1-\rho^2 \L \bar \L)} \label{xiV} \;, \;\;\;\;\;\; \xi^\rho = \rho (f'(u)+\bar f'(v))
\ee
where $f, \bar f$ are arbitrary functions of the field-dependent coordinates and $ \L_f$, $\bar \L_{\bar f}$ are defined as

\be
\L_f (u) = \int^{u}  \!\!\!\!du' \L (u')  f'(u')\;,\;\;\; \;\;\;\;\;\; \bar \L_{  \bar f}(v) =  \int^{v}  \!\!\!\!dv'\bar \L (v') \bar f'(v') \label{defLf}
\ee
As before, these primitives are only defined up to an overall Fourier zero mode, which will be denoted as $c_{\L_f}, c_{\bar \L_{\bar f}}$. Under these diffeomorphisms, the background values of $\L, \bar\L$ change as
\bea
\d_{f,\bar f} \L &= & 2 \L f'(u)  - \frac{\ell}{8\pi G} f'''(u) + \frac{\L'  [f(u) +\mu \bar \L_{\bar f} (v) +\mu \bar \L (\bar f +\mu \L_f) - \frac{\mu \ell}{8\pi G} (\bar f''+\mu \bar \L  f'')]}{1-\mu^2 \L \bar \L} \;\;\;\;\;\;\;\;\;\; \nonumber \\
\d_{f,\bar f} \bar \L &= & 2 \bar \L \bar f'(v)  - \frac{\ell}{8\pi G} \bar f'''(v) + \frac{\bar \L'  [\bar f(v) +\mu  \L_f (u) +\mu \L (f+\mu \bar \L_{\bar f}) - \frac{\mu \ell}{8\pi G} (f''+ \mu \L  \bar f'' )]}{1-\mu^2 \L \bar \L} \label{deltvv}
\eea
Note that general variations of $\L, \bar \L$ can be decomposed into an `intrinsic' part, which only depends on $u$ and, respectively, $v$,   and one induced by  the change  $\d u$, $\d v$ in  their arguments 

\be
\d \L (u)= \d \L_{int} (u) + \L' (u) \d u \hspace{1cm} \d \bar\L (v)= \d \bar\L_{int} (v) + \bar\L' (v) \d v \label{gendL}
\ee
Since the coordinates $u,v$ depend on the field configuration, we need to pay special attention to the quantities that are being held fixed in any given variation, i.e. whether it is the fields, the coordinates, or both. For example,  the values of $\d \L_{int}, \d u$ that correspond to \eqref{deltvv} are obtained by assuming that the field-independent $T\bar T$ coordinates $U,V$ are fixed in the variation; the result is given in \eqref{dLintdiff}. 
%

The conserved charges associated with these diffemorphisms were proposed in \cite{mirage} to be

\be
Q_{L,f}
 = \frac{1}{2} \oint d\s\, \frac{f(u) \L(u) (1+\mu \bar \L ) }{1-\mu^2 \L \bar \L}= \frac{1}{2} \oint d \s\,  f(u) \, \L(u)\, \p_\s u \label{QL}
\ee
and similarly for the right-movers, where $f$ was assumed to be a periodic function of $\hat u \equiv u/R_u$, as required by conservation. This proposal was based on the fact that the currents $T_{U\a} f(u)$ are conserved for arbitrary $f(u)$, and thus the above  amounted to a natural guess for the conserved charges. In this appendix, we will instead \emph{derive} a variant of this formula using the covariant phase space formalism and show that, whereas the proposal of \cite{mirage} is essentially correct, the relation between the periodic function that appears in \eqref{QL} and the function parametrizing the asymptotic symmetry generators is slightly more involved.


\subsection{Conserved charges from the covariant phase space formalism}
In this section, we compute the charges associated to the diffeomorphisms \eqref{xiV} using the covariant phase space formalism and show that they are finite, integrable and conserved. For this, we need to first properly set up the problem. 

\subsubsection*{Setup}

We start with a few important remarks. First,  note that since the physical $T\bar T$ metric corresponds, via \eqref{mixedbcdtr}, to the metric induced on the $\rho = -\mu$ surface (in the pure gravity picture), from the point of view of the $T\bar T$ coordinates $U,V$, the diffeomorphisms \eqref{xiV} act as 

\be
U \r U + f(u) + \mu \bar \L_{\bar f} (v) \;, \;\;\;\;\;\; V \r V + \bar f(v) + \mu \L_f (u)
\ee
We immediately observe  that $f, \bar f$ cannot be periodic functions of their arguments, lest the periodicity of the  coordinates $U,V$ - which should be fixed - would be affected. We can thus split $f, \bar f$ into a periodic and a winding part, assumed to be simply linear in the field-dependent coordinates
 
 \be
  f(u) = f_p( u) + w_f u \;, \;\;\;\;\; \bar f(v) = \bar f_p (v) + w_{\bar f} v \label{fdecomp}
 \ee
This assumption is supported by the fact that the winding of $f, \bar f$ around the $\s$ circle is constant. Using \eqref{holoHLR}, we find that the windings satisfy 
\be
2\pi w_f R_u + \mu \oint dv \bar \L \bar f_p'(v) + 2 \mu w_{\bar f} H_R =0 \;, \;\;\;\;\;2 \pi  w_{\bar f} R_v + \mu \oint du \L f_p'(u) + 2 \mu w_f H_L=0 \label{eqwffb}
\ee
and so they are determined by the periodic part of the function and the given background. 

%
%
The second remark concerns the periodicity properties of $\d \L_{int}$, defined in \eqref{gendL}. Remember that in the covariant formalism, one always computes the charge difference between two  nearby on-shell backgrounds. We thus consider two backgrounds of the form \eqref{g24UV} that differ in  the intrinsic values of $\L$ and $\bar \L$ by $\d \L_{int}, \d \bar \L_{int}$; the differences $\d u, \d v$ in the associated field-dependent coordinates simply follow from the definition \eqref{deffdcoord}, taking into account the fact that the $T\bar T$ coordinates $U,V$ are fixed in this variation. Assuming that $\L(u)$ is a periodic function of its argument, we can expand it in  Fourier modes as

\be
\L(u) = \sum_n \L_n \, e^{i n u/R_u}
\ee
Its varition reads
\be
\d \L(u) = \sum_n \d \L_n \, e^{i n u/R_u} + \left(\frac{\d u}{R_u}- \frac{u \d R_u}{R_u^2}\right)\sum_n i n \L_n e^{i n u/R_u} 
\ee
where the variation of the field-dependent radius  is generally non-zero, due to the fact that it is proportional, via \eqref{fdepradii}, to the difference in energy between the two backgrounds.
Comparing with \eqref{gendL}, it follows that $\d \L_{int}$ consists of both a periodic and a winding part
\be
\d \L_{int} (u) = \d \L_{int}^p(u) - \frac{u \d R_u}{R_u} \L'(u)  \label{dLintdecomp}
\ee
Changing the intrinsic value of $\L,\bar{\L}$ also induces a change in the field-dependent coordinates, which is determined by the requirement that the coordinates $U,V$ be fixed
\be
\d U = \d u - \mu \bar \L \d v - \mu \int^v \d \bar \L_{int} (v') dv'  =0 \;, \;\;\;\;\; \d V = \d v -\mu \L \d u - \mu \int^u \d \L_{int} (u') du' =0 \label{eqdudv}
\ee
These equations are easily solved to determine the associated $\d u, \d v$. 
It is again interesting to look into the windings
of these coordinate differences. Plugging \eqref{dLintdecomp} into \eqref{eqdudv} and taking the variation, we find that the windings
satisfy
\bea
w_{\d u} - \mu \bar \L(v)   w_{\d v}- \mu \oint \d \bar \L_{int}^p(v) dv + \frac{\mu \d R_v}{R_v} (2\pi R_v \bar \L(v) - \oint \bar \L (v) dv )&  = & 0 \nonumber \\
w_{\d v} - \mu  \L (u) w_{\d u}- \mu \oint \d \L_{int}^p (u) du + \frac{\mu \d R_u}{R_u} (2\pi R_u  \L(u) - \oint  \L(u) du ) & = & 0 \label{eqwdudv}
\eea
By requiring that the terms proportional to $\L (u), \bar \L (v)$ cancel separately, we find a natural solution where the  windings are constant, with

\be
w_{\d u} = 2\pi \d R_u =2 \mu \d H_R\;, \;\;\;\;\;w_{\d v} = 2\pi \d R_v = 2 \mu \d H_L \label{wdudv}
\ee 
where we have used \eqref{fdepradii}. Plugging these back into \eqref{eqwdudv}, we find the following constraint 

\be
2 \d H_L = \oint \d \L^p_{int} (u) du + 2 H_L \frac{\d R_u}{R_u}\;, \;\;\;\;\;\;    2 \d H_R =   \oint \d \bar \L_{int}^p(v) dv + 2  H_R \frac{\d R_v}{R_v}  \label{ctrdHL}
\ee
which should better be satisfied by the variations of the left- and right-moving Hamiltonian given in \eqref{holoHLR}. One can check this is indeed the case, since 

\be
\d H_L = \frac{1}{2} \, \d \! \! \oint d\s \p_\s u \, \L(u) =\frac{1}{2}  \oint d\s \left( \d \L_{int}^p + \L \frac{\d R_u}{R_u} \right) \p_\s  u + \frac{1}{2}  \oint d\s \p_\s \left( \L \d u - \frac{u \d R_u}{R_u}\, \L \right)
\ee
and similarly for the right-moving Hamiltonian. Note that, thanks to \eqref{wdudv}, the term in the second paranthesis has no winding, so the total derivative can be dropped. Thus, the  consistency requirements \eqref{ctrdHL} are indeed satisfied.

The solution to \eqref{eqdudv} is  found to be
\bea
\d u &=& \frac{\mu  \int^v \d\bar \L^p (v') dv'  + \mu^2 \bar \L \int^u \d \L^p(u') du'  - \mu \bar \L \hat v \d R_v - \hat u \d R_u }{1-\mu^2 \L \bar \L} +\hat u\, \d R_u \nonumber \\
\d v &= &\frac{\mu  \int^u \d \L^p (u') du' + \mu^2  \L \int^v \d \bar \L^p(v') dv'  - \mu  \L \hat u \d R_u - \hat v \d R_v }{1-\mu^2 \L \bar \L} + \hat v \, \d R_v \label{soldudv}
\eea
where we have defined two new periodic functions
\be
\d \L^p \equiv \d \L^p_{int} + \L \frac{\d R_u}{R_u} \;, \;\;\;\;\;\; \d \bar \L^p \equiv \d \bar \L^p_{int} + \bar \L \frac{\d R_v}{R_v} \label{defLp}
\ee
and we have introduced the normalized field-dependent coordinates $\hat u \equiv u/R_u$, $\hat v \equiv v/R_v$, which have fixed periodicity $2\pi$. Note that, according to this definition

\be
\d u = R_u \d \hat u + \hat u \d R_u \;, \;\;\;\;\;\; \d v = R_v \d \hat v + \hat v \d R_v \label{dudvdecomp}
\ee
so the entire winding of $\d u, \d v$ is concentrated in the last terms of \eqref{soldudv}, whereas the rest can be identified with $R_u \d \hat u$ and, respectively, $R_v \d \hat v$, both of which are periodic on the cylinder. Note the primitives of $\d \L^p$ and $\d \bar \L^p$ are still ambiguous up to a constant term that equals $\d c_\L$ and, respectively, $\d c_{\bar \L}$, where we assumed that the constant ambiguity in \eqref{deffdcoord} may depend on the field configuration.

\subsubsection*{Computing the  conserved charges}

We are  finally ready to compute the charges associated to the diffeomorphisms \eqref{xiV}, using the standard formula \eqref{keinst} in the covariant formalism. We find that the charge difference between two backgrounds that differ by $\d \L_{int}, \d \bar \L_{int}$ is given by 
\bea
 \not{\!\d} Q_{f,\bar f} &= & \frac{1}{2} \oint d\s (f \p_\s u \,\d \L_{int} - \bar f \p_\s v \, \d \bar \L_{int}) +\frac{1}{2} \oint d\s \p_\s  \bigg[\d u (\L f - \L_f) - \d v (\bar \L \bar f - \bar \L_{\bar f}) +  \nonumber \\
 && \hspace{0.5cm}+\;   \frac{1}{4} (\p_u \d u - \p_v \d v)(f'+\bar f')\bigg] \label{dnotQ}
\eea
where $\d u, \, \d v$ are determined by $\d \L_{int}$ via \eqref{soldudv}.
In order for this expression to be well-defined, it should be the integral of a periodic quantity. This is not obvious \emph{a priori}, since the integrand of the first term has winding - given in  \eqref{dLintdecomp} - that needs to be carefully taken into account, as do the terms under the total derivative. 

Let us start by analysing the term on the second line of \eqref{dnotQ}. Its only potentially non-zero contribution  to the charge  comes from winding terms in $\d u$ and $\d v$, whose form is given in \eqref{dudvdecomp}. It is then easy to see that their $u,v$ derivatives will just produce a periodic function under the total $\s$ derivative, which can be dropped.  

  Moving on  to the first term in \eqref{dnotQ}, we plug in the explicit form \eqref{fdecomp} of $f$ and  write it  as
\bea
\frac{1}{2}\oint d\s (f_p(u) + w_f u) \p_\s u \, \d \L_{int} (u) & = & \frac{1}{2} \oint d\s \left[f_p \p_\s u \,\left( \d\L_{int}^p+  \L \frac{\d R_u}{R_u}\right)  - \frac{\d R_u}{R_u} \p_\s(u f(u) \L ) + \right. \nonumber \\
&& \hspace{-2.3cm}+ \;\left. u \p_\s u \left(  w_f \d \L_{int}^p      + \frac{\d R_u}{R_u} \L f_p'  + \frac{2 w_f \d R_u}{R_u}  \L \right) \right]
\eea
where we have integrated by parts all terms containing $\L'$. Next, we write the contributions on the second line by introducing the primitives of the terms in  paranthesis, which are separated into a periodic part and a purely winding part proportional to the zero mode of the function, e.g. 
\be
\oint d\s \, u\, \p_\s u \, \d \L_{int}^p = \oint d\s \, u\, \p_\s \left( \int^u \d \L_{int}^p du - \frac{u}{2\pi R_u} \oint \d \L_{int}^p\right) + \frac{1}{2\pi R_u}\oint \d \L_{int}^p\oint d \s u\, \p_\s u  
\ee
Dropping the zero mode of the term in the first paranthesis, which does not contribute, and integrating by parts, the first term yields a total derivative minus a term whose circle integral vanishes. Applying such manipulations to all the terms, we find 
 \bea
 \frac{1}{2} \oint d\s f(u) \p_\s u \, \d \L_{int} \!\!\! & = &\!\!\! \d Q_{f_p} +  \frac{1}{2}  \oint \!\!d\s \p_\s \left[w_f  u \left(\int^u \!\!\!\d \L_{int}^p - \frac{\hat u}{2\pi} \oint \d \L_{int}^p \right)_{\!\!\cancel{zm}} + \frac{\d R_u}{R_u} u \left(\int^u \!\!\!\L f_p' - \frac{\hat u}{2\pi} \oint \L f_p'\right)_{\!\!\cancel{zm}}  +  \right. \nonumber \\
 && \hspace{-3.8 cm}+ \left. \frac{2 \d R_u w_f}{R_u} u \, \left(\int^u\!\!\! \L - \frac{\hat u}{2\pi} \oint \L\right)_{\!\!\cancel{zm}} + \left(w_f \!\!  \oint   \d \L_{int}^p + \frac{\d R_u}{R_u} \oint\L f'_p+\frac{2 \d R_u w_f}{R_u } \oint  \L\right)  \!\!  \frac{u^2}{4\pi R_u}  - \frac{\d R_u}{R_u} \L u f(u) \right] \label{manipft}
 \eea
where the subscript on the parantheses indicates the zero mode of the given quantity is not present. We also introduced the trivially integrable quantity
\bea
\d Q_{f_p} & \equiv &  \d\! \left[\frac{1}{2}\oint d\s f_p(\hat u) \p_\s u \, \L\right] = \frac{1}{2}\,\oint d\s f_p \p_\s u \left(\d \L^p_{int} +\L \frac{\d R_u}{R_u}\right) +\frac{1}{2}\, \oint d\s \p_\s (f_p \L R_u \d \hat u) \nonumber\\
&  = & \frac{1}{2}\,\oint d\s f_p(\hat u) \p_\s u\, \d \L^p \label{defdelQp}
\eea
The total derivative term was dropped because $\d \hat u$ has no winding and we have explicitly used the fact that the periodic function $f_p$ only depends on $u$ (and $R_u$) through the combination $\hat u = u/R_u$. Plugging \eqref{manipft} into \eqref{dnotQ} and  dropping all the terms without winding, we find
\bea
\not{\!\d} Q_{f,\bar f} &=& \d Q_{f_p} + \frac{1}{2}\! \oint \!\! d\s \p_\s \!\left[ w_f u \! \left(\int^u \!\!\!\d \L^p \right)_{\!\!\cancel{zm}}+ \left(w_f \L   u - \!\!\int^u\!\!\! \L (f_p' + w_f)\right) \! R_u \d \hat u   - u \frac{\d R_u}{R_u} c_{\L_f} -\right. \nonumber\\
&& \left.- \frac{u^2}{4\pi R_u}\! \left(\! w_f \oint \! \d \L^p + \frac{\d R_u}{R_u}\!\oint \! \L (f_p' + w_f)\right)\right]+ RM
\eea
where $c_{\L_f}$ is the integration constant appearing in the definition \eqref{defLf} of $\L_f$ and  ``$RM$'' stands for the right-moving counterpart of these terms, which is simply obtained via \emph{minus} the replacement $f \leftrightarrow \bar f , \L \leftrightarrow \bar \L$, etc. Note that we have a number of terms with double winding, which render the integral inconsistent i.e., dependent on the starting point on the circle. In the following, we would like to show that they  cancel. We start from \eqref{soldudv}, which implies that

\be
\L R_u \d \hat u + \int^u \!\!\!\d \L^p
 = \frac{\d v}{\mu} 
\ee
where we recall that the zero mode of  $\int^u \d \L_p$ is $\d c_{\L}$. Replacing the $\int^u \L(f_p'+w_f)$ term by $\frac{\hat u}{2\pi} \oint  \L (f_p'+w_f)$ - as their difference is a periodic function that can be dropped from the integral, including its zero mode -  and using \eqref{eqwffb} to write $\oint \L (f_p'+w_f)$ in terms of $w_{\bar f}$, we find
\bea
\not{\!\d} Q_{f,\bar f} &= & \d Q_{f_p} + \frac{1}{2} \!\oint \!\! d\s \p_\s \left[ w_f  \frac{u\d v}{\mu}   +\frac{w_{\bar f} R_v }{\mu}  u \d \hat u  - \frac{u^2}{4\pi R_u} \left(2 w_f \d H_L - \frac{2\pi w_{\bar f} R_v\d R_u}{\mu R_u}  \right)   \right] \nonumber \\[5pt]
&- & \pi (c_{\L_f} \d R_u + \d c_\L w_f R_u) +RM
\eea
At this point, it is useful to write
\be
\hat u = \s + \D \hat u \;, \;\;\;\;\; \hat v = \s + \D \hat v
\ee 
where $\D \hat u, \D \hat v$ are periodic. The expression for the charge difference becomes 
\bea
\not{\!\d} Q_{f,\bar f} & = & \d Q_{f_p} +  \frac{1}{2\mu}  \oint d\s  \p_\s \bigg[ \frac{\s^2}{2} (w_f R_u \d R_v+ w_{\bar f} R_v \d R_u) + \s  R_u R_v (w_f \d \D \hat v + w_{\bar f} \d \D \hat u) + \nonumber \\
&& \hspace{0.5 cm}+\;    \s ( w_f R_u \d R_v \D \hat v+ w_{\bar f} R_v \d R_u \D \hat u) \bigg]- \pi (c_{\L_f} \d R_u + \d c_\L w_f  R_u) +
RM 
\eea 
where we dropped all the periodic terms under $\p_\s$. Since the expression in paranthesis is entirely symmetric under $f \leftrightarrow \bar f, u \leftrightarrow v $ and the RM contribution is just \emph{minus} the left-moving one  with  $f \leftrightarrow \bar f, u \leftrightarrow v $, all the winding terms cancel and we are left with

\be
\not{\!\d} Q_{f,\bar f} = \d Q_{f_p} + \d \bar Q_{\bar f_p} + \pi \left(c_{\bar\L_{\bar f}} \d R_v + \d c_{\bar \L} w_{\bar f}  R_v- c_{\L_f} \d R_u - \d c_\L w_f  R_u\right) \label{finaldelQ}
\ee
where we have defined

\be
 \d \bar Q_{\bar f_p} =  \d \left[- \frac{1}{2}\oint d\s \p_\s v \bar f_p \bar \L \right] = - \frac{1}{2} \oint d \s \bar f_p(\hat v) \p_\s v \d \bar \L^p
 \ee
  It is clear that the charge \eqref{finaldelQ} will only be integrable for special values of the integration constants. One such value is of course zero; a more non-trivial choice would be to take $c_{\L} = 2 \a H_L$ and $c_{\bar \L_{\bar f}} = - \a \oint \bar \L \bar f'$ for some constant $\a$. In both cases, the contribution of the constants  to the conserved charges vanishes. As it is hard to imagine how one could obtain a non-zero contribution that is integrable for general variations, we will simply choose these constants to be zero, which amounts to a particular way to fix the ambiguities in the definition of $u,v$ and $\L_f$. 
  
\subsubsection*{Final result }

    Upon a trivial integration in phase space,  the final result for the holographic conserved charges of $T\bar{T}$-deformed CFTs is 
\begin{align}
Q_{f,\bar f}&=\frac{1}{2}\oint d\s f_p \partial_{\s}u\L -\frac{1}{2}\oint d\s \bar{f}_p \partial_{\s}v\bar{\L} \equiv Q_{f}+\bar{Q}_{\bar{f}}
\end{align}
which precisely agrees with the formula already proposed in \cite{mirage}. Note, however, that the derivation presented in this section is entirely from \emph{first principles} and the steps through which we arrived at this result were rather non-trivial:  first, we showed   that the functions $f(u), \bar f(v)$ that parametrize the asymptotic diffeomorphisms of  the space-time dual to $T\bar{T}$-deformed CFTs necessarily have winding, which is entirely fixed \eqref{eqwffb}  by the periodic part of the functions  to

\be
w_f =  \frac{\mu R_v}{\pi R R_H} \bar Q_{\bar f_p'} + \frac{\mu^2 H_R}{ \pi^2 R R_H} Q_{f_p'} \;, \;\;\;\;\; w_{\bar f} = - \frac{\mu R_u}{\pi R R_H}  Q_{ f_p'} - \frac{\mu^2 H_L}{\pi^2 R R_H} \bar Q_{\bar f_p'} \label{windingffb}
\ee
Then, we \emph{proved} that only the periodic  part of the functions enters in the charge formula, as required by conservation. For this reason,  the notation $Q_f$ and $Q_{f_p}$ can be used interchangeably. As we will see in the next subsection,  the non-trivial structure we have found does nevertheless have consequences for the charge algebra.

 Our derivation \emph{assumed} - specifically in \eqref{defdelQp}  - that the periodic part of the function $f$ only depends on $u$ and $R_u$ via the combination $\hat u = u/R_u$. Note that more possibilities to obtain integrable charges open up if we relax this requirement. For example, if we consider instead the rescaled function $ R_u f(\hat u)$, the charge variation given by  our general formula \eqref{finaldelQ} is 

\be
{\not\! \d} Q_{R_u f, R_v \bar f} = R_u \d Q_{f_p} -\pi c_{\L_{R_u f}} \d R_u -\d c_\L w_{R_u f} R_u + RM
\ee
This variation can be rendered integrable by choosing $\d c_\L=\d c_{\bar \L } =0$ as before, but now with

\be
c_{\L_{f}} = - \frac{Q_f}{\pi R_u} \;, \;\;\;\;\; c_{\bar \L_{\bar f} } = \frac{\bar Q_{\bar f}}{\pi R_v} \label{altsolcLf}
\ee
Using the fact that the charge is linear in the function $f$, this turns the right-hand-side into a trivially integrable $\d (R_u Q_{f_p})$ - the charges associated with the rescaled diffeomorphisms. 

%
%
%
%
%

%




\subsection{The charge algebra  \label{repthm}}

 In this subsection, we compute the algebra of the conserved charges we have found. 
As briefly reviewed at the beginning of section \eqref{asychalg}, in the covariant phase space formalism the charge algebra  
%
%
%
%
is given by 
$\d_\chi Q_\xi$: the change in the charge associated to an allowed diffeomorphism $\xi$ between two backgrounds that differ by another diffeomorphism, $\chi$. In our case, both diffeomorphisms are of the form \eqref{xiV}. We will take  $\chi$, which acts on the background, to be labeled by two functions $g,\bar g$.  The associated $\d_{\chi} \L$ is given in \eqref{deltvv} with $f$ replaced by $ g$, and the corresponding $\d \L_{int}$ and $\d u$ are given by\footnote{Note that $\d u$ is \emph{not} given by \eqref{xiV} translated to the $u,v$ coordinate system; rather,  \eqref{xiV}  describes a change in the coordinates $U,V$ with the intrinsic part of the $\L$ variation set to zero, 
 whereas \eqref{deltvv} represents  the \emph{same} transformation of the bulk metric produced from a change in the fields, but with the $T\bar T$ coordinates $U,V$ held fixed.    } 
\be
\d_\chi \L_{int} = 2 \L g' + \L' g  - \frac{\ell}{8\pi G} g''' - \L' \d_\chi c_u \nonumber \ee

\be \d_\chi u = \frac{\mu (\bar \L_{\bar g} +  \bar \L \bar g- \frac{\ell}{8\pi G} \bar g'') +\mu ^2 \bar \L (\L_g + \L g- \frac{\ell}{8\pi G} g'')}{1-\mu^2 \L \bar \L} +\d_\chi c_u \label{dLintdiff}
\ee
where $\d c_u$ (and its right-moving counterpart, $\d c_v$) represents an ambiguity in distributing $\d_\chi \L$   into an intrinsic variation and one due to the shift of its arguments, as in \eqref{gendL}. This ambiguity does not affect the previous discussion, where it was assumed that $\d \L_{int}$ was given, but it does enter here, where only the total variation \eqref{deltvv} is known. Upon distributing the $\d c_u$ factor between the two parantheses (and similarly for its right-moving counterpart), one can easily read off from this solution the zero modes $\d c_\L, \d c_{\bar \L}$ of $\int^u \! \d \L^p$ and $\int^v \! \d \bar \L^p$ 

\be
\d_{\chi} c_\L = c_{\L_g} + \frac{ Q_{g_p} }{\pi R_u}+ \frac{\d_\chi c_v}{\mu} - \frac{H_L}{\pi R_u} \d_\chi c_u \;, \;\;\;\;\d_{\chi} c_{\bar \L} = c_{\bar \L_{\bar g}} - \frac{\bar Q_{\bar g_p}}{\pi R_v} + \frac{\d_\chi c_u}{\mu} - \frac{H_R}{\pi R_v} \d_\chi c_v \label{eqdcuv}
\ee
Remember that integrability of the charges required that $\d c_\L = c_{\L_f} =0$. This, together with the equation above, determines the values of $\d_\chi c_{u,v}$ in \eqref{dLintdiff}. 

Plugging in the explicit form \eqref{fdecomp} of $g$, we find that $\d \L_{int}$ has a non-periodic part proportional to $w_g u \L'$. Comparing this with the general form \eqref{dLintdecomp}, we conclude that 
\be
w_g = - \frac{\d_\chi R_u}{R_u}  = - \frac{\mu }{\pi R_u} \, \d_\chi H_R \;, \;\;\;\;\;\;\; w_{\bar g} = - \frac{\d_\chi R_v}{R_v}  = - \frac{\mu}{\pi R_v} \, \d_\chi H_L
\ee
Comparing this with the general formula \eqref{windingffb} for $w_g, w_{\bar g}$,  we find a prediction for how $H_{L,R}$ change under the diffeomorphism, which should tell us about the corresponding commutators  of the Hamiltonians with the conserved charges 

\be
 \d_{\chi} H_L = \frac{R_u R_v}{R R_H}  Q_{ g_p'} + \frac{\mu H_L R_v}{\pi R R_H} \bar Q_{\bar g'_p} \;, \;\;\;\;\;\; \d_{\chi} H_R = - \frac{R_u R_v}{R R_H} \bar Q_{\bar g_p'} - \frac{\mu H_R R_u}{\pi RR_H} Q_{g'_p} \label{varHdiff}
\ee
%
Given the expression \eqref{dLintdiff} for $\d \L_{int}$, we can obtain - using \eqref{defLp} - the  periodic quantity $\d \L^p(u)$ that enters the formula for charge difference
\be
\d \L^p (u) =  2 \L g_p' + \L' g_p +w_g \L - \frac{\ell}{8\pi G}\, g_p''' -\L' \d c_u
\ee
where $g_p$ is the periodic part of $g$. A similar expression holds for the right-movers. 
The charge difference between the two backgrounds is given by 
\bea
 Q_f (\d_g \Phi) &= & \frac{1}{2} \oint d\s f_p \left(2 \L g_p' + \L ' g_p +  w_g \L   - \frac{\ell}{8\pi G} g_p'''\right) \p_\s u =  Q_{f_p g_p' - g_p f_p'} +   \nonumber \\
 && \hspace{-1.4 cm} +\; w_g Q_{f_p} +  \d_{\chi} c_u \,  Q_{f_p'}  - \frac{\ell}{16\pi G } \oint d u \, f_p( u)g_p'''( u) +\frac{1}{2} \oint d\s \p_\s (\L f_p g_p)  + RM \label{delgQf}
 \eea
The last term integrates to zero, as all the functions under the derivative are periodic, and the third term yields the usual central contribution. We can immediately check that for $f=1$, for which we obtain $ \d_\chi H_L$, this result agrees with \eqref{varHdiff}   above. The final result reads\footnote{Noting that $
 \d_\chi c_u \,  Q_{f_p'}  +  \d_\chi c_v \,  \bar Q_{\bar f_p'}   = w_f \left(  \frac{\pi R_u}{\mu} \d_\chi c_v - H_L \d_\chi c_u \right) + w_{\bar f} \left(H_R \d_\chi c_v - \frac{\pi R_v}{\mu} \d_\chi  c_u \right) = - w_f Q_g - w_{\bar f} Q_{\bar g}
$, where we re-expressed $Q_{f'}, \bar Q_{\bar f'}$ in terms of the winding using \eqref{windingffb}, the non-linear correction to the charge algebra can be written more symmetrically as $ (\d_\chi Q_{f,\bar f} )_{\text{non-lin.}} = w_g Q_{f} + w_{\bar g} Q_{\bar f} - w_f Q_g - w_{\bar f} Q_{\bar g}  $ .} 

\be \label{algebr}
\d_\chi Q_{f,\bar f} =  Q_{f_p g_p' - g_p f_p'} +  \bar Q_{\bar f_p \bar g_p' - \bar g_p \bar f_p'} + w_g Q_{f_p} + w_{\bar g_p} \bar Q_{\bar f_p} +  \d_\chi c_u \,  Q_{f_p'}  +  \d_\chi c_v \,  \bar Q_{\bar f_p'}   + \mathcal{K}_{f_p,g_p} + \mathcal{\bar K}_{\bar f_p,\bar g_p}
\ee
%
%
%
Plugging in the values of $\d_\chi c_{u,v}$ and $w_{g,\bar g}$, we find a \emph{non-linear algebra}. To present it, it is useful to split the commutation relation between left-movers and right-movers, by associating $Q_f$ to the case where $f_p \neq 0$ and  $\bar f_p =0$, and the reverse for $\bar Q_{\bar f}$; note that in both cases, the winding parts of $f$ and $\bar f$ are non-zero, being determined by the periodic parts via \eqref{windingffb}.  
%
The charge algebra that we find is  

\be
\{Q_f, Q_g\}  =   Q_{f_p g_p'-g_p f_p'}   + \frac{\mu^2 H_R}{\pi^2 R R_H } (Q_f Q_{g'} - Q_g Q_{f'}) +\mathcal{K}_{f,g} \nonumber
\ee

\be
\{ \bar{Q}_{\bar f}, \bar{Q}_{\bar g}\}  =   \bar{Q}_{\bar f_p \bar g_p'-\bar g_p \bar  f_p'}   - \frac{\mu^2 H_L}{\pi^2 R R_H } (\bar{Q}_{\bar f} \bar{Q}_{\bar g'} - \bar{Q}_{\bar g} \bar{Q}_{\bar f'}) +\bar{\mathcal{K}}_{\bar f,\bar g} \label{nonlineartt}
\ee
where $\mathcal{K}_{f,g}$ is the central extension in \eqref{delgQf}, and $\bar{ \mathcal{K}}_{\bar f,\bar g}$ is its right-moving counterpart. For the left-right commutator we find 

\be\label{non-linearttmixed}
\{Q_f,\bar Q_{\bar g}\} = \frac{2\mu}{2\pi R R_H} (R_u Q_{f'} \bar Q_{\bar g} + R_v Q_f \bar Q_{\bar g'} )
\ee
It is useful to rewrite these charges and their commutators in terms of a Fourier basis, where $f_n = e^{i n u/R_u}$ and $\bar f_n = -e^{-i n v/R_v}$. We find 

\be
i \{ Q_m, Q_n \} = \frac{1}{R_u} (m-n) Q_{m+n} + \frac{\mu^2 H_R}{\pi^2 R R_H R_u} (m-n) Q_m Q_n + \frac{c\, m^3}{12 R_u^2} \d_{m+n}\nonumber
\ee

\be
i \{ \bar Q_m, \bar Q_n \} = \frac{1}{R_v} (m-n) \bar Q_{m+n} + \frac{\mu^2 H_L}{\pi^2 R R_H R_v} (m-n)\bar  Q_m \bar  Q_n + \frac{c \, m^3}{12 R_v^2} \d_{m+n}\nonumber
\ee

\be
i \{  Q_m, \bar Q_n \} = - \frac{\mu (m-n)}{\pi R R_H} Q_m \bar Q_n
\ee
The semi-classical commutation relations are obtained by the replacement $i   \{\;, \} \r \hbar^{-1} [\; , ]$. However, we expect this to yield only the first term in the expansion of the full quantum commutator with respect to $\hbar$, as in principle a Poisson algebra, such as the one we find here, can contain arbitrary powers of the generators; in particular, in the full quantum theory the ordering of the generators that appear in the non-linear correction  would need to be specified. This issue is further discussed in \cite{clsqsymmttb}.

We would now like to match this algebra with the symmetry algebra of $T\bar T$ - deformed CFTs, which was argued in \cite{Guica:2021pzy} to be Virasoro $\times$ Virasoro in a certain basis.  For this, it is interesting to work out the algebra of the rescaled charges $Q_{R_u f}$. Remember that, due to the integrability constraint, the variation of this charge was not simply equal to $R_u \d Q_f$, which implies that the algebra of these charges will be different. The main difference is that now the value \eqref{altsolcLf} for the integration constants $c_{\L_f}, c_{\bar \L_{\bar f}}$ when plugged into \eqref{eqdcuv} with $\d_\chi c_{\L, \bar \L} =0$ sets $\d_\chi c_{u,v} =0$. This new solution affects the charge variation,  which now reads

\be
\d_{R_u g} Q_{R_u f} = R_u^2 Q_{f_p  g_p'-  g_p f_p'} - \frac{R_u^2 \ell}{16 \pi G} \oint du f_p g_p''' + w_g R_u Q_f + Q_f \d_g R_u
\ee
Since $\d_g R_u = - w_g R_u$, the last two terms cancel and we are left with  a Virasoro algebra with the same central extension as the undeformed CFT,  precisely reproducing the prediction of \cite{Guica:2021pzy}. It is to be expected that this algebra will not receive further quantum corrections. In the semiclassical limit, the Virasoro generators are simply rescaled versions of the  ones in the Fourier basis by a factor of $R_u$; however, their relation could  become more complicated in the full quantum theory.   

Note that from a field-theoretical perspcetive, the fact that multiplying the generators by a factor of the field-dependent radius, which depends on the Hamiltonian, changes the algebra of the corresponding generators is trivial. However, in the covariant phase space formalism, this multiplicative factor is simply seen as a $c$-number, and it is the integrability of the conserved charges that ensures that the correct algebra of the rescaled generators is reproduced, via the subtle contributions of integration constants that were constrained by integrability, as we explained.

Finally, let us note that it is not clear whether the representation theorem -  which in the most general situation that is currently known  links the variation of the charges  to the modified Lie bracket  \eqref{modbr} of the associated vector fields -   holds in our case. The reason is that the modified Lie bracket \eqref{modbr}, when applied to the vector fields \eqref{xiV}, does not yield a vector whose winding is constant. While it may still be the case that the charges associated with this more non-trivial vector only depend on its periodic part, a separate analysis would be needed to ascertain it. The more likely scenario is that the modified Lie bracket needs to be modified even further in order to hold in these situations. In appendix \eqref{commrepthm} we discuss an additonal modification that appears to be needed, independently of the winding.

%

%

\section{Perturbative analysis around constant backgrounds \label{pertanctb}}
In this section, we particularize our non-linear expressions for the field-dependent coordinates and allowed diffeomorphisms in  the spacetimes dual to double-trace $T\bar T$-deformed CFTs
to 
%
%
constant backgrounds and small perturbations thereof, in order to facilitate the comparison with the analogous quantities in the asymptotically linear dilaton backgrounds.  We also re-compute the charge algebra  from the ground up using these perturbative expressions, in order to mimick the analogous computations in the asymptotically linear dilaton backgrounds; the details of the two sets of computations turn out to be extremely similar. Finally, we check that the representation theorem - which is needed to determine the allowed diffeomorphisms of the perturbed asymptotically linear dilaton specetimes - does hold in the related double-trace $T\bar T$ dual at least to leading order in the perturbation, so we can still trust it for the purposes of this article.

To avoid confusion, in this section the constant backgrounds and their associated field-dependent coordinates  will be denoted as $\L^{(0)}, \bar \L^{(0)} $ and, respectively, $u^{(0)}, v^{(0)}$, whereas the superscript is dropped in the main body of the paper.

\label{appendixB1}
\subsection{Perturbative expansion of the symmetry generators}

\subsubsection*{Constant backgrounds}

 When $\L, \bar \L$ are constant, the equations \eqref{deffdcoord} are trivially solved for $u^{(0)},v^{(0)}$ in terms of the fixed $T\bar T$ coordinates

\be
u^{(0)} = \frac{U+\mu \bar \L^{(0)} V}{1-\mu^2 \L^{(0)} \bar \L^{(0)}} \;, \;\;\;\;\;\; v^{(0)} = \frac{V+ \mu \L^{(0)} U}{1-\mu^2 \L^{(0)} \bar \L^{(0)}}
\ee
In terms of  the natural  parametrization \eqref{ttbpar}   of the linear dilaton background, we find

\be
u^{(0)} =  \frac{(p+\b)U + 2 \mu L_v V}{2p}  \;, \;\;\;\;\; v^{(0)} = \frac{(p+\b)V + 2 \mu L_u V}{2p} \label{fdepcoordctb}
\ee
The allowed diffeomorphisms $\xi_{f,\bar f}$ upon such a background take the form 

\be
\xi^U = f(u^{(0)}) + \mu \bar \L^{(0)}\,  ( \bar f (v^{(0)}) -\bar f_0 )\;, \;\;\;\;\; \xi^V = \bar f (v^{(0)}) + \mu \L^{(0)}\,( f (u^{(0)}) -f_0) \label{alleta}
\ee
and $\xi^\rho = \rho (f'+\bar f')$, where we have specifically made the choice $c_{\L_h} =0$, which amounts to subtracting the Fourier zero modes $f_0$, $\bar f_0$  of the two functions, i.e. we are considering the diffeomorphisms associated with the ``unrescaled'' generators of the charge algebra. In addition, we have  chosen to ignore
the effect of the radial component of the diffeomorphism on $\eta^{U,V}$ by dropping all the terms proportional to $\ell/(8\pi G)$ from the diffeomorphism, which amounts to discarding the contributions proportional  to  the central charge. The reason behind this choice is that the radial component of the diffeomorphism in the spacetime dual to $T\bar T$ - deformed CFTs does not match the radial component of the Einstein radial gauge diffeomorphism in the asymptotically linear dilaton backgrounds, so a comparison cannot be made anyways. In addition, in  the    asymptotically linear dilaton backgrounds we set $F_r =0$
%
, which significantly simplifies the $\O(\e)$ computations. 


 Note also that since the functions are periodic to this order, there is no winding that needs to be compensated and the change of coordinates $U \r U' = U + \e \xi^U$, $V \r V' = V + \e \xi^V$ is well-defined. Another way to see this is that the extra winding in \eqref{windingffb} vanishes because $Q_{f'},\bar{Q}_{\bar{f}'}$ are zero for the constant background.

\subsubsection*{Perturbed backgrounds}

Let us now perturb the constant background by a diffeomorphism $\eta_{h, \bar h}$ of the form \eqref{alleta}, labeled by two periodic functions $h(u)$ and $\bar h (v)$. The $T\bar T$ coordinates change as

\be
U \r U' = U + \e \, \eta^U \;, \;\;\;\;\;\;\; V \r V' = V + \e \, \eta^V \label{shiftUVpert}
\ee
From the active point of view, acting with this diffeomorphism changes the  values of $\L,\bar{\L}$. Particularizing \eqref{deltvv} to  constant parameters and dropping the $f''',\bar{f}'''$ terms since they are proportional to $\ell/(8\pi G)$, we obtain
\be
\L(u) = \L^{(0)} + 2\e \L^{(0)} h' (u^{(0)}) + \O(\e^2) \;, \;\;\;\;\; \bar \L(v) =  \bar \L^{(0)} + 2 \e \bar \L^{(0)} \bar h' (v^{(0)}) +\O(\e^2)
\ee 
%
%
%
With these expressions for the parameters, we solve again the equations for the field-dependent coordinates up to $\O(\e)$
\be
u= u^{(0)} + \e\, u^{(1)} = \frac{U+ 2\mu \e \bar \L^{(0)} (\bar h - \bar h_0)+\mu \bar \L^{(0)} (V+2\mu \e \L^{(0)} (h -h_0) )}{1-\mu^2 \L^{(0)} \bar \L^{(0)}} \nonumber \ee
\be
 v=v^{(0)} + \e \, v^{(1)} = \frac{V+2\mu \e \L^{(0)} (h-h_0) + \mu \L^{(0)} (U+ 2\mu \e \bar \L^{(0)} (\bar h-\bar h_0))}{1-\mu^2 \L^{(0)} \bar \L^{(0)}}
\ee
where the zero mode of $h, \bar h$ has been removed because for $h, \bar h$ constant, the transformations\eqref{shiftUVpert}  are simply isometries of the background, which do not affect $\L^{(0)}, \bar \L^{(0)}$, and thus one should not have to change the definition of the field-dependent coordinate. This also agrees with
our general rule of setting to zero the integration constants $c_{\L}, c_{\bar \L}$. The allowed diffeomorphisms upon this background take the form 
\be
\xi^U = f(u) +\mu \int^v \bar \L (v)\bar f'  = f(u^{(0)} + \e u^{(1)} )+\mu \bar \L^{(0)} ( \bar f (v^{(0)} + \e v^{(1)})-\bar f_0) +2\mu \e \bar \L^{(0)} \int^{v^{(0)}} \!\!\!\! \bar h' \bar f'  \nonumber
\ee
\be
\xi^V = \bar f (v) +\mu \int^u \L(u) f'  = \bar f (v^{(0)} + \e v^{(1)}) +\mu  \L^{(0)} ( f(u^{(0)} + \e u^{(1)} )-f_0) +2\mu \e \L^{(0)} \int^{u^{(0)}}\!\!\!\! h' f' \label{difper2}
\ee
For the comparison with the asymptotically linear dilaton background, it is useful  to write down explicitly the $\O(1)$ and $\O(\e)$ terms in the expansion of these diffeomorphisms as in \eqref{pexpdiff}, where $\xi^{(0)}$ is given by \eqref{alleta} and $\xi^{(1)}$ reads

\begin{align}
\xi^{(1)U}&=\frac{2\mu\bar{\L}^{(0)}}{1-\mu^2\L^{(0)}\bar{\L}^{(0)}}\bigg[f'(\bar{h}-\bar{h}_0+\mu\L^{(0)}(h-h_0))+\mu\L^{(0)}\bar{f}'(h-h_0+\mu\bar{\L}^{(0)}(\bar{h}-\bar{h}_0))\bigg]+2\mu\bar{\L}^{(0)}\!\!\int^{v^{(0)}}\!\!\!\!\!\! \bar{h}'\bar{f}'\nonumber\\
\xi^{(1)V}&=\frac{2\mu\L^{(0)}}{1-\mu^2\L^{(0)}\bar{\L}^{(0)}}\bigg[\bar{f}'(h-h_0+\mu\bar{\L}(\bar{h}-\bar{h}_0))+\mu\bar{\L}^{(0)}f'(\bar{h}-\bar{h}_0+\mu\L^{(0)}(h-h_0))\bigg]+2\mu\L^{(0)}\!\!\int^{u^{(0)}}\!\! \!\!\!\!h'f' \label{vectorexp}
\end{align}
It is important to note that the functions $f, \bar f$ above do have winding, contrary to the homonymous functions in the main text, which are purely periodic. As  per our general discussion, this explicit winding is determined by the requirement 
 the periodicities of the $U,V$ coordinates be fixed. Given that the winding only appears at subleading order in perturbation theory, it appears useful to separate the diffeomorphisms \eqref{xiV} into  two parts

\be
\xi_{f,\bar f} = \xi_{(p)} + \xi_{(w)}
\ee
one determined by the periodic part of the functions $f,\bar f$, and  the other purely associated with the winding part, which we will sometimes call a ``compensating diffeomorphism'' 
\be
\xi^U_{(w)} = w_f u + w_{\bar f} (u-U)  
\;, \;\;\;\;\; \xi^V_{(w)} = w_{\bar f}\, v +  w_f (v-V)\;, \;\;\;\;\;\; \xi^\rho_{(w)} = \rho (w_f + w_{\bar f})
\ee
where the coefficients $w_{f,\bar f}$ are  fixed as in  \eqref{windingffb}. For the perturbative analysis at hand,  $\xi_{(p)}$ simply corresponds to \eqref{difper2} with the functions taken to be periodic and

\be
\xi^U_{(w)} = w_f u^{(0)} + \mu  w_{\bar f}\,\bar \L^{(0)}   v^{(0)} \;, \;\;\;\;\;\; \xi^V_{(w)} = w_{\bar f} v^{(0)} +  \mu w_f  \L^{(0)} \,  u^{(0)}
\ee
where 
the windings are given by 
\begin{align}
w_f&=\frac{2\mu\bar{\L}^{(0)}(-\oint dv^{(0)} \bar{f}'\bar{h}'+\mu\L^{(0)} \oint du^{(0)} h'f')}{2\pi R (1+\mu\bar{\L}^{(0)})}\, \e & w_{\bar{f}}&=\frac{2\mu\L^{(0)}(-\oint du^{(0)} f'h'+\mu\bar{\L}^{(0)} \oint dv^{(0)}\bar{h}'\bar{f}')}{2\pi R (1+\mu\L^{(0)})} \, \e \label{pertwind}
\end{align}
This is in perfect agreement with  \eqref{windingffb} evaluated at this order, where 

\be
Q_f' =  \e \L^{(0)} \oint du^{(0)}\,  h' f'\;, \;\;\;\;\;\; \bar Q_{\bar f'} = -   \e \bar \L^{(0)} \oint dv^{(0)}  \, \bar  h' \bar f'
\ee 
and we use the fact that  $ H_L= \L^{(0)} \pi  R_u$, $H_R= \bar \L^{(0)} \pi  R_v$, which follows from \eqref{holoHLR}, as well as $R_H/R_v = 1+\mu \bar \L^{(0)}$, etc. 
%

%
%
%

\subsection{The perturbative charge algebra \label{pertchalg}}

In this appendix, we compute the charge algebra  corresponding to the asymptotic vector fields \eqref{vectorexp} up to  $\O(\e)$.
For the comparison with the asymptotically linear dilaton analysis, it is useful   to split the computation of the charge difference at $\O(\e)$ according to the splitting of the vector fields into periodic and winding parts
\begin{align}
\delta_{\chi_{(p)}+\chi_{(w)}}Q_{\xi_{(p)}+\xi_{(w)}}&=\delta_{\chi_{(p)}}Q_{\xi_{(p)}}+\delta_{\chi_{(w)}}Q_{\xi_{(p)}}+\delta_{\chi_{(p)}}Q_{\xi_{(w)}}+\delta_{\chi_{(w)}}Q_{\xi_{(w)}}
\end{align}
Since the compensating diffeomorphisms are each at least $\O(\e)$, the last term does not contribute to the order we are interested in. 

Let us first compute  the  algebra of two left-moving diffeomorphisms, by setting all the periodic functions of $v^{(0)}$ to zero. As before, the diffeomorphism $\xi$ is given by the periodic function $f$ and $\chi$ by $g$. We start by computing the charge difference $\delta_{\chi_{(p)}}Q_{\xi_{(p)}}$. The zero modes $h_0,\bar{h}_0$ do not contribute to the result. 
After various integrations by parts, we checked that the function $\bar{h}$ and its derivatives do not contribute to the result and we are left with
\begin{align}
\delta_{\chi_{(p)}}Q_{\xi_{(p)}}&=\frac{r_u\L^{(0)}}{1-\mu^2\L^{(0)}\bar{\L}^{(0)}}\bigg[-2\mu^2\L^{(0)}\bar{\L}^{(0)}\bigg(\oint d\s g'\bigg(\int^{u^{(0)}}\!\!\!\!\!\! f'h'-u^{(0)}\frac{\oint d\s f'h'}{2\pi R}\bigg)+\frac{\oint d\s f'h'}{2\pi R}\oint d\s u^{(0)}g' \bigg)+\nonumber \\
&\hspace{-1cm}+\mu^2\L^{(0)}\bar{\L}^{(0)}\bigg(f_0\oint d\s g'h' + g_0\oint d\s f'h'\bigg)+\oint d\s\bigg((1-\mu^2\L^{(0)}\bar{\L}^{(0)})fg'-(1+\mu^2\L^{(0)}\bar{\L}^{(0)})gf'\bigg)h'\bigg]\label{perper}
\end{align}
where we split the first term into an integral of a  periodic function and one of a non-periodic one. We  integrate by parts the periodic part
\begin{align}
\delta_{\chi_{(p)}}Q_{\xi_{(p)}}&=r_u\L^{(0)}\oint d\s(fg'-gf')h'-\frac{r_u\L^{(0)}}{1-\mu^2\L^{(0)}\bar{\L}^{(0)}}2\mu^2\L^{(0)}\bar{\L}^{(0)}\frac{\oint d\s f'h'}{2\pi R}\oint d\s u^{(0)}g'+\nonumber\\
&+\frac{r_u\L^{(0)}}{1-\mu^2\L^{(0)}\bar{\L}^{(0)}}\mu^2\L^{(0)}\bar{\L}^{(0)}\bigg(f_0\oint d\s g'h'-g_0\oint d\s f'h'\bigg)
\end{align}
It is useful to rewrite the expression above using the 
winding, $w_f=\frac{2\mu^2\L^{(0)}\bar{\L}^{(0)}}{1-\mu^2\L^{(0)}\bar{\L}^{(0)}}\frac{\oint d\s h'f'}{2\pi R}$, which we particularized for $\bar{f}=0$. We obtain
\begin{align}
\delta_{\chi_{(p)}}Q_{\xi_{(p)}}&=r_u\L^{(0)}\oint d\s(fg'-gf')h'-r_u\L^{(0)}w_f \oint d\s u^{(0)}g'+ r_u\L^{(0)}\frac{1}{2}2\pi R(w_g f_0-w_f g_0)=\nonumber\\
&=Q_{fg'-gf'}+w_g Q_f^{(0)}-w_f Q_g^{(0)}-r_u \L^{(0)}w_f\oint d\s u^{(0)}g'
\end{align}
where in the last line we recognized the charges at $\O(\e)$ and $\O(1)$ respectively. Next, we computed the charge difference $\delta_{\chi_{(w)}}Q_{\xi_{(p)}}$, which vanishes
\begin{align}\label{perwind}
\delta_{\chi_{(w)}}Q_{\xi_{(p)}}&=w_g\L^{(0)}\frac{1+\mu\bar{\L}^{(0)}}{1-\mu^2\L^{(0)}\bar{\L}^{(0)}}\oint d\s (f-f_0)=0
\end{align}
because it is the integral on the circle of a periodic function without zero mode. Finally, we computed the charge difference $\delta_{\chi_{(p)}}Q_{\xi_{(w)}}$, which can be written as
\begin{align}
\delta_{\chi_{(p)}}Q_{\xi_{(w)}}&=w_f\L^{(0)}\frac{1+\mu\bar{\L}^{(0)}}{(1-\mu^2\L^{(0)}\bar{\L}^{(0)})^2}\oint d\s (U+\mu\bar{\L}^{(0)}V)g'=w_f\L^{(0)}r_u\oint d\s u^{(0)} g'
\end{align}
This contribution cancels exactly the non-periodic part of $\d_{\chi_{(p)}}Q_{\xi_{(p)}}$, leading to a well-defined result for the total charge difference
\begin{align}
\{Q_{\xi},Q_{\chi}\}=\d_{\chi}Q_{\xi}&=Q_{fg'-gf'}+w_g Q_f^{(0)}-w_f Q_g^{(0)}
\end{align}
Clearly, this matches the final double-trace result linearized around constant backgrounds.

We should point out that removing the zero modes from the vector fields as in \eqref{alleta} is of utmost importance in obtaining the correct \emph{non-linear} charge algebra. Had we not done so, it would have amounted to choosing constants in \eqref{eqdcuv} $c_{\L_f} = \L^{(0)} f_0$ and $\d_g c_\L = 2 \L^{(0)} g_0$. The na\"{i}ve calculation would have seemed to work, 
as the integrals of non-periodic functions cancel consistently with each other. The remaining terms yield however an \emph{incorrect} result that we can recover directly by setting to zero  $f_0$, $g_0$ in \eqref{perper} and \eqref{perwind} 
\begin{align}
\d_{\chi}Q_{\xi}&=r_u\L^{(0)}\oint d\s(fg'-gf')h'+2\pi R r_u  \L^{(0)}(w_g f_0-w_f g_0)+\O(\e^2)=\nonumber\\
&=Q_{fg'-gf'}+ 2 w_g Q_{f}^{(0)} -2 w_f Q_{g}^{(0)}+\O(\e^2)
\end{align}
Finally, by removing twice the zero modes, which corresponds to the choice $c_{\mathcal{L}_f}=2\L^{(0)}f_0$ and $\delta_g c_{\L}=0$, we obtain, letting $f_0\mapsto 2 f_0, g_0\mapsto 2 g_0$ in \eqref{perper}, \eqref{perwind}
\begin{align}
\d_{\chi}Q_{\xi}&=Q_{fg'-gf'}+\O(\e^2)
\end{align}
Multiplying the charges by $R_u$ as dictated by integrability and taking the derivatives with respect to $u/R_u$, we find the Witt algebra of the rescaled generators\footnote{Our analysis can no longer see the central charge because we have set the responsible terms to zero in the asymptotic diffeomorphisms \eqref{alleta} used in this appendix. }.

The algebra of two right-moving generators works in an exactly analogous way. In order to obtain the mixed commutator, we repeat the previous computation for one ``left" and one ``right" vector field. The periodic-periodic charge difference is
\begin{align}
\d_{\bar{\chi}_{(p)}}Q_{\xi_{(p)}}&=\frac{\mu\L^{(0)}\bar{\L}^{(0)}}{(1-\mu^2\L^{(0)}\bar{\L}^{(0)})^2}\bigg[-2(1+\mu\L^{(0)})\oint d\s \bar{g}'\bigg(\int^{u^{(0)}} f'h'-u^{(0)}\frac{\oint d\s f'h'}{2\pi R}\bigg)
+ \nonumber\\
&\hspace{-1.4cm}+f_0(1+\mu\L^{(0)})\oint d\s \bar{g}'\bar{h} +(1+\mu\bar{\L}^{(0)})\oint d\s (\bar{g}_0-2\bar{g})f'h'-2(1+\mu\L^{(0)})\frac{\oint d\s f'h'}{2\pi R}\oint d\s \bar{g}' u^{(0)}\bigg]
\end{align}
After integrating by parts the first term and recognizing the windings, we are left with
\begin{align}
\d_{\bar{\chi}_{(p)}}Q_{\xi_{(p)}}&=2\pi R\frac{w_{\bar{f}}\bar{g}_0\bar{\L}^{(0)}(1+\mu\L^{(0)})-w_g f_0 \L^{(0)}(1+\mu\bar{\L}^{(0)})}{2(1-\mu^2\L^{(0)}\bar{\L}^{(0)})}+w_{\bar{f}}\bar{\L}^{(0)}\frac{r_v^2}{r_u}\oint d\s u^{(0)} \bar{g}'\label{incons}
\end{align}
Next, we computed the charge difference 
\begin{align}\label{chdif2}
\d_{\bar{\chi}_{(w)}}Q_{\xi_{(p)}}&=w_g\frac{1+\mu\bar{\L}^{(0)}}{1-\mu^2\L^{(0)}\bar{\L}^{(0)}}\L^{(0)}2\pi R f_0
\end{align}
The last contribution to the charge difference is given by
\begin{align}
\d_{\bar{\chi}_{(p)}}Q_{\xi_{(w)}}&=-w_{\bar{f}}\bar{\L}^{(0)} r_v \oint d\s v^{(0)} \bar{g}'
\end{align}
Adding to it the last term in \eqref{incons} and splitting the field-dependent coordinates into a part proportional to $\s$ and one proportional to $t$, we see that the terms proportional to $\s$ cancel among themselves. The ones proportional to $t$ do not contribute either since
\begin{align}
w_{\bar{f}}\bar{\L}^{(0)}r_v^2\oint d\s \bar{g}'\bigg(\frac{u^{(0)}}{r_u}-\frac{v^{(0)}}{r_v}\bigg)=w_{\bar{f}}\bar{\L}^{(0)}r_v^2 \frac{2(1-\mu^2\L^{(0)}\bar{\L}^{(0)})}{(1+\mu\L^{(0)})(1+\mu\bar{\L}^{(0)})} t\oint d\s \bar{g}'=0
\end{align} 
The remaining terms in \eqref{incons} and \eqref{chdif2} give
\begin{align}
\d_{\bar{\chi}}Q_{\xi}&=\frac{1}{2}2\pi R(w_g f_0 r_u \L^{(0)}+w_{\bar{f}} g_0 r_v \bar{\L}^{(0)})=w_g Q^{(0)}_{f}-w_{\bar{f}}\bar{Q}^{(0)}_{\bar{g}}
\end{align}
which is the double-trace result linearized around constant backgrounds.

In order to see all the non-linear terms of the $T\bar{T}$ algebra, one needs to expand the final results up to $\O(\e^2)$. Such terms are proportional to the windings, which are at least $\O(\e)$, so one can na\"{i}vely expect to obtain them from the vector fields expanded up to $\O(\e)$ only. However, this is not true, as one can see from the expansion \eqref{difper2}, which is valid for the full functions $f,\bar{f}$ and not only their periodic parts. Particularizing the functions to $f=w_f u^{(0)},\bar{f}=w_{\bar{f}} v^{(0)}$, we obtain an $\O(\e^2)$ periodic contribution to the vector fields, which depends on the windings up to $\O(\e)$ and enters in the non-linear charge algebra. Thus, we conclude that without the knowledge of the vector fields up to $\O(\e^2)$, we cannot obtain all the non-linear terms in the charge algebra.

\subsection{Comments on the representation theorem \label{commrepthm}}

As discussed at the end of the previous appendix, the representation theorem in its most general currently known form \cite{Barnich:2010eb,Barnich:2010xq}
\be
\d_\chi Q_\xi = Q_{[\xi,\chi]_*}
\ee
where $[\xi,\chi]_*$ is the modified Lie bracket \eqref{modbr} of the two vector fields, may not hold for the spacetimes dual to double-trace $T\bar T$ - deformed CFTs. Given the similarities between the asymptotic symmetries of these spacetimes and those of asymptotically linear dilaton backgrounds, one may worry that the representation theorem could be violated in the  latter case, which may in turn invalidate our use of this theorem in section \eqref{alloweddiffs} to fix the allowed diffeomorphisms upon the perturbed backgrounds. 
In this appendix, we show that the representation theorem does hold perturbatively to leading order in the spacetimes dual to double-trace $T\bar T$ - deformed CFTs, which is sufficient for the purposes of this article.


More precisely,  at $\O(1)$ the winding terms vanisgh and the vector fields are given only by periodic functions. Thus, the problem with the representation theorem due to winding terms is not present at this order. In this section, we show that standard modified Lie bracket is almost enough for the theorem to hold for the ``unrescaled" vector fields; an additional modification, as imposed by integrability of charges, is required for the closure of the algebra. In the case of the ``rescaled" vector fields, which lead to a Virasoro $\times$ Virasoro charge algebra, the representation theorem is automatically satisfied. In both cases, for the computation of the modified Lie bracket, it is essential to take into consideration the origin of the terms appearing in the vector fields, which can only be inferred from knowledge of the full non-linear expressions in double-trace $T\bar T$ - deformed CFTs.

\subsubsection*{Representation theorem at $\O(1)$ for the ``unrescaled" generators}
We compute the modified Lie bracket of vector fields at $\O(1)$. We restrict to left vector fields by setting all functions of $v$ to zero. Even though we set to zero the radial component of the vector fields and  we drop the terms proportional to $\ell/(8\pi G)$ as explained, we write also the full result at the end of the computation.

For two vector fields of the form
\begin{align}
\xi_{f}&=f\p_U+\mu[\int^{u^{(0)}}\!\!\!\!\L^{(0)} f']_{\cancel{zm}}\p_V=f\p_U+\mu\L^{(0)}(f-f_0)\p_V
\end{align}
the Lie bracket is given by
\begin{align}
[\xi_f,\xi_g]_{L.B.}&=\bigg[\frac{1+\mu^2\L^{(0)}\bar{\L}^{(0)}}{1-\mu^2\L^{(0)}\bar{\L}^{(0)}}(fg'-gf')+\frac{\mu^2\L^{(0)}\bar{\L}^{(0)}}{1-\mu^2\L^{(0)}\bar{\L}^{(0)}}(g_0 f'-f_0 g')\bigg](\p_U+\mu\L^{(0)}\p_V)
\end{align}
The standard modification of the Lie bracket, taking into account the field-dependence in the vector fields, is given by \eqref{modbr}:
\begin{align}
\d_g\xi_f&=f'\d_g u^{(0)}(\p_U +\mu\L^{(0)} \p_V)+2\mu\L^{(0)} \int^{u^{(0)}}\!\!\!\!\! g'f'\p_V
\end{align}
The last term, which comes from the intrinsic variation of the parameter in the integral $[\int^{u^{(0)}}\!\!\!\!\!\!\!\d_g\L_{int}^{(0)}f']_{\cancel{zm}}$, is symmetric under $f\leftrightarrow g$, so it will drop out from the modified Lie bracket:
\begin{align}
[\xi_f,\xi_g]_*&=[\xi_f,\xi_g]_{L.B.}-(\d_f\xi_g-\d_g\xi_f)=[\xi_f,\xi_g]_{L.B.}+(f'\d_g u^{(0)} -g'\d_f u^{(0)})(\p_U+\mu\L^{(0)}\p_V)
\end{align}
We can already predict, before plugging in the expression for $\d_g u^{(0)}$, that the algebra does not close with this standard modified Lie bracket, because the $U$ and $V$ components of the resulting vector field are proportional. This is not the case for our $\O(1)$ vector fields, for which the $V$ component differs by a subtraction of the zero mode. Let us see that indeed this zero mode is not zero, preventing the algebra to close.

For constants backgrounds, $\d_g c_u=-\frac{\mu^2\L^{(0)}\bar{\L}^{(0)}}{1-\mu^2\L^{(0)}\bar{L}^{(0)}}g_0$, which implies:
\begin{align}\label{varucoord}
\d_g u^{(0)}&=\frac{2\mu^2\L^{(0)}\bar{\L}^{(0)}}{1-\mu^2\L^{(0)}\bar{L}^{(0)}}(g-g_0)
\end{align}
Plugging in this expression, we find:
\begin{align}
[\xi_f,\xi_g]_*&=\bigg[\bigg(f+\frac{\mu^2\L^{(0)}\bar{\L}^{(0)}}{1-\mu^2\L^{(0)}\bar{\L}^{(0)}}f_0\bigg)g'-\bigg(g+\frac{\mu^2\L^{(0)}\bar{\L}^{(0)}}{1-\mu^2\L^{(0)}\bar{\L}^{(0)}}g_0\bigg)f'\bigg](\p_U+\mu\L^{(0)}\p_V)
\end{align}
We can see now that the zero mode of the $V$ component, as computed from the Lie bracket and its standard modification, is $\mu\L^{(0)}(fg'-gf')_{zm}\neq 0$. For charge integrability, we need to remove this zero mode, by a supplementary modification of the Lie bracket:
\begin{align}
[\xi_f,\xi_g]_{**}&=[\xi_f,\xi_g]_{*}-([\xi_f,\xi_g]_{*}^V)_{zm}\p_V
\end{align}
With this modification, the algebra closes:
\begin{align}
[\xi_f,\xi_g]_{**}&=[(f-\d_f c_u)g'-(g-\d_g c_u)f'](\p_U+\mu\L^{(0)}\p_V)-\mu\L^{(0)}((f-\d_f c_u)g'-(g-\d_g c_u)f')_{zm}\p_V\nonumber\\
&=\xi_{(f-\d_f c_u)g'-(g-\d_g c_u)f'}=\xi_{fg'-gf'}+\d_g c_u\xi_{f'}-\d_f c_u\xi_{g'}
\end{align}
Mapping the vector fields to charges, this corresponds to:
\begin{align}
\{Q_{f},Q_{g}\}^{(0)}&=Q_{fg'-gf'}^{(0)}+\d_g c_u Q^{(0)}_{f'}-\d_f c_u Q^{(0)}_{g'}
\end{align}
Since on constants backgrounds $Q^{(0)}_{f',g'}=0$, we are left with the correct result at $\O(1)$:
\begin{align}
\{Q_{f},Q_{g}\}^{(0)}&=Q_{fg'-gf'}^{(0)}
\end{align}
We checked that including the radial components and the terms proportional to $\ell/(8\pi G)$, we obtain:
\begin{align}
[\xi_f,\xi_g]_{*}&=\xi_{(f-\d_f c_u)g'-(g-\d_g c_u)f'}+\mu\bigg(\L^{(0)}((f-\d_f c_u)g'-(g-\d_g c_u)f')-\frac{\ell}{16\pi G}(f''g'-g''f')\bigg)_{zm}\!\!\!\!\!\!\p_V
\end{align}
Hence, all the terms group to give the full vector field we obtained before in the partial computation, and the only modification is an extra ``central charge" contribution to the zero mode of the $V$ component that we need to extract by hand. The final result after the modification is:
\begin{align}
[\xi_f,\xi_g]_{**}&=\xi_{(f-\d_f c_u)g'-(g-\d_g c_u)f'}
\end{align}
leading to the correct charge algebra, up to the central extension that the algebra of vector fields cannot see.

Let us now notice that a naive computation, without taking into consideration the fact  that $\mu\L^{(0)}(f-f_0)$ comes actually from $\mu[\int^{u^{(0)}} \L^{(0)} f']_{\cancel{zm}}$ leads to a wrong result, that needs extra modifications of the Lie bracket:
\begin{align}
\d_g\xi_f&=f'\d_g u^{(0)}(\p_U+\mu\L^{(0)})+\mu (\d_g\L^{(0)}_{int}f-\d_g(\L^{(0)} f_0))\p_V=f'\d_g u^{(0)}(\p_U+\mu\L^{(0)})+\nonumber\\
&+\mu (2\L^{(0)} g'f-\d_g(\L^{(0)} f_0))\p_V
\end{align}
Hence, the modified Lie bracket would give:
\begin{align}\label{wronglie}
[\xi_f,\xi_g]_*&=\bigg[\bigg(f+\frac{\mu^2\L^{(0)}\bar{\L}^{(0)}}{1-\mu^2\L^{(0)}\bar{\L}^{(0)}}f_0\bigg)g'-\bigg(g+\frac{\mu^2\L^{(0)}\bar{\L}^{(0)}}{1-\mu^2\L^{(0)}\bar{\L}^{(0)}}g_0\bigg)f'\bigg](\p_U+\mu\L^{(0)}\p_V)+\nonumber\\
&+\mu\bigg(2\L^{(0)}(fg'-gf')+\d_f(\L^{(0)} g_0)-\d_g(\L^{(0)} f_0)\bigg)\p_V
\end{align}
For closure of the algebra, we need the result to be of the form:
\begin{align}
\mathcal{F}(\p_U+\mu\L^{(0)}\p_V)-\mu\L^{(0)}\mathcal{F}_{zm}\p_V
\end{align}
for some periodic function $\mathcal{F}(u^{(0)})$. The last term needs to be constant, which is not the case for our last term in \eqref{wronglie}, since $\d_f(\L^{(0)} g_0)-\d_g(\L^{(0)} f_0)$ is constant and $fg'-gf'$ is not. Thus, the algebra obtained from this naive modification of the Lie bracket does not close and we need to modify it further by extracting the nonconstant piece. The natural explanation for this modification is that $\mu\L^{(0)}(f-f_0)$ comes from $\mu[\int^{u^{(0)}} \L^{(0)} f']_{\cancel{zm}}$ and $\mu[\int^{u^{(0)}}\d_g\L^{(0)}_{int}f']_{\cancel{zm}}-\mu[\int^{u^{(0)}}\d_f\L^{(0)}_{int}g']_{\cancel{zm}}=0$, so that extra piece should not have been there in the first place.

\subsubsection*{Representation theorem at $\O(1)$ for the ``rescaled"  generators}
For the generators that lead to the Virasoro algebra, the representation theorem is satisfied at $\O(1)$ with the standard modified Lie bracket. Again, we will write explicitly the partial computation with radial component and terms proportional to $\ell/(8\pi G)$ to 0 and state the full result in the end. Let us remember that in this case 
\begin{align}
c_{\L_f}=-\L^{(0)} f_0=-\frac{Q^{(0)}_{f}}{\pi R_u}
\end{align}
Since windings are 0, $\d R_u=0$, so we can ignore the rescaling by the radius at this order because it will not lead to any extra terms, but just to an overall multiplication factor. Thus, we start with the vector fields of the form:
\begin{align}
\xi_f&=f \p_U+\mu\bigg([ \int^{u^{(0)}}\!\!\!\!\L^{(0)} f']_{\cancel{zm}} - \L^{(0)} f_0 \bigg)\p_V=f \p_U+\mu\L^{(0)}(f-2f_0)
\end{align}
However, writing the last term as in the rightmost equation might be misleading, because the origin of the two zero modes is different and this affects how we take their variations. The Lie bracket gives:
\begin{align}
[\xi_f,\xi_g]_{L.B.}&=\bigg[\frac{1+\mu^2\L^{(0)}\bar{\L}^{(0)}}{1-\mu^2\L^{(0)}\bar{\L}^{(0)}}(fg'-gf')+\frac{2\mu^2\L^{(0)}\bar{\L}^{(0)}}{1-\mu^2\L^{(0)}\bar{\L}^{(0)}}(g_0 f'-f_0 g')\bigg](\p_U+\mu\L^{(0)}\p_V)
\end{align}
Let us now compute the modification to the Lie bracket, as before:
\begin{align}
\d_g \xi_f&=f'\d_g u^{(0)}(\p_U+\mu\L^{(0)}\p_V)+\mu[\int^{u^{(0)}}\!\!\!\! \d_g\L^{(0)}_{int}f']_{\cancel{zm}}\p_V-\mu\d_g(\L^{(0)} f_0)\p_V
\end{align}
The term with the primitive again drops out in the modified Lie bracket from antisymmetry. The variation of the field-dependent coordinates is still \eqref{varucoord}. Hence we obtain
\begin{align}
[\xi_f,\xi_g]_*&=(fg'-gf')(\p_U+\mu\L^{(0)}\p_V)-\mu(\d_g(\L^{(0)} f_0)-\d_f(\L^{(0)} g_0))\p_V=\nonumber\\
&=(fg'-gf')(\p_U+\mu\L^{(0)}\p_V)-\frac{\mu}{\pi R_u}(\d_g Q_f^{(0)}-\d_f Q_g^{(0)})\p_V
\end{align}
We want to compare this modified Lie bracket with:
\begin{align}
\xi_{fg'-gf'}&=(fg'-gf')(\p_U+\mu\L^{(0)}\p_V)-\mu\L^{(0)}(fg'-gf')_{zm}+\mu c_{\L_{fg'-gf'}}=\nonumber\\
&=(fg'-gf')(\p_U+\mu\L^{(0)}\p_V)-2\frac{\mu}{\pi R_u}Q^{(0)}_{fg'-gf'}
\end{align}
We see that the two results agree if $\d_g Q_f^{(0)}-\d_f Q_g^{(0)}=2Q^{(0)}_{fg'-gf'}$, which is in agreement with the charge algebra, from which we dropped the central charge:
\begin{align}
\d_g Q_f^{(0)}=Q^{(0)}_{fg'-gf'}
\end{align}
With this condition satisfied by the constant terms, we see that we obtain precisely $[\xi_f,\xi_g]_*=\xi_{fg'-gf'}$. Taking into account the overall multiplication by $R_u$ and considering the derivatives with respect to $u/R_u$, we obtain:
\begin{align}\label{wittresc}
[R_u\xi_f,R_u\xi_g]_*&=R_u\xi_{fg'-gf'}
\end{align}
which leads to the correct charge algebra.

Finally, we performed the same computation for the full $\O(1)$ vector fields. The result \eqref{wittresc} is still valid, now for full vector fields,  while the constraint on the constants receives an extra term
\begin{align}
\d_g Q_f^{(0)}=Q^{(0)}_{fg'-gf'}-\frac{\ell}{16\pi G}\oint du^{(0)} fg'''
\end{align}
which is exactly the full charge algebra at $\O(1)$, including the central charge. With the constraint being automatically satisfied, we conclude that at least to this order, the algebra of vector fields closes with the standard modified Lie bracket.

\end{document}